\documentclass[preprints,article,accept,moreauthors,pdftex]{Definitions/mdpi}
\firstpage{1}
\pubvolume{1}
\issuenum{1}
\articlenumber{0}
\pubyear{2022}
\copyrightyear{2022}
\externaleditor{Academic Editor:  Sergei B. Popov}
\datereceived{4 May 2022}
\dateaccepted{14 June 2022}
\datepublished{16 June 2022}
\hreflink{https://doi.org/10.3390/\linebreak universe8070355} 
\pdfoutput=1

\Title{The Statistical Similarity of Repeating and Non-Repeating Fast Radio Bursts}

\TitleCitation{The Statistical Similarity of Repeating and Non-Repeating Fast Radio Bursts}

\Author{Kongjun Zhang 
 $^{1}$\orcidA{}, Longbiao Li $^{2,}$*\orcidB{},  Zhibin Zhang $^{1,}$*\orcidC{}, Qinmei Li $^{1}$, Juanjuan Luo $^{3}$ and Min Jiang $^{1}$}

\AuthorNames{Kongjun Zhang, Longbiao Li,  Zhibin Zhang, Qinmei Li, Juanjuan Luo and Min Jiang}

\AuthorCitation{Zhang, K.; Li, L.; Zhang, Z.; Li, Q.; Luo, J.; Jiang, M.}

\address{%
$^{1}$ \quad Department of Physics, College of Physics, Guizhou University, Guiyang 550025, China; \mbox{k\_j\_zhang@163.com (K.Z.)}; qinmli@163.com (Q.L.); \mbox{j17385197986@163.com (M.J.)}\\
$^{2}$ \quad School of Mathematics and Physics, Hebei University of Engineering, Handan 056005, China\\
$^{3}$ \quad College of Physics and Electronic Science, Qiannan Normal University, Duyun 55800, China; j\_j\_luo@sina.com (J.L.)}

\corres{Correspondence: lilongbiao@hebeu.edu.cn (L.L.); zbzhang@gzu.edu.cn (Z.Z.);}

\abstract{In this paper, we present a sample of 21 repeating fast radio bursts (FRBs) detected by different radio instruments before September 2021. Using the Anderson--Darling test, we compared the distributions of extra-Galactic dispersion measure ($DM_{\rm E}$) of  non-repeating 
 FRBs, repeating FRBs and all FRBs. It was found that the $ DM_{\rm E}$ values of three sub-samples are log-normally distributed. The $DM_{\rm E}$ of repeaters and non-repeaters were drawn from a different distribution on basis of the Mann--Whitney--Wilcoxon test. In addition, assuming that  the non-repeating FRBs identified currently may be potentially repeators, i.e., the repeating FRBs to be universal and representative, one can utilize the averaged fluence of repeating FRBs as an indication from which to derive an apparent intensity distribution function (IDF) with a power-law index of $a_1=$ $1.10\pm 0.14$ ($a_2=$ $1.01\pm 0.16$, the observed fluence as a statistical variant), which is  in good agreement with the previous IDF of 16 non-repeating FRBs found by Li et al. 
Based on the above statistics of repeating and non-repeating FRBs, we propose that both types of FRBs may have different cosmological origins, spatial distributions and circum-burst environments. Interestingly, the differential luminosity distributions of repeating and non-repeating FRBs can also be well described by a broken power-law function with the same power-law index \mbox{of $-$1.4.}}


\usepackage{longtable}
\usepackage{threeparttable}
\usepackage{booktabs}
\makeatletter
\let\c@lofdepth\relax
\let\c@lotdepth\relax
\makeatother
\usepackage{subfigure}
 \usepackage{epstopdf}
\begin{document}

\section{Introduction}           
\label{sect:intro}

Fast radio bursts (FRBs) are mysterious millisecond-duration radio pulses which occur randomly on the sky (e.g.,~\cite{lorimer2007aBailes,thornton2013aStappers,cordes2019fastChatterjee}). FRBs were first found
 in archived pulsar survey data more than a decade ago~\citep{lorimer2007aBailes}, and more than 600 FRBs were
 reported as of April 2022 (e.g.,~\mbox{\cite{petroff2016frbcatBarr,2021arXiv210604352T,li2021}}).
However, the detection rate of FRBs is considered to be \mbox{$\sim\!\!10^3$--$10^4 \,\rm sky^{-1}\,day^{-1}$~\citep{thornton2013aStappers,
keane2015fastPetroff,law2015aBower,champion2016fivePetroff,oppermann2016euclidean,
rane2016aLorimer,Bhandari2018TheSF,connor2018detecting,patel2018palfa,Shannon2018TheDR,
Farah2019FiveNR,Parent2020FirstDO,2021arXiv210604352T}},
which means that FRBs are not uncommon in the Universe; e.g., 1652 repeating events from FRB 20121102A~\citep{Lidi2021}
and 1863 bursts from FRB 20201124A~\citep{xu2021fast} were recently detected by the Five-hundred-meter Aperture Spherical
radio Telescope (FAST,~\citet{Lidi2018}). The observed dispersion measures ($DM$) 
 are larger than the $DM$ contributions
from the Milk Way ($DM_{MW}$) (except FRB 200428, which was confirmed to be from the Galactic magnetar SGR 1935+2154
\citep{chime2020aAndersen,bochenek2020fast,lin2020no,mereghetti2020integral,li2021hxmt,ridnaia2021peculiar,tavani2021x}),
which suggests an extragalactic, even cosmological origin in \mbox{most cases.}

It should be mentioned that there is no unambiguous physical origin for FRBs now. In general, FRBs' millisecond duration and high $DM$ indicate the high brightness temperature ($10^{32}\,$K -- $3.5 \times 10^{35}\,$ K)~\citep{petroff2019fast} and the corresponding isotropic energy
($E$) released (\mbox{$10^{35} {\rm\, erg}$--$10^{44}$ erg})~\citep{zhang2018fast,ravi2019fast,bochenek2020fast,li2021}. Many progenitor models have been proposed to figure out what FRBs are (for a review, see, e.g.,~\cite{platts2019living}),
such as mergers of compact \mbox{objects~\citep{yamasaki2018repeatingTotani},}
flaring magnetars~\citep{popov2013millisecond},
young magnetars in supernova remnants~\citep{metzger2019fastMargalit},
collisions between neutron star/magnetar and asteroids~\citep{geng2015fast,dai2016repeating,xiao2020double,geng2020frb,geng2021repeating},
collisions between episodic magnetic \mbox{blobs~\citep{li2018model},}
and massive black hole model~\citep{zhang2018frb}.
However, none of the current models can explain all the observational properties of FRBs.
Fortunately, great progress in observations can help to constrain the progenitor model of FRBs; there has been a great leap forward in the research of FRBs.
For example, the periodic activity of repeating FRB 20180916B suggests that
the source is modulated by the orbital motion of a binary system~\citep{liyang2021}. Additionally, the discovery of Galactic FRB 200428 indicates the origin of a magnetar (e.g.,~\cite{yang2020}). In addition, some other models, e.g., the precession like a gyroscope model
\citep{levin2020precessing,tong2020periodicity,yang2020,li2021emission,sridhar2021periodic}
and the spin period of isolated neutron star/magnetars model~\citep{zhang2020unexpected,xu2021periodic}, can also work.

\textls[-15]{With the increase in FRBs monitoring by many radio telescopes, such as Very Large Array
(VLA,~\citet{Thompson19080vla}), Australian Square Kilometer Array Pathfinder (ASKAP)~\citep{bannister2017detection},} Canadian Hydrogen Intensity Mapping Experiment (CHIME,~\citet{chime2018}) and FAST,
it was found that some events are apparently ``non-repeating,''
i.e., they did not repeat within a monitoring period.
On the other hand, some sources have been repeating (e.g.,~\cite{spitler2016aScholz,
chime2019aAmiriBandura,chime2019chimeAndersenBandur,kumar2019faintShannon,luo2020diverse}),
which led to the successful identification of host galaxies and the precise mensuration of redshifts.
\citet{li2021} analyzed 133 FRBs, including 110 non-repeating and 23 repeating ones, and proposed to
classify FRBs into short and long groups according to pulse duration less than 100 ms or not.
Interestingly, they found long FRBs are on average more energetic than short ones about two orders of magnitude.
Moreover, they pointed out that FRBs could be used as a standard candle because peak luminosity
becomes weakly dependent of the cosmological distance at higher redshift.
 Some observational properties of FRBs are similar to those of short and
long gamma-ray bursts~\citep{zhang2008analysis,zhang2018spectrum,zhang2020spectral}.
However, it is still uncertain whether this classification of FRBs is derived
from the intrinsic physical characteristics, and whether repeating or ``non-repeating''
is due to observational selection bias---e.g., the monitor is not in the most active window for the ``non-repeating'' FRBs.
Although the astrophysical origins of repeating and non-repeating FRBs
are considered to be different
(e.g.,~\cite{luo2018onLee,caleb2019areStappers,li2019frb,zhang2020the}),
another viewpoint that most FRBs may be repeating sources has also been proposed
\citep{ravi2019aCatha,lu2020implications,luo2020frb,li2021}.
It is now accepted that there are usually two types of FRBs, i.e.,  repeating FRBs and non-repeating FRBs.
It is worth noting that the luminostiy function can place important constraints on the physical origins of FRBs and possible progenitors
(e.g.,~\cite{caleb2016distributions,niino2018fast,luo2018onLee,lu2019implications, platts2019living,zhang2019energy,bhattacharya2020population,luo2020frb}).
\citet{niino2018fast} has hypothesized three luminosity distribution function models (i.e., standard candle, power-law, and \mbox{power-law + exponential cutoff}) to investigate how differences in luminosity functions (LF) affect the observation properties of FRBs,
and used the LF model and the cosmic FRB rate density to examine the distribution of $DM_{\rm E}$, the $\log N-\log S$ distribution, and the $DM_{\rm {E}}-S_{\rm{peak}}$ correlation.
\citet{luo2018onLee} constructed a Schechter luminosity function of 33 FRBs samples with a power-law index ranging from $-$1.8 to $-$1.2.
Unfortunately, the LF of repeating FRBs has not been constructed yet due to the limit of numbers in the past. Hence, it is important and necessary to investigate their statistical characteristics and build the apparent intensity distribution function (IDF) and the LF of these repeating FRBs in order to disclose more natural differences from those non-repeating ones.

Our article is organized as follows. In Section \ref{sec2}, we introduce the FRB sample and
present the statistical analyses of their parameters.
The intensity distribution function of our repeating FRB sample is derived in Section \ref{sec3}.
In Section \ref{sec4}, we construct the differential broken power-law LF of repeating FRBs
and non-repeating FRBs based on the $k$-corrected isotropic luminosity ($L$).
In Section \ref{sec5}, our conclusion and discussion are presented. The flat $\Lambda$CDM cosmological parameters $H_{0}$ = 67.74  km\,s$^{-1}$ Mpc$^{-1}$, $\Omega_{b}$ = 0.0486, $\Omega_{m}$ = 0.3089, \mbox{$\Omega_{\Lambda}$ = 0.6911} have been adopted throughout the paper
\citep{ade2016planck}.


\section{{The Statistical Properties of Repeating FRBs}} \label{sec2}
\subsection{{Distributions of Extra-Galactic Dispersion Measure and Total Energy}}
\setlength{\tabcolsep}{1.5mm}{
\begin{table}[H]
\caption{Key 
 physical parameters of 21 repeating FRBs before September 2021. \label{tab:repeating FRBs data}}
	\begin{adjustwidth}{-\extralength}{0cm}
		\newcolumntype{C}{>{\centering\arraybackslash}X}
		\begin{tabularx}{\fulllength}{lcccccccccc}
			\toprule
            \multirow{2}{*}{ \textbf{\boldmath{TNS Name}}} & \textbf{\boldmath{$\nu_{\mathrm{c}}\ ^{\mathrm{a}}$}}&\textbf{\boldmath{ $DM$ }}& \textbf{\boldmath{$ DM_{MW}$}} & \textbf{\boldmath{$ DM_{\rm E}\ ^{\mathrm{b}}$ }}& \textbf{\boldmath{$W_{\rm obs}$}}
             &\textbf{\boldmath{$S_{\rm peak}$ }}&\textbf{\boldmath{ $F_{\rm obs}$ }}&\textbf{\boldmath{ $z$ }}&\textbf{\boldmath{$E$ }}&\textbf{\boldmath{ $\overline{F}\ ^{\mathrm{d}}$}}\\
             &\textbf{\boldmath{ MHz}}&\textbf{\boldmath{ (${\rm pc\,cm^{-3}}$) }}& \textbf{\boldmath{(${\rm pc\,cm^{-3}}$)}}  &\textbf{\boldmath{ (${\rm pc\,cm^{-3}}$) }} &\textbf{{  (ms) }} & \textbf{\boldmath{ (Jy) }}& \textbf{\boldmath{ (Jy ms)  }}&   & \textbf{\boldmath{($10^{39}\,\rm erg$)}}  & \textbf{\boldmath{ (Jy ms)}}\\
             \midrule
FRB 20121102A &1375& $557.00\pm 2.00$  & 188.00   & 369.00   & $3.00\pm 0.50$    & $0.40^{+0.40}_{-0.10}$   & $1.20^{+1.60}_{-0.55}$   & 0.31  & 0.13  & 0.36         \\
FRB 20171019A  &1297& $460.80\pm 1.10$  & 37.00    & 423.80   & $5.40\pm 0.30$    & 40.50                    & 219.00 & 0.35  & 34.01 & 101.62       \\
FRB 20180814A  &600 & $189.38\pm 0.09$  & 87.00    & 102.38   & $2.60\pm 0.20$    & 8.08 $^{\mathrm{c}}$                 & 21.00  & 0.09  & 0.13  & 22.57        \\
FRB 20180908B  &600 & $195.70\pm 0.90$  & 38.00    & 157.70   & $1.91\pm 0.10$    & $0.60\pm 0.40$           & 2.70   & 0.13  & 0.04  & 2.03         \\
FRB 20180916B  &600 & $349.70\pm 0.70$  & 200.00   & 149.70   & $1.06\pm 0.05$    & 7.64 $^{\mathrm{c}}$                 & 8.10   & 0.12  & 0.12  & 10.26        \\
FRB 20181017A  &600 & $1281.00\pm 0.60$ & 43.00    & 1238.00  & $20.20\pm 1.70$   & 0.79 $^{\mathrm{c}}$                 & 16.00  & 1.03  & 60.07 & 8.50         \\
FRB 20181030A  &600 & $103.50\pm 0.70$  & 40.00    & 63.50    & $0.59\pm 0.08$    & 12.37 $^{\mathrm{c}}$                & 7.30   & 0.05  & 0.02  & 4.75         \\
FRB 20181119A  &600 & $364.00\pm 0.30$  & 34.00    & 330.00   & $2.66\pm 0.10$    & 0.94 $^{\mathrm{c}}$                 & 2.50   & 0.28  & 0.24  & 1.77         \\
FRB 20181128A  &600 & $450.20\pm 0.30$  & 112.00   & 338.20   & $2.43\pm 0.16$    & 1.81 $^{\mathrm{c}}$                 & 4.40   & 0.28  & 0.46  & 3.45         \\
FRB 20190116B  &600 & $443.60\pm 0.80$  & 20.00    & 423.60   & $1.50\pm 0.30$    & 1.87 $^{\mathrm{c}}$                 & 2.80   & 0.35  & 0.52  & 1.80         \\
FRB20190117A   &600 & $393.30\pm 0.10$  & 48.00    & 345.30   & $1.44\pm 0.03 $   & $1.70\pm 0.60$           & 5.90   & 0.29  & 0.64  & 6.36         \\
FRB 20190208A  &600 & $580.20\pm 0.20$  & 72.00    & 508.20   & $1.31\pm 0.14$    & $0.60\pm 0.30$           & 2.00   & 0.42  & 0.60  & 1.70         \\
FRB 20190209A  &600 & $424.60\pm 0.60$  & 46.00    & 378.60   & $3.70\pm 0.50$    & 0.54 $^{\mathrm{c}}$                 & 2.00   & 0.32  & 0.28  & 1.25         \\
FRB 20190213A  &600 & $651.50\pm 0.40$  & 43.00    & 608.50   & 4.00              & $0.50\pm 0.30$           & 3.00   & 0.51  & 1.46  & 1.80         \\
FRB 20190212A  &600 & $301.40\pm 0.20$  & 49.00    & 252.40   & $2.10\pm 0.30$    & $1.10\pm 0.60$           & 2.50   & 0.21  & 0.13  & 2.67         \\
FRB 20190222A  &600 & $460.60\pm 0.10$  & 87.00    & 373.60   & $2.97\pm 0.90$    & 2.53 $^{\mathrm{c}}$                 & 7.50   & 0.31  & 1.00  & 5.45         \\
FRB 20190303A  &600 & $221.80\pm 0.50$  & 29.00    & 192.80   & $2.00\pm 0.30$    & $0.50\pm 0.30$           & 2.30   & 0.16  & 0.06  & 2.47         \\
FRB 20190417A  &600 & $1378.50\pm 0.30$ & 78.00    & 1300.50  & $1.19\pm 0.02$    & $0.70\pm 0.20$           & 1.70   & 1.08  & 7.40  & 3.10         \\
FRB 20190604A  &600 & $552.60\pm 0.20$  & 32.00    & 520.60   & $3.00\pm 0.40$    & $0.90\pm 0.40$           & 8.30   & 0.43  & 2.64  & 5.00         \\
FRB 20190711A  &23.8& $593.10\pm 0.40$  & 56.40    & 536.70   & $6.50\pm 0.50$    & 5.23 $^{\mathrm{c}}$                 & 34.00  & 0.45  & 9.88  & 17.70        \\
FRB 20190907A  &600 & $309.50\pm 0.30$  & 53.00    & 256.50   & $0.54\pm 0.14$    & $0.40\pm 0.20$           & 0.90   & 0.21  & 0.05  & 1.10         \\
        \bottomrule
		\end{tabularx}
	\end{adjustwidth}
	\noindent{\footnotesize{Note: The all data were taken from \url{http://www.frbcat.org} \citep{petroff2016frbcatBarr} (accessed on 3 May 2022), and we report the properties of the brightest bursts for each repeater.
                            \textsuperscript{a} $\nu_c$ is the central frequency within the observational bandwith for an FRB.
                            \textsuperscript{b} $ DM_{\rm E}$ is the extra-Galactic dispersion measure, which is defined as $DM_{\rm E}=DM - DM_{MW}$.
                            \textsuperscript{c} For $S_{\rm peak}$ which is not given in FRBCAT, the corresponding value is estimated by $S_{\rm peak} = F_{\rm obs}/W_{\rm obs}$.
                            \textsuperscript{d} $\overline{F}$ is the average fluence, which is calculated based on all archived $F_{\rm obs}$ of one repeating FRB.}}
\end{table}}
Currently, about 600 FRBs have been reported (including repeating bursts
and apparent non-repeating bursts). Here, we collect the observation data of 21 repeating FRBs and 571 non-repeating FRBs.
The sample of repeating FRBs was extracted from FRBCAT and several reported observational datasets
(e.g.,~\cite{spitler2016aScholz,chime2019aAmiriBandura, chime2019chimeAndersenBandur,kumar2019faintShannon,fonseca2020nineAndersen}).
We list the key physical parameters of 21 repeating FRBs in Table \ref{tab:repeating FRBs data}.
It is worth mentioning that for the observed peak flux density ($S_{\rm peak}$)
listed in Table \ref{tab:repeating FRBs data},
which is used as a representative of the brightness of an FRB,
we chose the brightest one in every monitoring period for one repeating event.
Additionally, the last column of Table \ref{tab:repeating FRBs data} is the average fluence $\overline{F}$,
which is calculated based on all archived observed fluence ($F_{\rm obs}$) for one repeating FRB. The reason is that multiple observations at different duty-circles inevitably suffer from the observational biases. Another reason is that number of sub-bursts within a repeater varies among distinct FRBs even for a comparable observation time, and thus is very difficult to count accurately. For instance, 1652 sub-bursts from FRB 20121102A~\citep{Lidi2021} and 1863 sub-bursts from FRB 20201124A~\citep{xu2021fast} were recently detected by FAST. In addition, the fluence fluctuations between different sub-bursts for one repeating FRB could be very large. For example, the fluences of FRB 20171019A were found to range from 0.37 to 388 Jy ms~\citep{kumar2019faintShannon}.
Although it is uncertain whether repeating or ``non-repeating'' is due to
the observational select bias
(e.g.,~\cite{connor2018detecting,caleb2019areStappers,kumar2019faintShannon} for a discussion of repeating FRBs),
here we ignore the probability that the present non-repeating events
may be found to be repeating in future,
and treat them as real non-repeating events,
whose corresponding parameters are not presented in the text
because of their large number. Note that in Table  \ref{tab:repeating FRBs data}, we consider the observed pulse width instead of the intrinsic one.
The reason is that the intrinsic pulses of the narrow FRBs would be broadened due to scattering,
and the scattering time scale is affected by the local environment of the FRB and is model-dependent,
which results in a larger uncertainty.
An appendix of non-repeating FRBs samples has been added separately at the end of the paper (for more details, see \mbox{Table \ref{table:tbA1}}). Moreover, we did not consider FRB 200428 in our non-repeating burst sample,
which is the only event detected in the Milky Way~\citep{bochenek2020fast,chime2020aAndersen,lin2020no,
kirsten2021detection,li2021hxmt}.

\begin{figure}[H]
\begin{adjustwidth}{-\extralength}{0cm}
\centering
\includegraphics[width=0.33\linewidth, angle=0,scale=1]{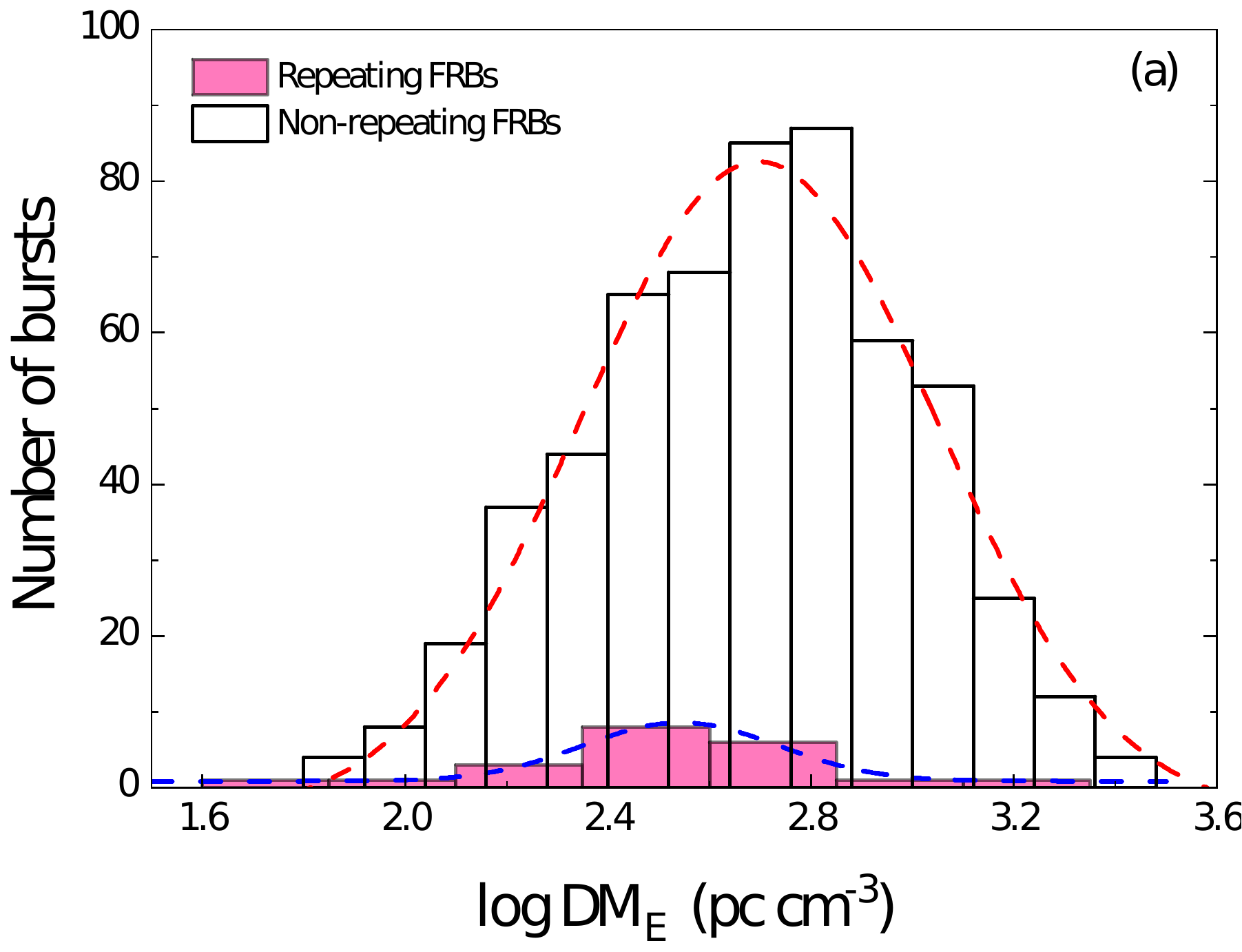}
\includegraphics[width=0.321\linewidth, angle=0,scale=1]{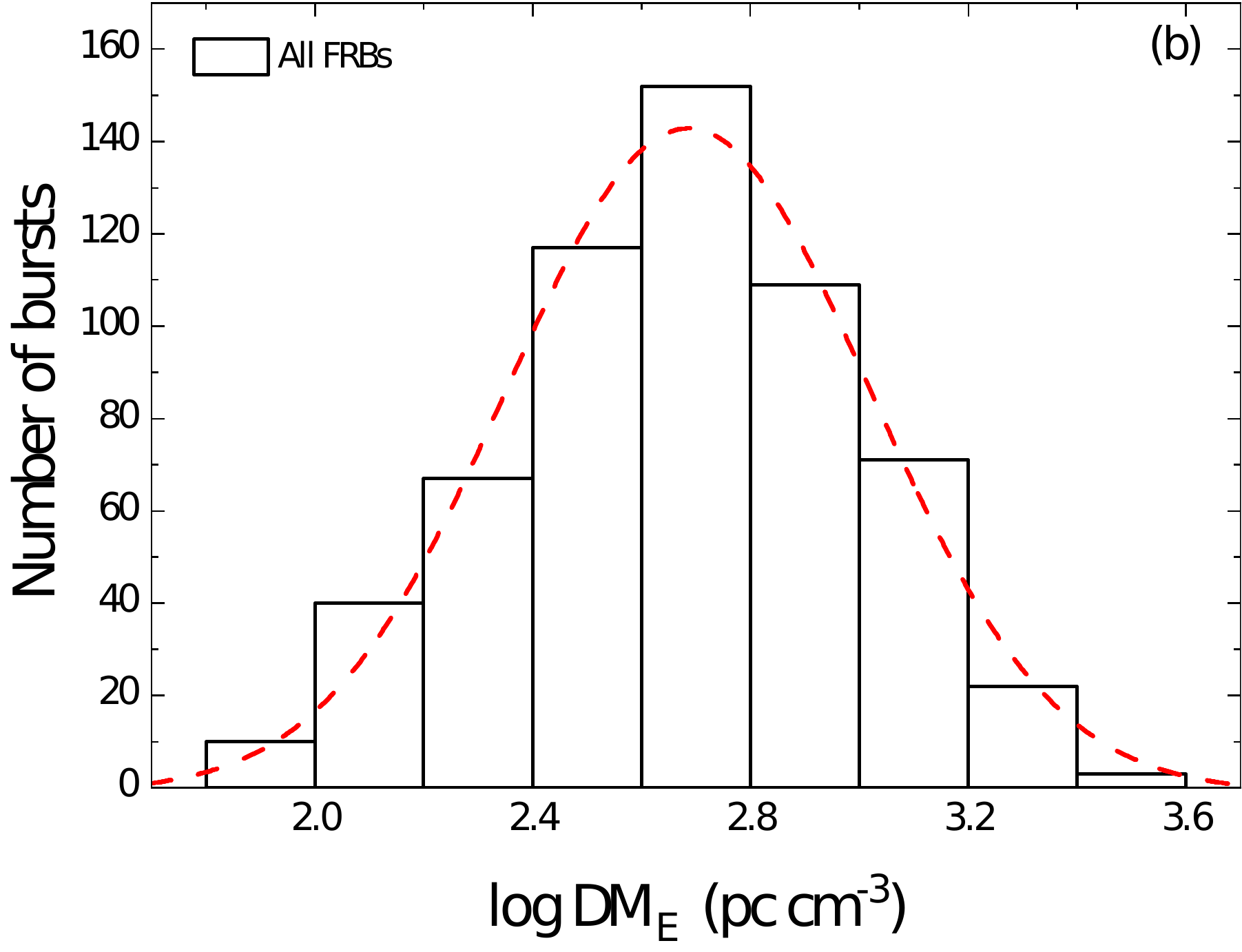}
\includegraphics[width=0.33\linewidth, angle=0,scale=1]{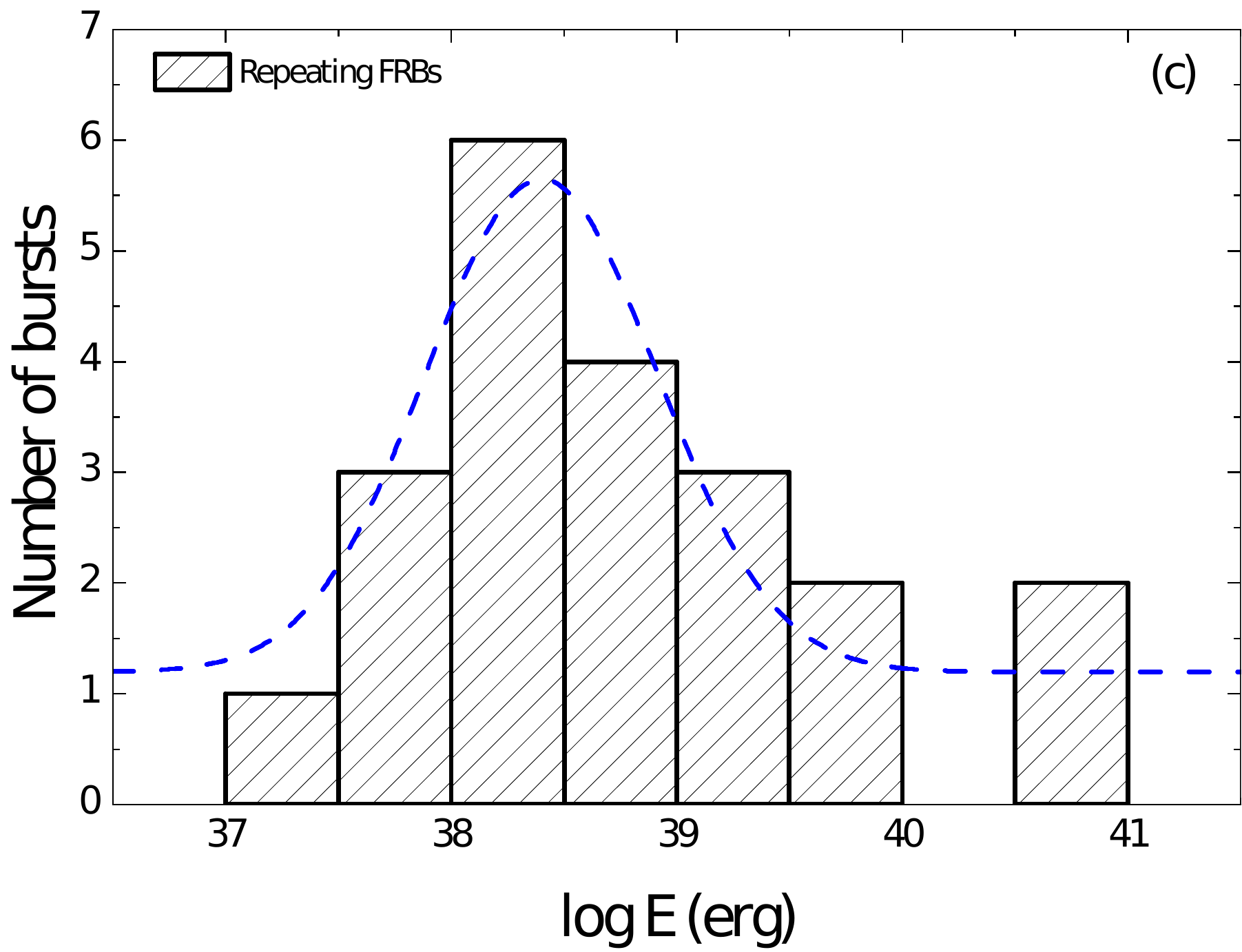}
\end{adjustwidth}
\caption{The logarithmic distributions of $DM_{\rm E}$ and radio energy of FRBs.
Panel (\textbf{a}) presents the $DM_{\rm E}$ distributions of non-repeating and repeating FRBs and panel (\textbf{b}) displays the $DM_{\rm E}$ distributions of all FRBs. Panel (\textbf{c}) shows the radio energy distribution of repeating events. The dash lines are the best fits to data with a Gauss function. \label{fig:general 1}}
\end{figure}

The $DM$ is a key quantity in the study of FRBs.
\mbox{\citet{chime2019chimeAndersenBandur}} argued that FRBs can be divided into two subclasses
(repeating and non-repeating FRBs) according to the current observations. If repeating FRBs indeed differ from the apparent non-repeating FRBs---for example, the two subclasses have different host or local environments, or one population is intrinsically more bright---the $DM$ distributions of the sub-samples could be obviously different. Meanwhile, their extra-Galactic dispersion measures ($DM_{\rm E}=DM-DM_{MW}$) could be
distinctly distributed. We applied the Anderson--Darling (A--D) test to different kinds of FRBs, see  Figure \ref{fig:general 1}a,b, and list the statistical results in Table \ref{tab:A-D test}, where it is shown that the $DM_{\rm E}$ distributions of the non-repeating, repeating and all FRBs can be well described by a log-normal function. Furthermore, we found the mean values of $ DM_{\rm E}$ are $496.9 \pm 16.9\, \rm pc\,cm^{-3}$ (0.70 dex) for non-repeating FRBs and $349.1 \pm 8.4\, \rm pc\,cm^{-3}$ (0.39 dex) for repeating FRBs via a Gauss fit. Simultaneously, we found that the mean values of $ DM_{\rm E}$ of all FRBs are $485.1\pm 15.2\, \rm pc\,cm^{-3}$ (0.67 dex).
It is notable that the mean $ DM_{\rm E}$ value of non-repeating FRBs is evidently larger than that
of repeating FRBs. To check whether the $DM_{\rm E}$ distributions of non-repeaters and repeaters are same or not, we used a Mann--Whitney--Wilcoxon (M--W--W) test~\citep{bauer1972constructing,hollander2013nonparametric} and obtained the statistic $W=4328.5$ with a \emph{p}-value of 0.031, less than the significance threshold of 0.05, which demonstrates that the $DM_{\rm E}$ distributions of the two kinds of FRBs may have different progenitors or different physical mechanisms~\cite{platts2019living,zhang2020the,xiao2021physics}. Notably, the $DM_{\rm E}$ is mainly contributed by the intergalactic medium, the host galaxies or the local environments.  Unfortunately, what we have learned about the host galaxies is so little that the $DM_{\rm E}$ distributions are still uncertain and need to be verified by more observations in future.
\setlength{\tabcolsep}{4.3mm}
{
\begin{table}[H]
 \small
\caption{The statistical results of $DM_{\rm E}$  for non-repeating FRBs and repeating FRBs.\label{tab:A-D test}}
\begin{adjustwidth}{-\extralength}{0cm}
		\newcolumntype{C}{>{\centering\arraybackslash}X}
\begin{tabular}{clcccc}
\toprule
 \textbf{\boldmath{$DM$}}&\textbf{FRB Sample} &   \textbf{Statistic Value }& \textbf{The Critical Value }  & \textbf{\emph{p}-Value}& \textbf{Methods}\\
\midrule
$DM_E$&Non-repeating   &0.814 &1.084  & 0.035 $^{\rm b}$  &A--D test $^{\rm a}$   \\
$DM_E$&Repeating     & 0.427 & 0.963  & 0.285 $^{\rm b}$ &A--D test $^{\rm a}$ \\
$DM_E$&All            & 0.727 &1.085  & 0.058 $^{\rm b}$&A--D test $^{\rm a}$   \\
\midrule
$DM_{\rm E}$& Repeating and Non-repeating FRBs&4328.5&\dots&0.031 $^{\rm d}$ &M--W--W test $^{\rm c}$ \\
\bottomrule
\end{tabular}
\end{adjustwidth}
\noindent{\footnotesize {$^{\rm a}$ The A--D test was executed in logarithmic scale with a significance level of $\alpha=0.01$.
$^{\rm b}$ A \emph{p}-value larger than $\alpha$ indicates a log-normal distribution is favored~\citep{thode2002testing}.
$^{\rm c}$  The M--W--W test was used to check different distributions with a significance Level of $\alpha=0.05$.
$^{\rm d}$  A \emph{p}-value greater than $\alpha$ means that the two distributions are the \mbox{same~\citep{bauer1972constructing}}.}}

\end{table}}

In addition, we found that the radio energy of repeating FRBs ranges from \mbox{$2.00\times10^{37}$}
to $6.01\times10^{40}$ erg. We also used the Gauss function to fit the energy statistical distribution of repeating FRBs.
As shown in panel  Figure \ref{fig:general 1}c, the total energy of repeating FRBs roughly follows a log-normal
distribution with a mean value of $2.51_{-0.59}^{+0.81}\times 10^{38}$ erg and a scatter of 1.03 dex.
Interesting, our result is roughly consistent with the early estimates of \mbox{\citet{li2021}} ($10^{39}$--$10^{42}$ erg) for the repeating FRBs.
Meanwhile, it is worth {mentioning} that this result is different from the bimodal energy distribution (a log-normal function and a generalized Cauchy function, with the peak of the energy distribution of $4.8 \times 10^{37} $ erg) of a sub-sample of all the bursts from FRB 20121102A found by~\citet{Lidi2021} with FAST. Alternatively, it is possible that the bimodal energy distribution could be existent for a single burst but disappear when many bimodal distributions of diverse repeating FRBs are randomly mixed.
However,~\citet{li2021} noticed that the total energy of non-repeaters is on average larger than that of repeaters about one order of magnitude, which may demonstrate that at least some of repeating and non-repeating FRBs have different physical origins.

\subsection{{Correlations between Some Characteristic Parameters}}



\begin{figure}[h]
\begin{adjustwidth}{-\extralength}{0cm}
\centering 
\includegraphics[width=\textwidth, angle=0]{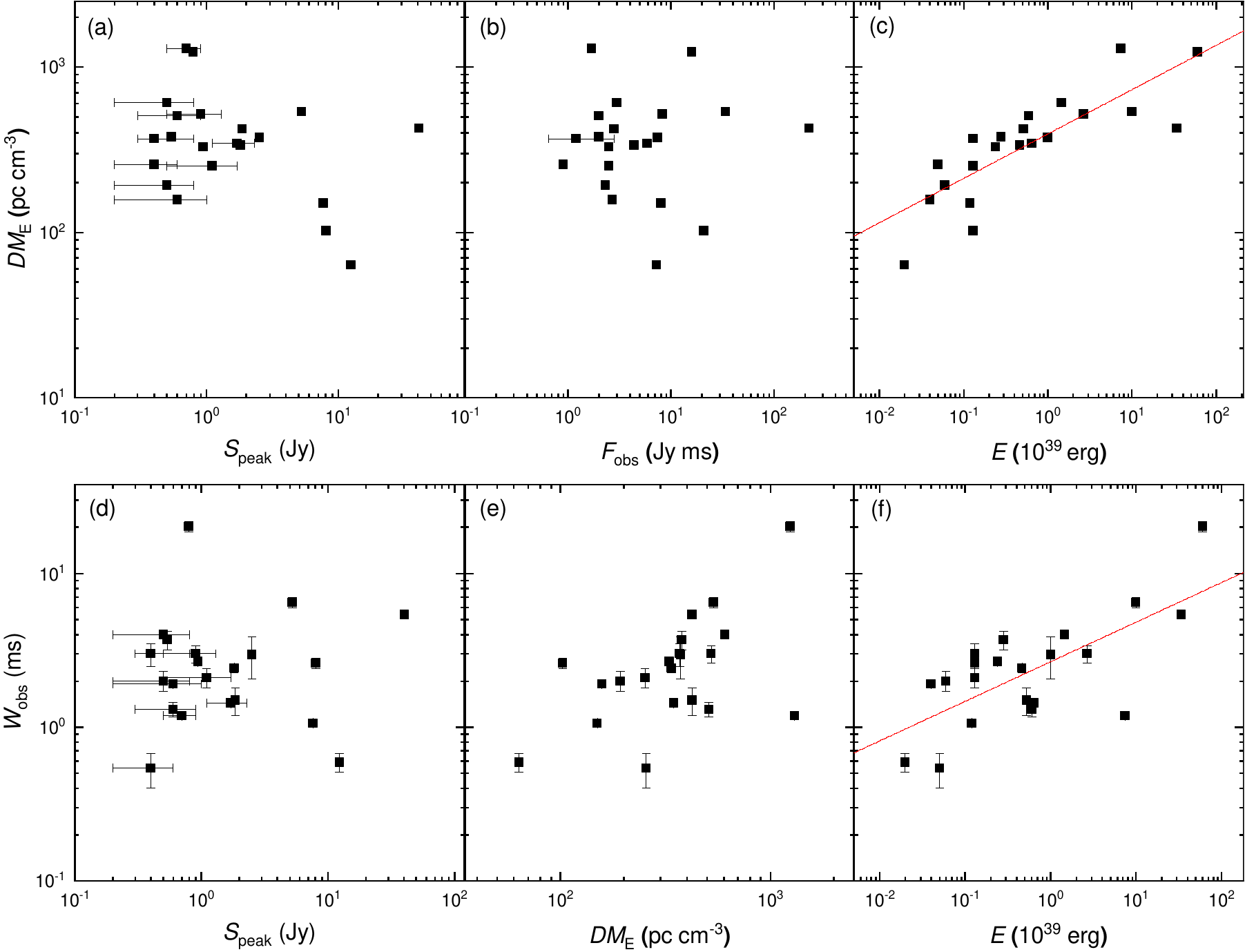}
\end{adjustwidth}
\caption{The correlations of $DM_{\rm E}$ with $S_{\rm{peak}}$, $F_{\rm{obs}}$ and $E$ are shown in Panels (\textbf{a}), (\textbf{b}) and (\textbf{c}) respectively. The $W_{\rm {obs}}$ is plotted against $S_{\rm{peak}}$ in Panel (\textbf{d}), $DM_{\rm{E}}$ in Panel (\textbf{e}) and $E$ in Panel (\textbf{f}). The observed data are  symbolized with the filled squares. The solid lines in Panels (\textbf{c}) and (\textbf{f}) stand for the best fits to data with a power-law function. \label{fig:general 2}}
\end{figure}

In Figure \ref{fig:general 2}, the relationships of the
\mbox{$DM_{\rm E}-S_{\rm {peak}},DM_{\rm E}-F_{\rm {obs}},DM_{\rm E}-E,$} \linebreak \mbox{$W_{\rm {obs}}-S_{\rm {peak}},W_{\rm {obs}}-DM_{\rm E},W_{\rm {obs}}-E$} of repeating FRBs are illustrated. Figure \ref{fig:general 2}a shows that $S_{\rm{peak}}$ is not correlated with $DM_{\rm E}$. Note that the intrinsic width of the narrow FRBs is difficult to discern due to dispersion smearing
and scattering broadening. Scattering is model dependent, and the assumptions of the model introduce high uncertainty.
Therefore, we used the observed pulse width instead of the intrinsic pulse width in our work.  \mbox{Figure \ref{fig:general 2}b} shows that $F_{\rm obs}$ does not correlate with $DM_{\rm E}$. As shown in Figure \ref{fig:general 2}c, the radio energy and the $DM_{\rm E}$ are positively correlated with a Pearson correlation coefficient of \mbox{0.81 and} can be well-fitted. The power-law relation is $E \propto  DM_{\rm E}^{2.60 \pm 0.04}$, which is roughly consistent with the tight correlation of non-repeating FRBs found by~\citet{li2017intensityHuang}, who argued that the positive correlation may be attributed to an observation selection effect---i.e., a fainter event is easier to be observed at a nearer distance. We found from \mbox{Panels (d) and (e)} that $W_{\rm obs}$ is not correlated with $S_{\rm peak}$ but obviously correlated with the $DM_{\rm E}$. Note that our $W_{\rm obs}-S_{\rm peak}$ relation is inconsistent with that of non-repeaters in~\citep{li2017intensityHuang}. The $S_{\rm{peak}}$ of the repeated FRBs spans three magnitudes and is more dispersive than the $S_{\rm{peak}}$ of the non-repeating FRBs sample of~\citet{li2017intensityHuang}, so that $W_{\rm{obs}}$ and $S_{\rm {peak}}$ are not correlated. Meanwhile, we found that there is a positive correlation between $W_{\rm obs}$ and the radio energy, as shown in Figure \ref{fig:general 2}f.~\citet{li2017intensityHuang} analyzed the correlations of key parameters of 16 non-repeating FRBs and found no clear correlation between energy and pulse width. For our repeating FRBs, the radio energy $E$ and the pulse width $W_{\rm obs}$,
respectively, span four and three orders of magnitude. The best-fitted power-law
relation is $E \propto W_{\rm obs}^{0.43\pm0.06}$, with a correlation coefficient of 0.71 and a chance probability of $1.49 \times 10^{-6}$. This means that the repeating FRB pulse with a longer duration has a greater energy release generally, which is similar to those non-repeaters reported by~\citet{li2017intensityHuang}, and both of which are consistent with the findings in~\cite{li2021}, where the averaged isotropic energies of long FRBs were found to be larger than those of short FRBs by at least two orders of magnitude, not only for non-repeating FRBs but also for repeating FRBs.

\begin{figure}[H]
\includegraphics[width=0.5\linewidth, angle=0,scale=1]{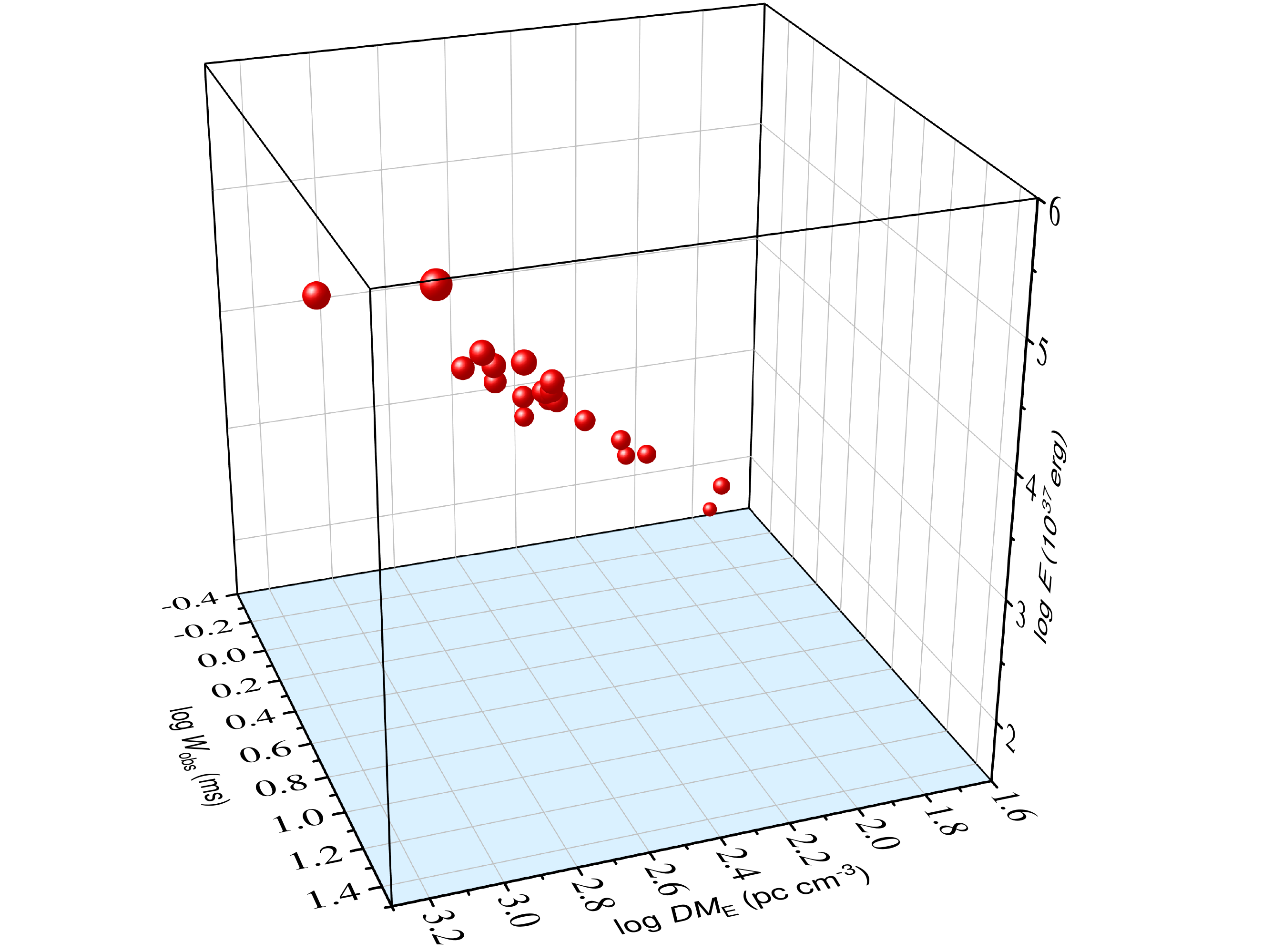}
\includegraphics[width=0.5\linewidth, angle=0,scale=1]{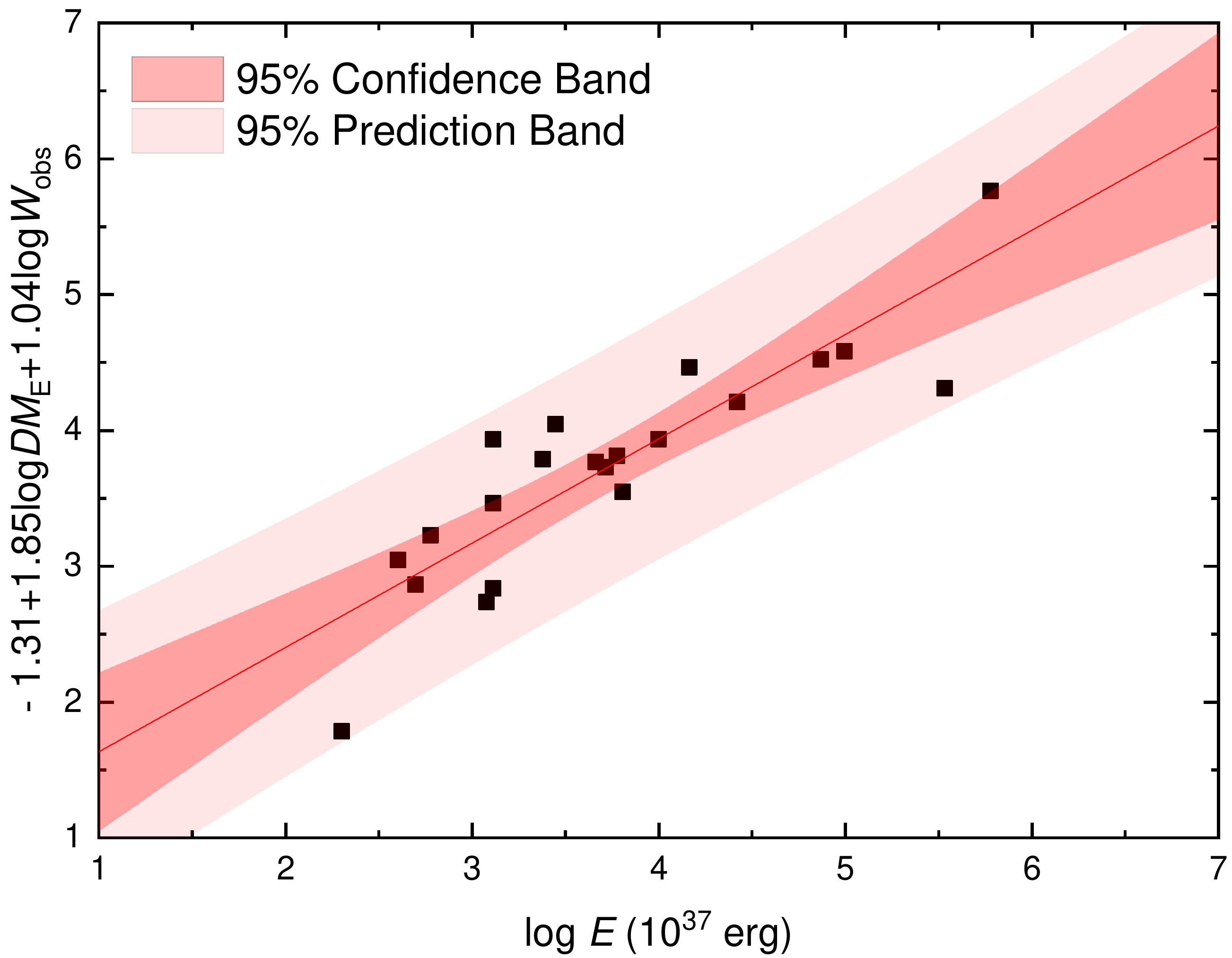}
\caption{Left panel:  
 Correlation among $E$, $DM_{\rm E}$ and $W_{\rm obs}$ of repeating FRBsis
illustrated by the 3D scatter plot. Right panel: The isotropic energy $E$ is plotted against the  energy estimated by
Equation (\ref{eq.1-E-DM-W}). The solid line is the best fit to the data. The light and the heavy shaded regions are the $95\%$
confidence and the $95\%$ prediction ranges, respectively.\label{fig:general 3} }
\end{figure}

Figure \ref{fig:general 2}e,f seems to show that $DM_{\rm E}$, $W_{\rm obs}$ and $E$ may be related. This motivated us to perform multiple linear regression fitting for these three parameters in Figure \ref{fig:general 3}. The three-parameter relation can be well described by a binary linear regression function as
\begin{equation}
\log E=(-1.31\pm0.97)+(1.85\pm0.41)\log DM_{\rm E}+(1.04\pm0.36)\log{W_{\rm obs}}
\label{eq.1-E-DM-W}
\end{equation}
with a Pearson correlation coefficient of 0.88, which implies that wider FRBs usually hold larger energy outputs and higher $DM_E$ values, and vice versa.
\section{Apparent Intensity Distribution Function of Repeating FRBs} \label{sec3}

\citet{li2017intensityHuang} applied the $F_{\rm obs}$ of 16
non-repeating FRBs to derive an intensity distribution function (IDF). We considered the IDF of
repeating FRBs to be determined by the average fluence ($\overline{F}$) and observed fluence ($F_{\rm {obs}}$) of the brightest burst for each repeater.
We then utilized the average fluence $\overline{F}$ presented in Table \ref{tab:repeating FRBs data} to investigate the IDF of \mbox{21 repeating FRBs}. This was done for several reasons: (1) For one repeating burst, multiple monitor observations in different time periods may lead to the observational bias. (2) The repeaters contain multiple bursts, and the number of monitored subbursts between different repeaters varies---e.g., FAST has recently detected 1652 repeating events from FRB 20121102A~\citep{Lidi2021} and 1863 bursts from FRB 20201124A~\citep{xu2021fast}. (3) There is a certain degree of fluctuation between the fluences of repetitions
for one repeating FRB; for example, the fluence of the repetitions from faint FRB 20171019A have a large range of 0.37  to $\sim$ \!388 $\rm{Jy\,ms}$~\citep{kumar2019faintShannon}. Therefore, we took the mean fluence as a statistical variable for each repeater in order to build an apparent IDF of the repeating FRBs. In addition, we got rid of the faint FRB 20171019A from our repeating FRB sample because of its large fluctuations of fluence in the following calculations.
\begin{figure}[h]
\centering
\subfigure{
\begin{minipage}[]{1\textwidth}
\includegraphics[width=0.49\textwidth, angle=0]{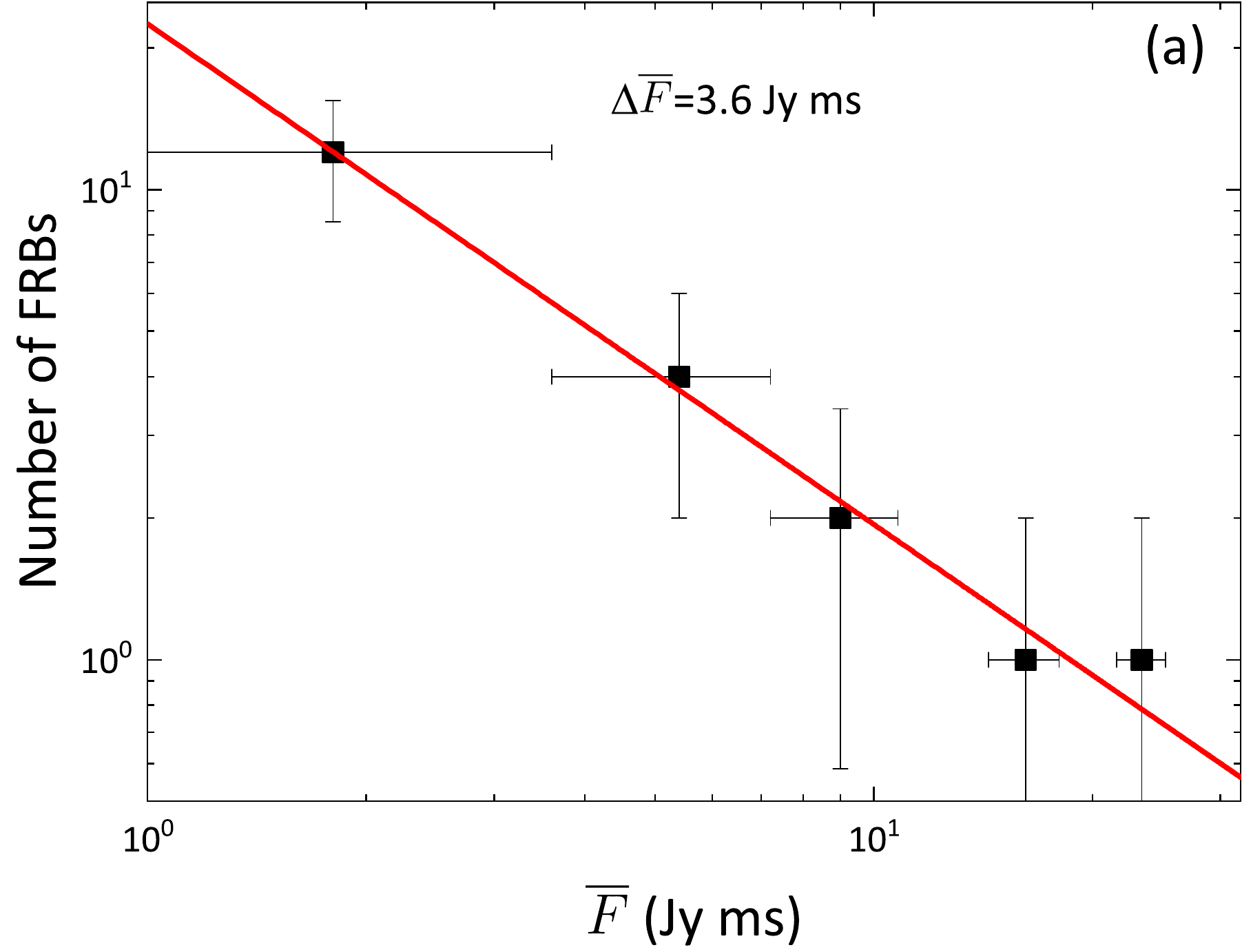}
\includegraphics[width=0.50\textwidth, angle=0]{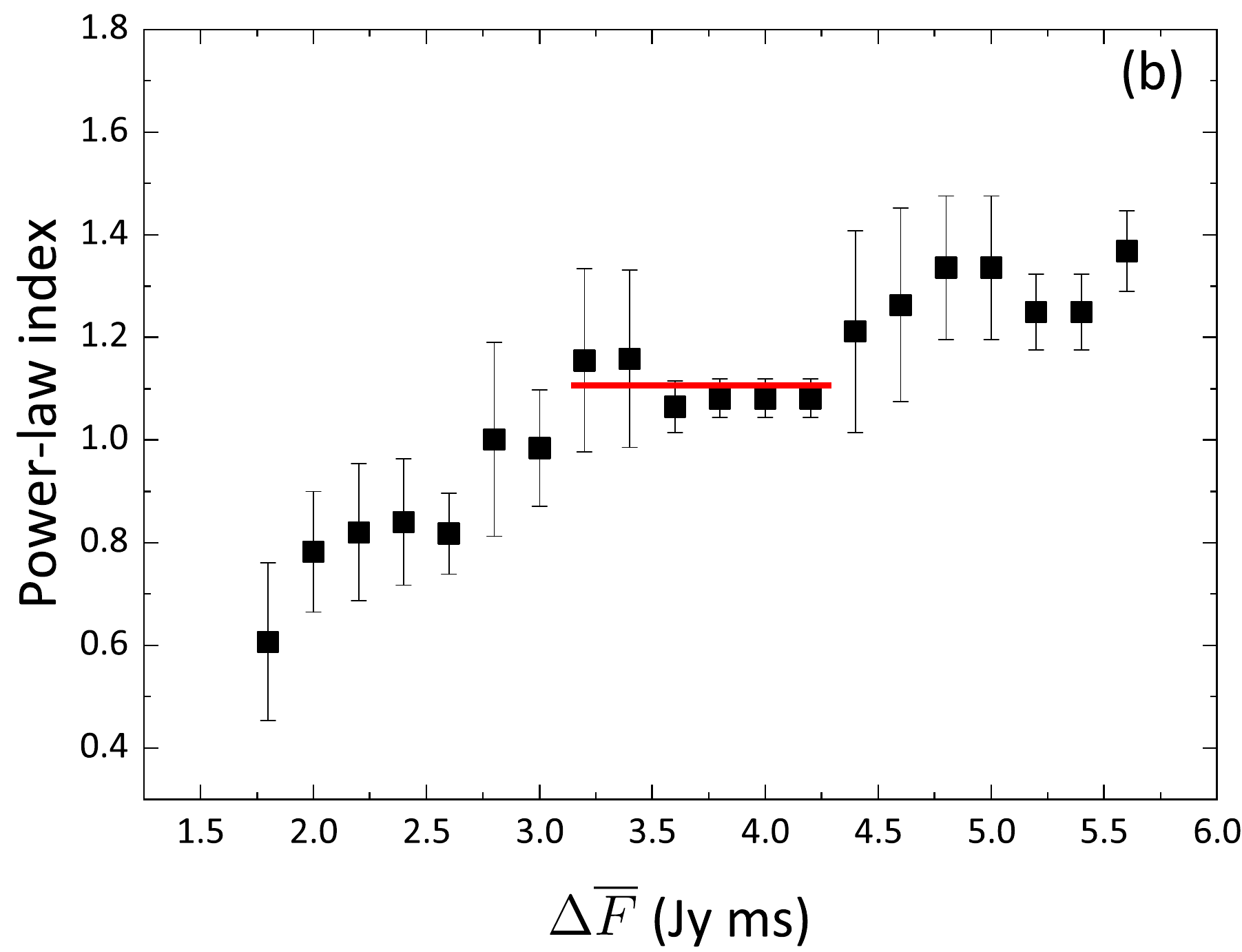}
\end{minipage}
}
\subfigure{
\begin{minipage}[]{1\textwidth}
\includegraphics[width=0.49\textwidth, angle=0]{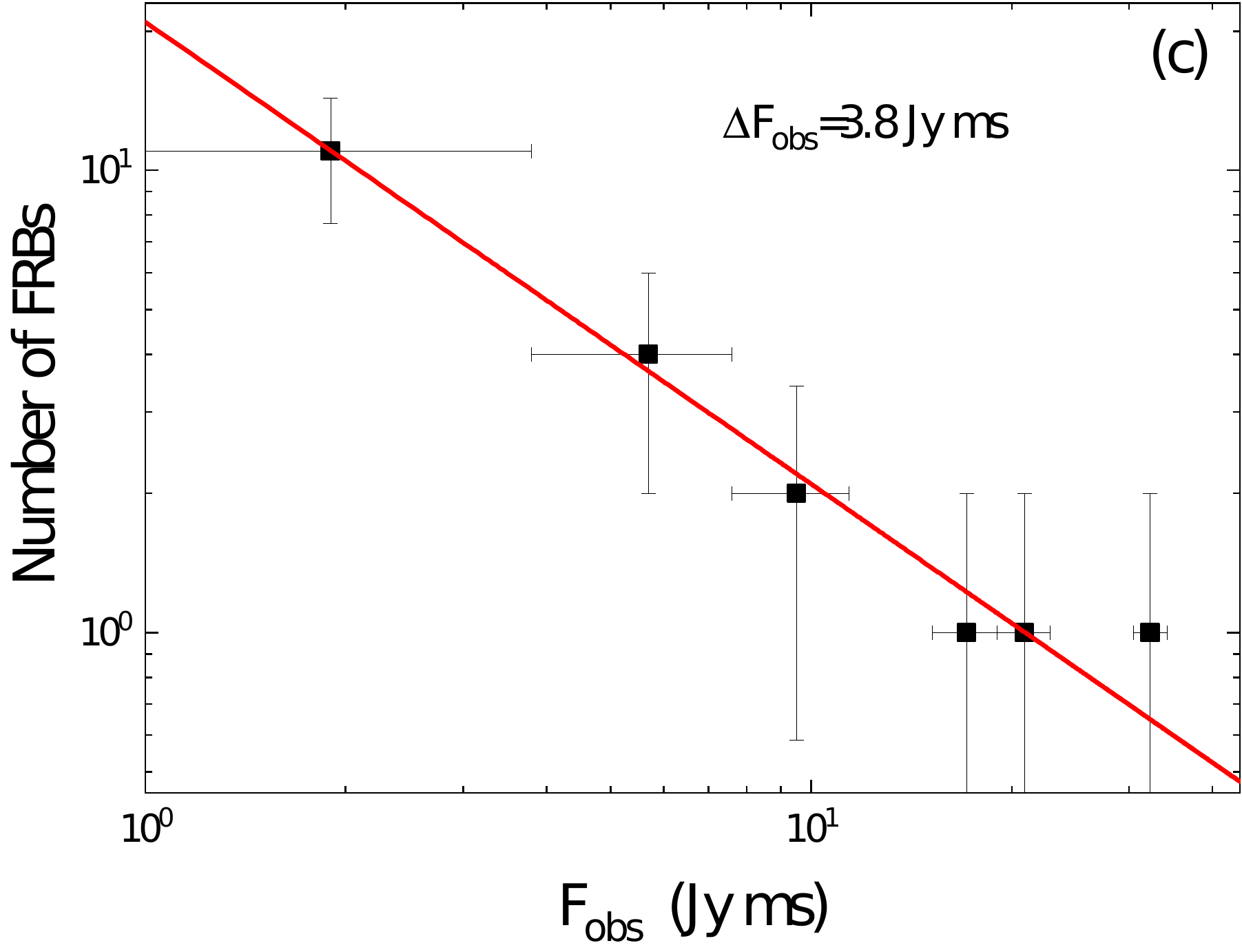}
\includegraphics[width=0.49\textwidth, angle=0]{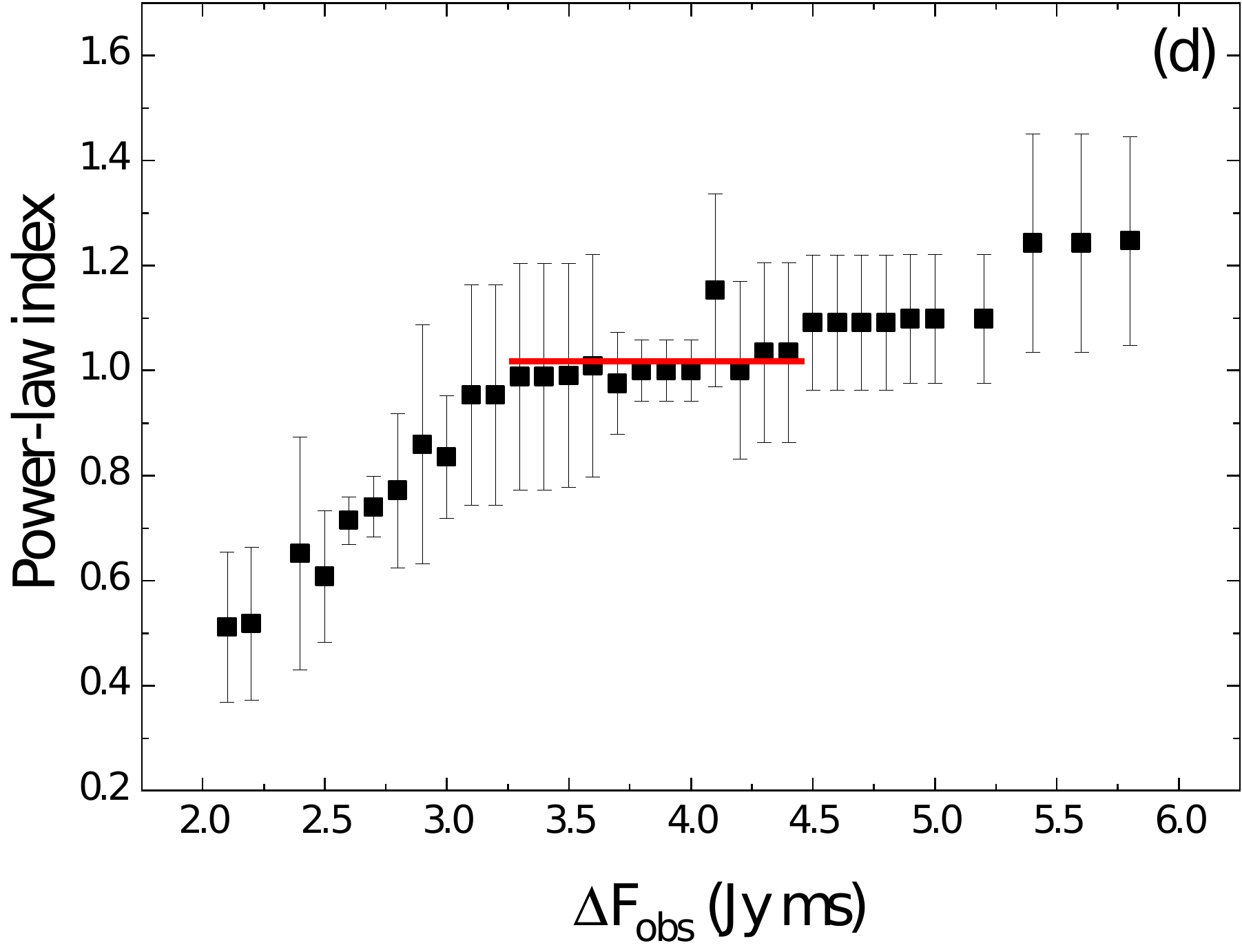}
\end{minipage}
}
\caption{Panel (\textbf{a}) shows an exemplary distribution of $\overline{F}$ for a bin width of
$\Delta \overline{F} = 3.6$ Jy ms. The solid line is the best-fit curve.
Panel (\textbf{b}) illustrates the best-fit power-law values $a_1$ for each bin width;
the solid short horizontal line shows the relatively stable range of
$\Delta \overline{F} = 3.2$--4.2 Jy ms for $a_1$.  Panel (\textbf{c}) shows an example distribution of $F_{\rm{obs}}$ for a bin width of $F_{\rm {obs}} = 3.8$ Jy ms. The solid line is the fitted curve. Panel (\textbf{d}) presents the best-fit power-law values $a_2$ for each bin width;
the solid short horizontal line shows the relatively stable range of
$\Delta F_{\rm{obs}} = 3.3$--4.4 Jy ms for $a_2$.\label{fig:general 4}}
\end{figure}

We assumed that FRBs can be considered as standard candles, and defined that the apparent IDF
of repeating FRBs is in the same form as the function given by~\citet{li2017intensityHuang}:
$dN/d F_{\rm obs} = A F_{\rm obs}^{-a}$ (or $N(>F_{\rm obs})\propto F_{\rm obs}^{-a+1}$), where $\alpha$ is the power-law index, $a=2.5$ is expected for a uniform distribution
and $A$ is a constant which depends on the observations and the event rates of FRBs.
We group the repeating FRB sample into several fluence intervals with a bin width of
$\Delta \overline{F}$, and created the exemplary distributions of $\overline{F}$,
as presented in Figure \ref{fig:general 4}.
Then we acquired a best-fit power-law curve for $d N/d F_{\rm obs}$.
In Figure \ref{fig:general 4}a, best fit results are shown for a bin width
of $\Delta \overline{F} = 3.6\,{\rm Jy\,ms}$; the corresponding power-law index and the coefficient of
determination (R-Square) are $a_1 = 1.07\pm0.05$ and 0.99, respectively.
Note that the error bars along the $x$-axis and $y$-axis are, respectively, $\Delta \overline{F}/2$
and the square root of FRB count in each interval.
Meanwhile, we also investigated the influence of bin width selection on
the fitted results. In our work, the range of the bin width $\Delta \overline{F}$ was selected to be from
0.2 to 6.0 $\,{\rm Jy\,ms}$, and the best-fit power-law indices
corresponding to each $\Delta \overline{F}$ are shown in Figure \ref{fig:general 4}b, where one can see that when $\Delta \overline{F}$ is in a range of $3.2$--$4.2$ Jy~ms,
the fitted $a_1$ has smaller fluctuations and the corresponding error bar is also small.
Hence, the  $a_1$ values of $\Delta \overline{F} = 3.2-4.2\,{\rm Jy\,ms}$ were selected to
calculate the final average value, which is $a_1 = 1.10 \pm 0.14$.
To constrain the unknown constant $A$, we adopted similar calculation processes to those
described in detail in \mbox{\citet{li2017intensityHuang}} and compare them with the published event rates of FRBs in the literature in
Table \ref{tab:event rate}. The $A$ value in our IDF was obtained to be
$(2.37 \pm 0.76) \times 10^3\,\rm sky^{-1} day^{-1}$. As a result, the final IDF of 20 repeating FRBs can be written as
\begin{equation}
\frac{d N}{d F}=(2.37 \pm 0.76) \times 10^{3} F_{\rm obs}^{-1.10 \pm 0.14}\,\rm sky^{-1} day^{-1}.
\end{equation}

Similarly, we derived, while adopting $F_{\rm{obs}}$, the power-law $a_2 = 1.01 \pm 0.16$ of IDF for repeating FRBs (see Figure \ref{fig:general 4}d). Furthermore, we got $A=(1.96 \pm 0.41) \times 10^3\,\rm sky^{-1} day^{-1}$ for our IDF. The IDF of repeating FRBs is given by
\begin{equation}
\frac{d N}{d F}=(1.96 \pm 0.41) \times 10^{3} F_{\rm obs}^{-1.01 \pm 0.16}\,\rm sky^{-1} day^{-1}.
\label{eq:3}
\end{equation}
In particular, we found that power-law indices $a_1$ and $a_2$ have small differences, which indicates that our results are stable and reliable.
If assuming the minimum fluence of 0.36 Jy ms as the threshold, one can estimate that the detection rate of repeating FRBs is at least $(9.64\pm3.09)\times10^3$ sky$^{-1}$ day$^{-1}$ for $a_1$ and $(1.38\pm0.19)\times10^4$ sky$^{-1}$ day$^{-1}$ for $a_2$. Combined with  all parameters of FAST~\citep{nan2011theLi,li2013theNan,Zhangzb2015},
we utilized the method of~\citet{li2017intensityHuang}, and estimate the detection rate for the 1000 h observation time is
about $3\pm 1$; that is slightly smaller than that for non-repeating FRBs predicted by~\citet{li2017intensityHuang}.
It is reasonable, since the current detection rate of non-repeating FRBs is obviously larger than that of repeating FRBs observationally.

\setlength{\tabcolsep}{5mm}{
\begin{table}
\small
\caption{Comparison of the deduced event rates of different FRB samples.}\label{tab:event rate}
\begin{adjustwidth}{-\extralength}{0cm}
		\newcolumntype{C}{>{\centering\arraybackslash}X}
\begin{tabular}{ccccc}
\toprule
\textbf{\boldmath{$F_{\rm \text {\bf{Limit} }}$}} &  \textbf{\boldmath{$R\left(>F_{\rm \text { Limit }}\right)$}} & \multirow{2}{*}{\textbf{\boldmath{Reference} }}& \textbf{\boldmath{Coefficient $A$ }}& \multirow{2}{*}{\textbf{\boldmath{FEB Class}}} \\
\textbf{\boldmath{(Jy~ms) }}& \textbf{\boldmath{($\rm sky^{-1}\,day^{-1}$) }}& & \textbf{\boldmath{$(10^{3}\, \rm sky^{-1}\,day^{-1}$) }}&\\
\midrule
3            & $1.0^{+0.6}_{-0.5}\times10^4$           &~\citet{thornton2013aStappers}    & $5.02 \pm 1.62$&non-repeaters \\
0.35         & $3.1^{+12}_{-3.1}\times10^4$            &~\citet{spitler2014fastCordes}    & $7.49 \pm 1.15$&non-repeaters \\
2            & $2.5\times10^3$                         &~\citet{keane2015fastPetroff}     & $1.06 \pm 0.32$&non-repeaters\\
1.8          & $1.2\times10^4$                         &~\citet{law2015aBower}            & $4.88 \pm 1.38$&non-repeaters \\
4            & $4.4^{+5.2}_{-3.1}\times10^3$           &~\citet{rane2016aLorimer}         & $2.53 \pm 0.87$&non-repeaters \\
0.13--1.5   & $7^{+5}_{-3}\times10^3$                 & ~\citet{champion2016fivePetroff} & $2.87 \pm 0.82$&non-repeaters\\
3.8          & $3.3^{+3.7}_{-2.2}\times10^3$           &~\citet{Crawford2016ASF}          & $1.85 \pm 0.63$&non-repeaters \\
0.03         & $(3.03\pm 1.56)\times10^4$              &~\citet{li2017intensityHuang}     & $4.19 \pm 0.22$&non-repeaters \\
6            & 587                                     & ~\citet{Lawrence2017TheNP}       & $0.42 \pm 0.16$&non-repeaters \\
2            & $1.7^{+1.5}_{-0.9}\times10^3$           &~\citet{Bhandari2018TheSF}        & $0.72 \pm 0.21$&non-repeaters \\
26           & $37\pm 8$                               & ~\citet{Shannon2018TheDR}        & $0.15 \pm 0.01$&non-repeaters \\
8            & $98^{+59}_{-39}$                        & ~\citet{Farah2019FiveNR}         & $0.01 \pm 0.003$&non-repeaters \\
2            & $3.4^{+15.4}_{-3.3}\times10^3$          & ~\citet{Parent2020FirstDO}       & $1.44 \pm 0.42$&non-repeaters \\
5            & $818\pm 64$                             &  ~\citet{2021arXiv210604352T}     & $0.53 \pm 0.19$&non-repeaters\\
0.36         &   $(9.64\pm3.09)\times10^3$             &   this work                      & $2.37 \pm 0.76$&repeaters\\
\bottomrule
\end{tabular}
\end{adjustwidth}
\end{table}}

\section{\textbf{Differential Bolometric Luminosity Distributions}}\label{sec4}

The luminosity functions of FRBs can be applied to reveal the origins of FRBs, design the optimal searching plan, guide future observations~\citep{bera2016onBhattacharyya,li2017intensityHuang,luo2018onLee,luo2020frb}, etc. The detection rate of the telescope can be calculated by the luminosity function, and the event rate density at different luminosity ranges sheds light on the origins of FRBs~\citep{platts2019living,xiao2021physics}.
Building the LF requires not only the flux and distance,
but also a k-correction of luminosity because of the different observational central
frequencies ($\nu_c$) in the mixed sample of FRBs.
According \mbox{to~\citet{zhang2018fast}}, the isotropic peak luminosity of a FRB can be
calculated with $L_{p} \simeq 4 \pi D_{\mathrm{L}}^{2}(z) S_{\rm peak} \nu_{c}$,
where $D_{\rm L}(z)$ is the lumonisity distance calculated with the redshift given by FRBCAT (\url{https://www.frbcat.org/}). (accessed on 3 May 2022).
We utilized $\nu_c$ to substitute the observational bandwidth of radio telescopes,
which was proposed \mbox{by~\citet{petroff2016frbcatBarr} and~\citet{Aggarwal2021}}.
\mbox{Additionally,~\citet{zhang2018fast}} also suggested to use the central frequency instead of
the receiver bandwidth since the FRB spectrum must be unknown and emissions may
extend beyond the receiver bandwidth.
In addition, when the observed flux densities of cosmological objects at different $\nu_c$
are considered, k-correction is also an important effect.
After the k-correction, the bolometric luminosity can be given by $L\equiv L_{p}\simeq 4\pi D_{\mathrm{L}}^{2}(z) S_{\rm peak} \nu_{c} k$, in which the $k$-correction factor is taken as $k=(1+z)^{\alpha_t-\beta}$ with a temporal index of $\alpha_t\sim0$ and a spectral index $\beta\sim1/3$~\citep{zhang2018redshift,li2021}, which are
similar to those parameters of normal pulsar spectra~\citep{macquart2018frb,macquart2019spectral,petroff2019fast}.
\begin{figure}[H]
\includegraphics[width=0.49\textwidth, angle=0]{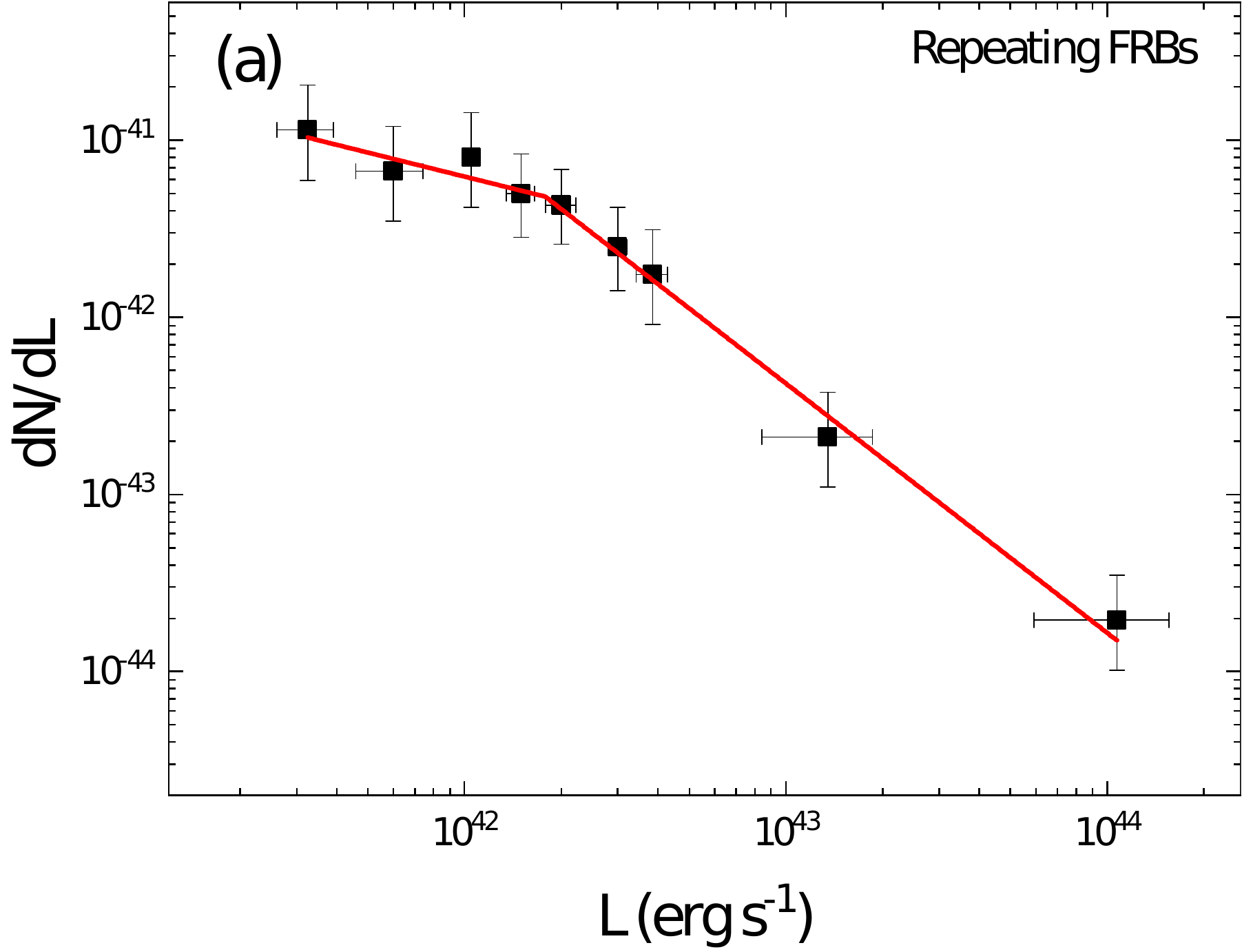}
\includegraphics[width=0.49\textwidth, angle=0]{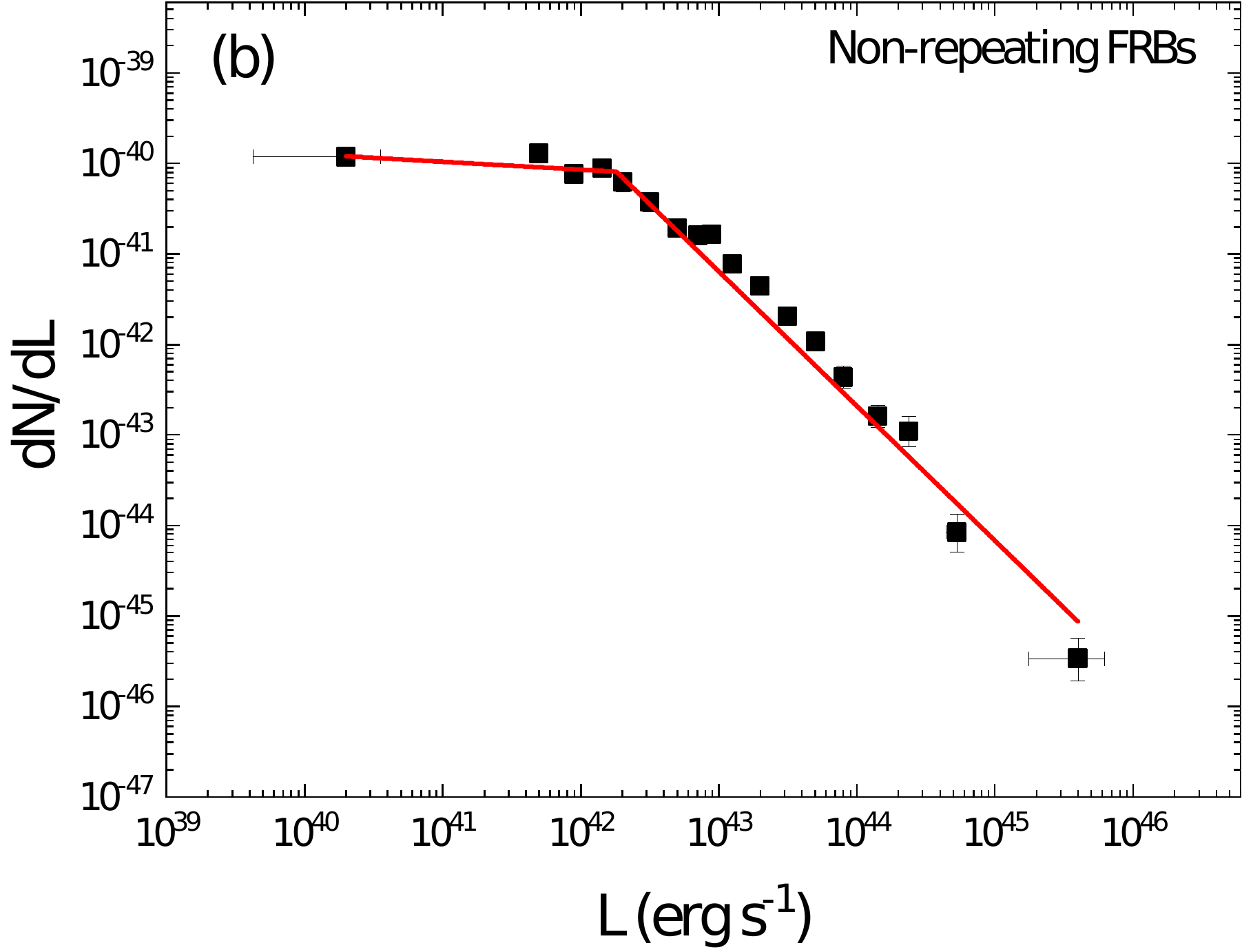}
\caption{
The luminosity functions of the repeating (Panel \textbf{a}) and non-repeating (Panel \textbf{b}) FRBs. The solid lines stand for the best fits with the power-law function of Equation (\ref{eq:3}). In each panel, the error bars along the $x$-axis are simply the standard errors in the corresponding luminosity bins,
and the $y$-axis error bars are the 1-$\sigma$ Poisson errors~\citep{gehrels1986confidence}.
 \label{fig:general 5}}
\end{figure}
\setlength{\tabcolsep}{3.75mm}{
\begin{table}[H]
\caption{{{The}
 best-fit parameters of a broken power-law LF for repeating and non-repeating FRBs.}\label{tab:Results}}
\begin{adjustwidth}{-\extralength}{0cm}
		\newcolumntype{C}{>{\centering\arraybackslash}X}
\begin{tabular}{lccccc}
\toprule
\textbf{Sample}& \textbf{\boldmath{$A$}} &\textbf{\boldmath{$\alpha_{1}$}} & \textbf{\boldmath{$\alpha_{2}$ }}&\textbf{\boldmath{$L_{b}~\rm {(erg\,s^{-1})}$}}&\textbf{\boldmath{$\chi_{\nu}^{2}$}} \\
\midrule
Repeating FRBs & $(4.80\pm1.41)\times 10^{-42}$ & $-0.45 \pm 0.20$  & $-1.41 \pm 0.08$  & $(1.78 \pm 0.44)\times 10^{42}$& 0.17\\
Non-repeating FRBs & $(8.59\pm2.01)\times 10^{-41}$ & $-0.09 \pm 0.10$  & $-1.49 \pm 0.05$  & $(1.74 \pm 0.29)\times 10^{42}$&3.17   \\
\bottomrule
\end{tabular}
\end{adjustwidth}
\end{table}}

The differential distributions of the k-corrected luminosity for 21 repeaters and 571 non-repeaters are shown in Figure \ref{fig:general 5}, where one can notice that the repeaters are normally less luminous than the non-repeaters by about two orders of magnitude. Subsequently, we adopted a broken power-law from~\citep{salvaterra2012complete,pescalli2016rate,deng2016cosmic} to build the LF as
\begin{equation}
\phi(L) = \begin{cases}A\left(\frac{L}{L_{\mathrm{b}}}\right)^{\alpha_{1}}, & L \leq L_{\mathrm{b}}
\\ A\left(\frac{L}{L_{\mathrm{b}}}\right)^{\alpha_{2}}, & L>L_{\mathrm{b}}\end{cases},
\end{equation}
where $A$ is the normalized factor; $L_b$ is the break luminosity; $\alpha_1$ and $\alpha_2$ are two power-law indices corresponding to lower and higher luminosity portions. In Figure \ref{fig:general 5}, we can see that the luminosity distributions of both repeating and non-repeating FRBs can be well described by the broken power-law function, and the best-fit parameters are provided in Table \ref{tab:Results}.
We excitingly found that the power-law indices of repeating and non-repeating bursts in the luminous end are the same as $\approx\!\!-1.4$ within the range from $-$1.8 to $-$1.2, constrained by the Schechter luminosity function by~\citet{luo2018onLee}. However, we noticed that non-repeaters decline slower that repeaters in the less luminous end instead. In addition, after k-correction for observational frequency, the distributions of luminosity  for repeating and non-repeating FRBs  are similar at their brightest ends. The characteristic luminosity values $L_b$ are not significantly different.

\section{Conclusions and Discussion}\label{sec5}

In this study, we analyzed the observed data of 21 repeating and 571 non-repeating FRBs published and drew the following conclusions:
\begin{itemize}
\item[$\bullet$] The extra-Galactic dispersion measures $ DM_{\rm E}$ of non-repeating and repeating FRBs were found to be log-normally distributed with mean values of $496.9 \, \rm pc\,cm^{-3}$ and \mbox{349.1 ${\rm \,pc\,cm^{-3}}$}, respectively. The M--W--W test showed that the $DM_{\rm E}$ is drawn from a different distribution.
\item[$\bullet$] It was found that the total radio energies of repeating FRBs are log-normally distributed with a mean value of $2.51\times10^{38}$ ergs (1.03 dex), which is smaller than to that of non-repeating FRBs, which is consistent with the conclusion in~\cite{li2021}. Surprisingly, the bimodal energy distribution discovered in FRB 20121102A by~\citet{Lidi2021} with FAST was not recovered in our repeating FRB sample any more.
\item[$\bullet$] We statistically analyzed the relationships among $DM_{\rm E}$, $S_{\rm{peak}}$, $F_{\rm{obs}}$, $E$ and $W_{\rm{obs}}$, and found that most correlations between them are similar to those of non-repeaters given by~\citet{li2017intensityHuang}, except that the $E-W_{\rm{obs}}$ relation of repeaters is tighter. The statistical results hint that the spatial distribution and the local environments of two samples of FRBs may be different, although more samples are needed to verify our argument, as noted by~\citet{chime2019chimeAndersenBandur} and
\mbox{\citet{fonseca2020nineAndersen}}.
\item[$\bullet$] We constructed a three-parameter relation to be $\log E\simeq-1.31+1.85\log DM_{\rm E}+1.04$ $\log{W_{\rm obs}}$, indicating that longer FRBs usually have larger extra-galactic dispersion measures and more energy releases, which supports the early findings by~\citet{li2021}, no matter whether a FRB is repeating or not.
\item[$\bullet$] We assume that FRBs can be considered as standard candles homogeneously in a flat Euclidean space, with an IDF of $dN/d F_{\rm obs} = A F_{\rm obs}^{-a}$ (or $N(>F_{\rm obs})\propto F_{\rm obs}^{-a+1}$), where $a$ should be theoretically 2.5. Using the averaged fluence as a characteristic quantity, we built the IDF of repeating FRBs as $d N/d F=(2.37 \pm 0.76) \times 10^{3} F_{\rm obs}^{-1.10 \pm 0.14}\,\rm sky^{-1} day^{-1}$. Likewise, using the observed fluence as a statistical indicator, we obtained the IDF as $d N/d F=(1.96 \pm 0.41) \times 10^{3} F_{\rm obs}^{-1.01 \pm 0.16}\,\rm sky^{-1} day^{-1}$. The power-law index ($a_1,a_2$) of IDF deviates from the theoretical value of 2.5 in a flat Euclidean space, and shows that the repeating FRBs may be not uniformly distributed. Assuming the averaged minimum fluence as the threshold, we predicted the detection rate of repeaters to be about $(9.64\pm3.09)\times10^3$ sky$^{-1}$ day$^{-1}$ or $(1.38\pm0.19)\times10^4$ sky$^{-1}$ day$^{-1}$; that is slightly lower in contrast with those non-repeating ones.
\item[$\bullet$] Finally, we constructed and compared the luminosity functions of repeating and non-repeating FRBs. Interestingly, we found that the luminosity functions for both kinds of FRBs can be well characterized by a broken power-law relation; their power-law indices at their luminous ends are equal to $\approx\!\!-1.4$, despite the discrepancy at their less-luminous ends.
\end{itemize}

Observationally, the power-law relation of $N\propto F^{\alpha}$, especially its power-law index $a$, indeed changes largely for different telescopes, owing to many factors, such as the selection effects/biases, the instrumental sensitivity, the sample incompleteness, the poor localizations, the small sample size or the cosmological effect (e.g.,~\cite{macquart2018frb,Bhandari2018TheSF,James2019}), which may cause the relation to deviate from the theoretical form of $N\propto F^{-3/2}$ in a Euclidean space. ~\citet{caleb2016distributions} pointed out by simulations that the slope of the logarithmic relation is mainly determined by cosmological effect and got $a=-0.9\pm0.3$, which matches a uniform distribution of FRBs roughly. Even for the same telescope, the deduced power-law indices could be inconsistent. For instance,~\citet{macquart2018frb} found $a=-2.6^{+0.7}_{-1.3}$ and \mbox{\citet{James2019} obtained $a=-1.18\pm0.24$} for Parkes FEBs. However, these power-law indices cannot show that the FRBs are not distributed in a Euclidean space. It is also why we took the average fluence as a characteristic quantity. Considering all above complex situations, we have assumed a universal slope in order to reduce the influences of uncertain factors on the $logN-logF$ relation.

In addition, we have constructed the luminosity functions for our samples of repeating and non-repeating FRBs on the basis of the k-corrected isotropic luminosities. It was found that the k-corrected luminosity of repeaters and non-repeaters span three and \textbf{six} orders of magnitude, respectively. On average, the repeaters are less energetic than the non-repeaters, which is coincident with~\cite{li2021}. The luminosity distributions of the two samples of FRBs can be well described by a broken power-law function with a break luminosity of $L_{\rm b}=1.74 \times 10^{42} \,\rm erg\,s^{-1}$ for non-repeaters and $L_{\rm b}=1.78 \times 10^{42} \,\rm erg\,s^{-1}$ for repeaters. This may imply that either repeaters or non-repeaters can be redivided into low- and high-luminosity types originating from different progenitors.

As addressed above, the $DM_{\rm E}$ distributions of the two samples have different distributions,
which suggests that the majority of FRBs might have different astrophysical origins. Recent observations indicate that both kinds of FRBs may have distinct environments, such as dense, offset and star formation distribution (e.g.,~\cite{Heintz2020,Pleunis2021}). Additionally, there exist some special observed properties (such as polarization, rotation measures and the complex frequency--time structure) and a difference between FRB host galaxies~\citep{marcote2020repeating,piro2021fast,bhardwaj2021nearby}.  Thus, these
observational differences imply the emergence of subpopulations of FRBs.
However, there are several recent works claiming no significant
differences in the host galaxy properties (e.g.,~\cite{Ravi2021,Bhandari2022}).
One possible reason is that the number of FRBs with known host galaxies is still limited currently.
 There is a debate about repeating and non-repeating (or more subclasses) or having a single population~\citep{caleb2018one,palaniswamy2018there,caleb2019areStappers}.
This means that the current classification results from the observations may be due to
observational selection bias, e.g., if the observational period is not in the most
active window for the individual FRBs~\citep{palaniswamy2018there,li2019frb},
and one non-repeating burst identified currently may be a potential repeating FRB,
as was the case for FRB 20171019A and FRB 20180301A.
It may be just a few lucky
cases, but we have probably missed many faint bursts of
other FRBs. Nonetheless, the large sample of known FRBs is starting to show trends
that suggest subclasses based on burst morphology and frequency--time structure
(the downward frequency drift of \mbox{sub-bursts~\citep{chime2019chimeAndersenBandur,hessels2019frb,fonseca2020nineAndersen}},
$\sim\!\!100\%$ linear polarization and negligible circular \mbox{polarization~\citep{nimmo2021highly}};
and some sources show periodicity~\citep{amiri2020periodic,rajwade2020possible,cruces2021repeating}).
The observations in future should be focused on the
FRB events which have relatively high probabilities of becoming repeating FRBs (see also~\cite{li2021}).
When more FRBs are accurately localized, more useful information (such as $DM$, redshift, hosts and local environments of repeaters and
apparently one-off FRBs) can be obtained from the direct observations, and the true LF will be accurately built as a cosmological probe to reveal the nature of FRBs.
\vspace{6pt}

\authorcontributions{Conceptualization, Z.Z.; methodology, Z.Z. and L.L.; software, K.Z.; validation, K.Z., L.L. and Z.Z.; formal analysis, Z.Z.; investigation,  K.Z.; resources, Z.Z.; data curation, Q.L., J.L. and M.J.; writing---original draft preparation,  K.Z.; writing---review and editing, Z.Z. and L.L.; visualization, K.Z. and Q.L.; supervision, Z.Z.; project administration, Z.Z.; funding acquisition, Z.Z. Correspondence and requests for materials should be addressed to Z.Z. (zbzhang@gzu.edu.cn). All authors have read and agreed to the published version of the manuscript.}

\funding{This research was funded by by the National Natural Science Foundation of China (grant numbers U2031118 and U1431126) and the science
research grants from the China Manned Space Project, CMS-CSST-2021-B11. L. B. Li acknowledges support from the Natural Science Foundation of Hebei Province of China (grant number A2020402010). J. J. Luo acknowledges the Youth Science \& Technology Talents Development Project of Guizhou Education Department (No.KY[2022]098).}

\dataavailability{The data presented in this study are openly available in the following websites: \url{https://www.frbcat.org/} and \url{https://www.chime-frb.ca/catalog}.}

\acknowledgments{We appreciate the referees for their constructive and helpful comments and suggestions that have improved the paper greatly. This study was partly supported by the National Natural Science Foundation of China (grant numbers U2031118 and U1431126) and the science research grants from the China Manned Space Project, CMS-CSST-2021-B11. L. B. Li acknowledges support from the Natural Science Foundation of Hebei Province of China (grant number A2020402010). J. J. Luo acknowledges the Youth Science \& Technology Talents Development Project of Guizhou Education Department (No.KY[2022]098).}

\conflictsofinterest{The authors declare no conflict of interest.}

\appendixtitles{no} 
\appendixstart
\appendix
\section[\appendixname~\thesection]{}
The appendix Table \ref{table:tbA1} contains supplementary non-repeating FRBs details and data. Note: The FRBCAT sample were taken from FRB Catalogue~\citep{petroff2016frbcatBarr} (\url{https://www.frbcat.org/} ). CHIME samples of non-repeating FRBs were extracted from CHIME/FRB Catalog 1~\citep{2021arXiv210604352T} (\url{https://www.chime-frb.ca/catalog}). The redshifts were calculated using the method of~\citet{petroff2016frbcatBarr} and~\citet{caleb2016distributions}.
\setlength{\tabcolsep}{1.1mm}{
\renewcommand\arraystretch{0.8}

\begin{table}[H]
\caption{ Extended data.}\label{table:tbA1}
\small
\setlength{\tabcolsep}{1.35mm}

\begin{tabular}{lcccccccc}
\toprule
\multirow{2}{*}[-.2em] {\textbf{\boldmath{TNS Name}}}  & \textbf{\boldmath{$\nu_c$}}   &\textbf{\boldmath{ DM}}    & \textbf{\boldmath{$DM_{\rm{MW}}$  }} & \textbf{\boldmath{$DM_{\rm {E}}$}}  &\textbf{\boldmath{ $S_{\rm{peak}}$ }} &\multirow{2}{*}{ z} & \textbf{\boldmath{$D_{\rm L}$}} & \textbf{\boldmath{\boldmath$L_{\rm p}$ }}\\
              & \textbf{\boldmath{(MHz)}}   & \textbf{\boldmath{($\rm{pc\,cm^{-3}}$) }}&\textbf{\boldmath{($\rm{pc\,cm^{-3}}$)  }}&\textbf{\boldmath{($\rm{pc\,cm^{-3}}$) }} &\textbf{\boldmath{(Jy\,ms) }}&    &\textbf{\boldmath{(Gpc)}} &\textbf{\boldmath{($\rm{erg\,s^{-1}}$)}}\\
              \midrule
FRBCAT sample\\
FRB 20010125A & 1372.5     & 790.30             & 110.00 & 680.30  & 0.54   & 0.57 & 3.42          & 8.91~$\times \ 10^{42}$  \\
FRB 20010305A & 1374       & 350.00             & 36.00  & 314.00  & 4.20   & 0.30 & 1.60          & 1.62~$\times \ 10^{43}$  \\
FRB 20010312A & 1374       & 1187.00            & 51.00  & 1136.00 & 0.25   & 1.40 & 10.29         & 3.25~$\times \ 10^{43}$  \\
FRB 20010621A & 1374       & 745.00             & 523.00 & 225.00  & 0.53   & 0.19 & 0.96          & 7.50~$\times \ 10^{41}$  \\
FRB 20010724A & 1374       & 375.00             & 44.58  & 330.42  & 30.00  & 0.28 & 1.48          & 9.94~$\times \ 10^{43}$  \\
\bottomrule
\end{tabular}
\end{table}

\begin{table}[H]\ContinuedFloat
\caption{ \em{Cont.}}
\setlength{\tabcolsep}{1.43mm}
\small
\begin{tabular}{lcccccccc}
\toprule
\multirow{2}{*}[-.2em] {\textbf{\boldmath{TNS Name}}}  & \textbf{\boldmath{$\nu_c$}}   &\textbf{\boldmath{ DM}}    & \textbf{\boldmath{$DM_{\rm{MW}}$  }} & \textbf{\boldmath{$DM_{\rm {E}}$}}  &\textbf{\boldmath{ $S_{\rm{peak}}$ }} &\multirow{2}{*}{ z} & \textbf{\boldmath{$D_{\rm L}$}} & \textbf{\boldmath{\boldmath$L_{\rm p}$ }}\\
              & \textbf{\boldmath{(MHz)}}   & \textbf{\boldmath{($\rm{pc\,cm^{-3}}$) }}&\textbf{\boldmath{($\rm{pc\,cm^{-3}}$)  }}&\textbf{\boldmath{($\rm{pc\,cm^{-3}}$) }} &\textbf{\boldmath{(Jy\,ms) }}&    &\textbf{\boldmath{(Gpc)}} &\textbf{\boldmath{($\rm{erg\,s^{-1}}$)}}\\
              \midrule

FRB 20170906A & 1297       & 390.30             & 39.00  & 351.30  & 29.60  & 0.29 & 1.54          & 1.00~$\times \ 10^{44}$  \\
FRB 20170922A & 835        & 1111.00            & 45.00  & 1066.00 & 5.19   & 1.20 & 8.51          & 2.88~$\times \ 10^{44}$  \\
FRB 20171003A & 1297       & 463.20             & 40.00  & 423.20  & 40.50  & 0.35 & 1.91          & 2.08~$\times \ 10^{44}$  \\
FRB 20171004A & 1297       & 304.00             & 38.00  & 266.00  & 22.00  & 0.22 & 1.13          & 4.05~$\times \ 10^{43}$  \\
FRB 20171019A & 1297       & 114.10             & 38.00  & 76.10   & 117.60 & 0.06 & 0.28          & 1.38~$\times \ 10^{43}$  \\
FRB 20171020A & 1297       & 618.50             & 36.00  & 582.50  & 19.60  & 0.49 & 2.85          & 2.16~$\times \ 10^{44}$  \\
\bottomrule
\end{tabular}
\end{table}

\begin{table}[H]\ContinuedFloat
\caption{ \em{Cont.}}\label{table:tbA1}
\small
\setlength{\tabcolsep}{1.46mm}
\begin{tabular}{lcccccccc}
\toprule
\multirow{2}{*}[-.2em] {\textbf{\boldmath{TNS Name}}}  & \textbf{\boldmath{$\nu_c$}}   &\textbf{\boldmath{ DM}}    & \textbf{\boldmath{$DM_{\rm{MW}}$  }} & \textbf{\boldmath{$DM_{\rm {E}}$}}  &\textbf{\boldmath{ $S_{\rm{peak}}$ }} &\multirow{2}{*}{ z} & \textbf{\boldmath{$D_{\rm L}$}} & \textbf{\boldmath{\boldmath$L_{\rm p}$ }}\\
              & \textbf{\boldmath{(MHz)}}   & \textbf{\boldmath{($\rm{pc\,cm^{-3}}$) }}&\textbf{\boldmath{($\rm{pc\,cm^{-3}}$)  }}&\textbf{\boldmath{($\rm{pc\,cm^{-3}}$) }} &\textbf{\boldmath{(Jy\,ms) }}&    &\textbf{\boldmath{(Gpc)}} &\textbf{\boldmath{($\rm{erg\,s^{-1}}$)}}\\
              \midrule

FRB20180727A  & 600        & 642.13             & 21.23   & 620.90  & 0.58   & 0.52 & 3.04          & 3.35~$\times \ 10^{42}$  \\
FRB20180729A  & 600        & 109.59             & 30.79   & 78.80   & 11.70  & 0.07 & 0.30          & 7.65~$\times \ 10^{41}$  \\
FRB20180729B  & 600        & 317.22             & 94.02   & 223.20  & 0.92   & 0.19 & 0.93          & 5.43~$\times \ 10^{41}$  \\
FRB20180730A  & 600        & 848.90             & 59.20   & 789.70  & 5.20   & 0.66 & 4.07          & 5.22~$\times \ 10^{43}$  \\
FRB20180801A  & 600        & 655.73             & 90.13   & 565.60  & 1.11   & 0.47 & 2.72          & 5.18~$\times \ 10^{42}$  \\
FRB20180806A  & 600        & 739.95             & 40.65   & 699.30  & 1.90   & 0.58 & 3.51          & 1.44~$\times \ 10^{43}$  \\
\bottomrule
\end{tabular}
\end{table}

\begin{table}[H]\ContinuedFloat
\caption{ \em{Cont.}}\label{table:tbA1}
\setlength{\tabcolsep}{1.46mm}
\small
\begin{tabular}{lcccccccc}
\toprule
\multirow{2}{*}[-.2em] {\textbf{\boldmath{TNS Name}}}  & \textbf{\boldmath{$\nu_c$}}   &\textbf{\boldmath{ DM}}    & \textbf{\boldmath{$DM_{\rm{MW}}$  }} & \textbf{\boldmath{$DM_{\rm {E}}$}}  &\textbf{\boldmath{ $S_{\rm{peak}}$ }} &\multirow{2}{*}{ z} & \textbf{\boldmath{$D_{\rm L}$}} & \textbf{\boldmath{\boldmath$L_{\rm p}$ }}\\
              & \textbf{\boldmath{(MHz)}}   & \textbf{\boldmath{($\rm{pc\,cm^{-3}}$) }}&\textbf{\boldmath{($\rm{pc\,cm^{-3}}$)  }}&\textbf{\boldmath{($\rm{pc\,cm^{-3}}$) }} &\textbf{\boldmath{(Jy\,ms) }}&    &\textbf{\boldmath{(Gpc)}} &\textbf{\boldmath{($\rm{erg\,s^{-1}}$)}}\\
              \midrule

FRB20181019B  & 600        & 725.18             & 159.78       & 565.40  & 0.72   & 0.47 & 2.72          & 3.36~$\times \ 10^{42}$  \\
FRB20181019C  & 600        & 501.64             & 39.24       & 462.40  & 1.16   & 0.39 & 2.14          & 3.43~$\times \ 10^{42}$  \\
FRB20181020A  & 600        & 1112.47            & 72.17      & 1040.30 & 0.80   & 0.87 & 5.70          & 1.51~$\times \ 10^{43}$  \\
FRB20181022C  & 600        & 528.47             & 49.87       & 478.60  & 0.91   & 0.40 & 2.23          & 2.91~$\times \ 10^{42}$  \\
FRB20181022D  & 600        & 514.33             & 20.13       & 494.20  & 2.90   & 0.41 & 2.32          & 9.97~$\times \ 10^{42}$  \\
FRB20181022E  & 600        & 285.99             & 21.79      & 264.20  & 0.69   & 0.22 & 1.13          & 5.88~$\times \ 10^{41}$  \\
\bottomrule
\end{tabular}
\end{table}

\begin{table}[H]\ContinuedFloat
\caption{ \em{Cont.}}\label{table:tbA1}
\setlength{\tabcolsep}{1.46mm}
\small
\begin{tabular}{lcccccccc}
\toprule
\multirow{2}{*}[-.2em] {\textbf{\boldmath{TNS Name}}}  & \textbf{\boldmath{$\nu_c$}}   &\textbf{\boldmath{ DM}}    & \textbf{\boldmath{$DM_{\rm{MW}}$  }} & \textbf{\boldmath{$DM_{\rm {E}}$}}  &\textbf{\boldmath{ $S_{\rm{peak}}$ }} &\multirow{2}{*}{ z} & \textbf{\boldmath{$D_{\rm L}$}} & \textbf{\boldmath{\boldmath$L_{\rm p}$ }}\\
              & \textbf{\boldmath{(MHz)}}   & \textbf{\boldmath{($\rm{pc\,cm^{-3}}$) }}&\textbf{\boldmath{($\rm{pc\,cm^{-3}}$)  }}&\textbf{\boldmath{($\rm{pc\,cm^{-3}}$) }} &\textbf{\boldmath{(Jy\,ms) }}&    &\textbf{\boldmath{(Gpc)}} &\textbf{\boldmath{($\rm{erg\,s^{-1}}$)}}\\
              \midrule

FRB20181214B  & 600        & 1120.75            & 42.05      & 1078.70 & 0.41   & 0.90 & 5.96          & 8.43~$\times \ 10^{42}$  \\
FRB20181214C  & 600        & 632.78             & 33.28      & 599.50  & 1.20   & 0.50 & 2.92          & 6.40~$\times \ 10^{42}$  \\
FRB20181214D  & 600        & 1177.32            & 26.02      & 1151.30 & 0.55   & 0.96 & 6.45          & 1.31~$\times \ 10^{43}$  \\
FRB20181214F  & 600        & 2105.76            & 40.26      & 2065.50 & 0.31   & 1.72 & 13.26         & 2.80~$\times \ 10^{43}$  \\
FRB20181215A  & 600        & 412.63             & 128.53       & 284.10  & 0.34   & 0.24 & 1.22          & 3.40~$\times \ 10^{41}$  \\
FRB20181215B  & 600        & 494.01             & 40.61      & 453.40  & 1.90   & 0.38 & 2.09          & 5.37~$\times \ 10^{42}$  \\
\bottomrule
\end{tabular}
\end{table}

\begin{table}[H]\ContinuedFloat
\caption{ \em{Cont.}}\label{table:tbA1}
\setlength{\tabcolsep}{1.46mm}
\small
\begin{tabular}{lcccccccc}
\toprule
\multirow{2}{*}[-.2em] {\textbf{\boldmath{TNS Name}}}  & \textbf{\boldmath{$\nu_c$}}   &\textbf{\boldmath{ DM}}    & \textbf{\boldmath{$DM_{\rm{MW}}$  }} & \textbf{\boldmath{$DM_{\rm {E}}$}}  &\textbf{\boldmath{ $S_{\rm{peak}}$ }} &\multirow{2}{*}{ z} & \textbf{\boldmath{$D_{\rm L}$}} & \textbf{\boldmath{\boldmath$L_{\rm p}$ }}\\
              & \textbf{\boldmath{(MHz)}}   & \textbf{\boldmath{($\rm{pc\,cm^{-3}}$) }}&\textbf{\boldmath{($\rm{pc\,cm^{-3}}$)  }}&\textbf{\boldmath{($\rm{pc\,cm^{-3}}$) }} &\textbf{\boldmath{(Jy\,ms) }}&    &\textbf{\boldmath{(Gpc)}} &\textbf{\boldmath{($\rm{erg\,s^{-1}}$)}}\\
              \midrule

FRB20190103E  & 600        & 736.25             & 45.65      & 690.60  & 0.46   & 0.58 & 3.46          & 3.39~$\times \ 10^{42}$  \\
FRB20190104A  & 600        & 549.43             & 151.53       & 397.90  & 1.45   & 0.33 & 1.80          & 3.06~$\times \ 10^{42}$  \\
FRB20190104B  & 600        & 530.14             & 49.84      & 480.30  & 2.70   & 0.40 & 2.24          & 8.70~$\times \ 10^{42}$  \\
FRB20190105A  & 600        & 383.55             & 46.35      & 337.20  & 0.60   & 0.28 & 1.49          & 8.75~$\times \ 10^{41}$  \\
FRB20190106A  & 600        & 340.06             & 88.86      & 251.20  & 0.27   & 0.21 & 1.06          & 2.06~$\times \ 10^{41}$  \\
FRB20190106B  & 600        & 316.59             & 141.69      & 174.90  & 1.70   & 0.15 & 0.71          & 5.94~$\times \ 10^{41}$  \\
\bottomrule
\end{tabular}
\end{table}

\begin{table}[H]\ContinuedFloat
\caption{ \em{Cont.}}\label{table:tbA1}
\setlength{\tabcolsep}{1.46mm}
\small
\begin{tabular}{lcccccccc}
\toprule
\multirow{2}{*}[-.2em] {\textbf{\boldmath{TNS Name}}}  & \textbf{\boldmath{$\nu_c$}}   &\textbf{\boldmath{ DM}}    & \textbf{\boldmath{$DM_{\rm{MW}}$  }} & \textbf{\boldmath{$DM_{\rm {E}}$}}  &\textbf{\boldmath{ $S_{\rm{peak}}$ }} &\multirow{2}{*}{ z} & \textbf{\boldmath{$D_{\rm L}$}} & \textbf{\boldmath{\boldmath$L_{\rm p}$ }}\\
              & \textbf{\boldmath{(MHz)}}   & \textbf{\boldmath{($\rm{pc\,cm^{-3}}$) }}&\textbf{\boldmath{($\rm{pc\,cm^{-3}}$)  }}&\textbf{\boldmath{($\rm{pc\,cm^{-3}}$) }} &\textbf{\boldmath{(Jy\,ms) }}&    &\textbf{\boldmath{(Gpc)}} &\textbf{\boldmath{($\rm{erg\,s^{-1}}$)}}\\
              \midrule

FRB20190202A  & 600        & 307.36             & 40.76       & 266.60  & 41.00  & 0.22 & 1.14          & 3.56~$\times \ 10^{43}$  \\
FRB20190202B  & 600        & 464.91             & 70.11     & 394.80  & 1.57   & 0.33 & 1.78          & 3.25~$\times \ 10^{42}$  \\
FRB20190203A  & 600        & 420.57             & 45.87       & 374.70  & 1.21   & 0.31 & 1.68          & 2.23~$\times \ 10^{42}$  \\
FRB20190203B  & 600        & 582.22             & 45.82       & 536.40  & 0.49   & 0.45 & 2.55          & 2.03~$\times \ 10^{42}$  \\
FRB20190203C  & 600        & 370.46             & 29.66      & 340.80  & 1.40   & 0.28 & 1.50          & 2.09~$\times \ 10^{42}$  \\
FRB20190204A  & 600        & 449.64             & 36.14      & 413.50  & 0.24   & 0.34 & 1.88          & 5.52~$\times \ 10^{41}$  \\
\bottomrule
\end{tabular}
\end{table}

\begin{table}[H]\ContinuedFloat
\caption{ \em{Cont.}}\label{table:tbA1}
\setlength{\tabcolsep}{1.46mm}
\small
\begin{tabular}{lcccccccc}
\toprule
\multirow{2}{*}[-.2em] {\textbf{\boldmath{TNS Name}}}  & \textbf{\boldmath{$\nu_c$}}   &\textbf{\boldmath{ DM}}    & \textbf{\boldmath{$DM_{\rm{MW}}$  }} & \textbf{\boldmath{$DM_{\rm {E}}$}}  &\textbf{\boldmath{ $S_{\rm{peak}}$ }} &\multirow{2}{*}{ z} & \textbf{\boldmath{$D_{\rm L}$}} & \textbf{\boldmath{\boldmath$L_{\rm p}$ }}\\
              & \textbf{\boldmath{(MHz)}}   & \textbf{\boldmath{($\rm{pc\,cm^{-3}}$) }}&\textbf{\boldmath{($\rm{pc\,cm^{-3}}$)  }}&\textbf{\boldmath{($\rm{pc\,cm^{-3}}$) }} &\textbf{\boldmath{(Jy\,ms) }}&    &\textbf{\boldmath{(Gpc)}} &\textbf{\boldmath{($\rm{erg\,s^{-1}}$)}}\\
              \midrule

FRB20190227B  & 600        & 331.23             & 23.93      & 307.30  & 0.48   & 0.26 & 1.34          & 5.70~$\times \ 10^{41}$  \\
FRB20190228A  & 600        & 419.08             & 20.18      & 398.90  & 1.79   & 0.33 & 1.80          & 3.80~$\times \ 10^{42}$  \\
FRB20190228B  & 600        & 1115.25            & 71.35       & 1043.90 & 4.80   & 0.87 & 5.72          & 9.16~$\times \ 10^{43}$  \\
FRB20190301B  & 600        & 621.33             & 82.83      & 538.50  & 0.40   & 0.45 & 2.57          & 1.67~$\times \ 10^{42}$  \\
FRB20190301C  & 600        & 802.91             & 20.51      & 782.40  & 0.34   & 0.65 & 4.02          & 3.34~$\times \ 10^{42}$  \\
FRB20190301D  & 600        & 1160.69            & 53.09       & 1107.60 & 0.39   & 0.92 & 6.15          & 8.53~$\times \ 10^{42}$  \\
\bottomrule
\end{tabular}
\end{table}

\begin{table}[H]\ContinuedFloat
\caption{ \em{Cont.}}\label{table:tbA1}
\setlength{\tabcolsep}{1.46mm}
\small
\begin{tabular}{lcccccccc}
\toprule
\multirow{2}{*}[-.2em] {\textbf{\boldmath{TNS Name}}}  & \textbf{\boldmath{$\nu_c$}}   &\textbf{\boldmath{ DM}}    & \textbf{\boldmath{$DM_{\rm{MW}}$  }} & \textbf{\boldmath{$DM_{\rm {E}}$}}  &\textbf{\boldmath{ $S_{\rm{peak}}$ }} &\multirow{2}{*}{ z} & \textbf{\boldmath{$D_{\rm L}$}} & \textbf{\boldmath{\boldmath$L_{\rm p}$ }}\\
              & \textbf{\boldmath{(MHz)}}   & \textbf{\boldmath{($\rm{pc\,cm^{-3}}$) }}&\textbf{\boldmath{($\rm{pc\,cm^{-3}}$)  }}&\textbf{\boldmath{($\rm{pc\,cm^{-3}}$) }} &\textbf{\boldmath{(Jy\,ms) }}&    &\textbf{\boldmath{(Gpc)}} &\textbf{\boldmath{($\rm{erg\,s^{-1}}$)}}\\
              \midrule

FRB20190403A  & 600        & 518.83             & 54.23      & 464.60  & 1.12   & 0.39 & 2.15          & 3.35~$\times \ 10^{42}$  \\
FRB20190403B  & 600        & 292.47             & 54.07      & 238.40  & 2.70   & 0.20 & 1.00          & 1.84~$\times \ 10^{42}$  \\
FRB20190403C  & 600        & 935.01             & 38.91      & 896.10  & 0.37   & 0.75 & 4.74          & 4.96~$\times \ 10^{42}$  \\
FRB20190403D  & 600        & 613.46             & 62.96      & 550.50  & 0.65   & 0.46 & 2.63          & 2.85~$\times \ 10^{42}$  \\
FRB20190403E  & 600        & 226.20             & 49.90      & 176.30  & 3.90   & 0.15 & 0.72          & 1.39~$\times \ 10^{42}$  \\
FRB20190403F  & 600        & 664.18             & 67.38       & 596.80  & 0.58   & 0.50 & 2.90          & 3.06~$\times \ 10^{42}$  \\
\bottomrule
\end{tabular}
\end{table}

\begin{table}[H]\ContinuedFloat
\caption{ \em{Cont.}}\label{table:tbA1}
\setlength{\tabcolsep}{1.46mm}
\small
\begin{tabular}{lcccccccc}
\toprule
\multirow{2}{*}[-.2em] {\textbf{\boldmath{TNS Name}}}  & \textbf{\boldmath{$\nu_c$}}   &\textbf{\boldmath{ DM}}    & \textbf{\boldmath{$DM_{\rm{MW}}$  }} & \textbf{\boldmath{$DM_{\rm {E}}$}}  &\textbf{\boldmath{ $S_{\rm{peak}}$ }} &\multirow{2}{*}{ z} & \textbf{\boldmath{$D_{\rm L}$}} & \textbf{\boldmath{\boldmath$L_{\rm p}$ }}\\
              & \textbf{\boldmath{(MHz)}}   & \textbf{\boldmath{($\rm{pc\,cm^{-3}}$) }}&\textbf{\boldmath{($\rm{pc\,cm^{-3}}$)  }}&\textbf{\boldmath{($\rm{pc\,cm^{-3}}$) }} &\textbf{\boldmath{(Jy\,ms) }}&    &\textbf{\boldmath{(Gpc)}} &\textbf{\boldmath{($\rm{erg\,s^{-1}}$)}}\\
              \midrule

FRB20190430B  & 600        & 2619.40            & 36.40       & 2583.00 & 0.38   & 2.15 & 17.43         & 5.65~$\times \ 10^{43}$  \\
FRB20190430C  & 600        & 400.56             & 99.06      & 301.50  & 2.17   & 0.25 & 1.31          & 2.47~$\times \ 10^{42}$  \\
FRB20190501B  & 600        & 784.07             & 43.57       & 740.50  & 0.88   & 0.62 & 3.76          & 7.61~$\times \ 10^{42}$  \\
FRB20190502A  & 600        & 625.77             & 34.97      & 590.80  & 3.18   & 0.49 & 2.87          & 1.64~$\times \ 10^{43}$  \\
FRB20190502B  & 600        & 918.61             & 33.81      & 884.80  & 2.58   & 0.74 & 4.67          & 3.36~$\times \ 10^{43}$  \\
FRB20190502C  & 600        & 396.84             & 47.44      & 349.40  & 3.60   & 0.29 & 1.55          & 5.68~$\times \ 10^{42}$  \\
\bottomrule
\end{tabular}
\end{table}

\begin{table}[H]\ContinuedFloat
\caption{ \em{Cont.}}\label{table:tbA1}
\setlength{\tabcolsep}{1.46mm}
\small
\begin{tabular}{lcccccccc}
\toprule
\multirow{2}{*}[-.2em] {\textbf{\boldmath{TNS Name}}}  & \textbf{\boldmath{$\nu_c$}}   &\textbf{\boldmath{ DM}}    & \textbf{\boldmath{$DM_{\rm{MW}}$  }} & \textbf{\boldmath{$DM_{\rm {E}}$}}  &\textbf{\boldmath{ $S_{\rm{peak}}$ }} &\multirow{2}{*}{ z} & \textbf{\boldmath{$D_{\rm L}$}} & \textbf{\boldmath{\boldmath$L_{\rm p}$ }}\\
              & \textbf{\boldmath{(MHz)}}   & \textbf{\boldmath{($\rm{pc\,cm^{-3}}$) }}&\textbf{\boldmath{($\rm{pc\,cm^{-3}}$)  }}&\textbf{\boldmath{($\rm{pc\,cm^{-3}}$) }} &\textbf{\boldmath{(Jy\,ms) }}&    &\textbf{\boldmath{(Gpc)}} &\textbf{\boldmath{($\rm{erg\,s^{-1}}$)}}\\
              \midrule

FRB20190614A  & 600        & 1064.04            & 52.34      & 1011.70 & 0.83   & 0.84 & 5.51          & 1.47~$\times \ 10^{43}$  \\
FRB20190614B  & 600        & 581.91             & 54.91      & 527.00  & 4.50   & 0.44 & 2.50          & 1.79~$\times \ 10^{43}$  \\
FRB20190614C  & 600        & 589.16             & 57.46      & 531.70  & 0.50   & 0.44 & 2.53          & 2.03~$\times \ 10^{42}$  \\
FRB20190616A  & 600        & 212.59             & 25.49      & 187.10  & 0.73   & 0.16 & 0.77          & 2.95~$\times \ 10^{41}$  \\
FRB20190617A  & 600        & 195.77             & 46.67      & 149.10  & 5.80   & 0.12 & 0.60          & 1.44~$\times \ 10^{42}$  \\
FRB20190617B  & 600        & 273.51             & 43.81       & 229.70  & 0.99   & 0.19 & 0.96          & 6.22~$\times \ 10^{41}$  \\
\bottomrule
\end{tabular}
\end{table}
\begin{adjustwidth}{-\extralength}{0cm}

\begin{thebibliography}{999}

\end{thebibliography}


\begin{thebibliography}{999}



\bibitem[{Lorimer} {et~al.}(2007){Lorimer}, {Bailes}, {McLaughlin},
  {Narkevic}, and {Crawford}]{lorimer2007aBailes}
{Lorimer}, D.R.; {Bailes}, M.; {McLaughlin}, M.A.; {Narkevic}, D.J.;
  {Crawford}, F.
\newblock {A Bright Millisecond Radio Burst of Extragalactic Origin}.
\newblock {\em Science} {\bf 2007}, {\em 318},~777.
  \href{http://xxx.lanl.gov/abs/0709.4301}{{
 }}
\newblock {{https://doi.org/10.1126/science.1147532}}.

\bibitem[{Thornton} {et~al.}(2013){Thornton}, {Stappers}, {Bailes},
  {Barsdell}, {Bates}, {Bhat}, {Burgay}, {Burke-Spolaor}, {Champion}, {Coster},
  {D'Amico}, {Jameson}, {Johnston}, {Keith}, {Kramer}, {Levin}, {Milia}, {Ng},
  {Possenti}, and {van Straten}]{thornton2013aStappers}
{Thornton}, D.; {Stappers}, B.; {Bailes}, M.; {Barsdell}, B.; {Bates}, S.;
  {Bhat}, N.D.R.; {Burgay}, M.; {Burke-Spolaor}, S.; {Champion}, D.J.;
  {Coster}, P.;  et~al.
\newblock {A Population of Fast Radio Bursts at Cosmological Distances}.
\newblock {\em Science} {\bf 2013}, {\em 341},~53--56.
  \href{http://xxx.lanl.gov/abs/1307.1628}{{
 }}
\newblock {{https://doi.org/10.1126/science.1236789}}.

\bibitem[{Cordes} and {Chatterjee}(2019)]{cordes2019fastChatterjee}
{Cordes}, J.M.; {Chatterjee}, S.
\newblock {Fast Radio Bursts: An Extragalactic Enigma}.
\newblock {\em Annu. Rev. Astron. Astrophys.} {\bf 2019}, {\em
  57},~417--465.  \href{http://xxx.lanl.gov/abs/1906.05878}{{
 }}
\newblock {{https://doi.org/10.1146/annurev-astro-091918-104501}}.

\bibitem[{Petroff} {et~al.}(2016){Petroff}, {Barr}, {Jameson}, {Keane},
  {Bailes}, {Kramer}, {Morello}, {Tabbara}, and {van
  Straten}]{petroff2016frbcatBarr}
{Petroff}, E.; {Barr}, E.D.; {Jameson}, A.; {Keane}, E.F.; {Bailes}, M.;
  {Kramer}, M.; {Morello}, V.; {Tabbara}, D.; {van Straten}, W.
\newblock {FRBCAT: The Fast Radio Burst Catalogue}.
\newblock {\em Publ. Astron. Soc. Aust.} {\bf
  2016}, {\em 33},~e045.
  \href{http://xxx.lanl.gov/abs/1601.03547}{{
 }}
\newblock {{https://doi.org/10.1017/pasa.2016.35}}.

\bibitem[{Li} {et~al.}(2021){Li}, {Dong}, {Zhang}, and {Li}]{li2021}
{Li}, X.J.; {Dong}, X.F.; {Zhang}, Z.B.; {Li}, D.
\newblock {Long and Short Fast Radio Bursts Are Different from Repeating and
  Nonrepeating Transients}.
\newblock {\em  Astrophys. J.} {\bf 2021}, {\em 923},~230.
  \href{http://xxx.lanl.gov/abs/2110.07227}{{
 }}
\newblock {{https://doi.org/10.3847/1538-4357/ac3085}}.

\bibitem[{Keane} and {Petroff}(2015)]{keane2015fastPetroff}
{Keane}, E.F.; {Petroff}, E.
\newblock {Fast radio bursts: Search sensitivities and completeness}.
\newblock {\em Mon. Not. R. Astron. Soc.} {\bf 2015},
  {\em 447},~2852--2856.  \href{http://xxx.lanl.gov/abs/1409.6125}{{
 }}
\newblock {{https://doi.org/10.1093/mnras/stu2650}}.

\bibitem[{Law} {et~al.}(2015){Law}, {Bower}, {Burke-Spolaor}, {Butler},
  {Lawrence}, {Lazio}, {Mattmann}, {Rupen}, {Siemion}, and
  {VanderWiel}]{law2015aBower}
{Law}, C.J.; {Bower}, G.C.; {Burke-Spolaor}, S.; {Butler}, B.; {Lawrence}, E.;
  {Lazio}, T.J.W.; {Mattmann}, C.A.; {Rupen}, M.; {Siemion}, A.; {VanderWiel},
  S.
\newblock {A Millisecond Interferometric Search for Fast Radio Bursts with the
  Very Large Array}.
\newblock {\em  Astrophys. J.} {\bf 2015}, {\em 807},~16.
  \href{http://xxx.lanl.gov/abs/1412.7536}{{
 }}
\newblock {{https://doi.org/10.1088/0004-637X/807/1/16}}.

\bibitem[{Champion} {et~al.}(2016){Champion}, {Petroff}, {Kramer}, {Keith},
  {Bailes}, {Barr}, {Bates}, {Bhat}, {Burgay}, {Burke-Spolaor}, {Flynn},
  {Jameson}, {Johnston}, {Ng}, {Levin}, {Possenti}, {Stappers}, {van Straten},
  {Thornton}, {Tiburzi}, and {Lyne}]{champion2016fivePetroff}
{Champion}, D.J.; {Petroff}, E.; {Kramer}, M.; {Keith}, M.J.; {Bailes}, M.;
  {Barr}, E.D.; {Bates}, S.D.; {Bhat}, N.D.R.; {Burgay}, M.; {Burke-Spolaor},
  S.;  et~al.
\newblock {Five new fast radio bursts from the HTRU high-latitude survey at
  Parkes: First evidence for two-component bursts}.
\newblock {\em Mon. Not. R. Astron. Soc. Lett.} {\bf
  2016}, {\em 460},~L30--L34.
  \href{http://xxx.lanl.gov/abs/1511.07746}{{
 }}
\newblock {{https://doi.org/10.1093/mnrasl/slw069}}.

\bibitem[{Oppermann} {et~al.}(2016){Oppermann}, {Connor}, and
  {Pen}]{oppermann2016euclidean}
{Oppermann}, N.; {Connor}, L.D.; {Pen}, U.L.
\newblock {The Euclidean distribution of fast radio bursts}.
\newblock {\em Mon. Not. R. Astron. Soc.} {\bf 2016},
  {\em 461},~984--987.  \href{http://xxx.lanl.gov/abs/1604.03909}{{
 }}
\newblock {{https://doi.org/10.1093/mnras/stw1401}}.

\bibitem[{Rane} {et~al.}(2016){Rane}, {Lorimer}, {Bates}, {McMann},
  {McLaughlin}, and {Rajwade}]{rane2016aLorimer}
{Rane}, A.; {Lorimer}, D.R.; {Bates}, S.D.; {McMann}, N.; {McLaughlin}, M.A.;
  {Rajwade}, K.
\newblock {A search for rotating radio transients and fast radio bursts in the
  Parkes high-latitude pulsar survey}.
\newblock {\em Mon. Not. R. Astron. Soc.} {\bf 2016},
  {\em 455},~2207--2215.
  \href{http://xxx.lanl.gov/abs/1505.00834}{{
 }}
\newblock {{https://doi.org/10.1093/mnras/stv2404}}.

\bibitem[{Bhandari} {et~al.}(2018){Bhandari}, {Keane}, {Barr}, {Jameson},
  {Petroff}, {Johnston}, {Bailes}, {Bhat}, {Burgay}, {Burke-Spolaor}, {Caleb},
  {Eatough}, {Flynn}, {Green}, {Jankowski}, {Kramer}, {Krishnan}, {Morello},
  {Possenti}, {Stappers}, {Tiburzi}, {van Straten}, {Andreoni}, {Butterley},
  {Chandra}, {Cooke}, {Corongiu}, {Coward}, {Dhillon}, {Dodson}, {Hardy},
  {Howell}, {Jaroenjittichai}, {Klotz}, {Littlefair}, {Marsh}, {Mickaliger},
  {Muxlow}, {Perrodin}, {Pritchard}, {Sawangwit}, {Terai}, {Tominaga}, {Torne},
  {Totani}, {Trois}, {Turpin}, {Niino}, {Wilson}, {Albert}, {Andr{\'e}},
  {Anghinolfi}, {Anton}, {Ardid}, {Aubert}, {Avgitas}, {Baret},
  {Barrios-Mart{\'\i}}, {Basa}, {Belhorma}, {Bertin}, {Biagi}, {Bormuth},
  {Bourret}, {Bouwhuis}, {Br{\^a}nza{\c{s}}}, {Bruijn}, {Brunner}, {Busto},
  {Capone}, {Caramete}, {Carr}, {Celli}, {Moursli}, {Chiarusi}, {Circella},
  {Coelho}, {Coleiro}, {Coniglione}, {Costantini}, {Coyle}, {Creusot},
  {D{\'\i}az}, {Deschamps}, {De Bonis}, {Distefano}, {Palma}, {Domi},
  {Donzaud}, {Dornic}, {Drouhin}, {Eberl}, {Bojaddaini}, {Khayati},
  {Els{\"a}sser}, {Enzenh{\"o}fer}, {Ettahiri}, {Fassi}, {Felis}, {Fusco},
  {Gay}, {Giordano}, {Glotin}, {Gregoire}, {Gracia-Ruiz}, {Graf}, {Hallmann},
  {van Haren}, {Heijboer}, {Hello}, {Hern{\'a}ndez-Rey}, {H{\"o}{\ss}l},
  {Hofest{\"a}dt}, {Hugon}, {Illuminati}, {James}, {de Jong}, {Jongen},
  {Kadler}, {Kalekin}, {Katz}, {Kie{\ss}ling}, {Kouchner}, {Kreter},
  {Kreykenbohm}, {Kulikovskiy}, {Lachaud}, {Lahmann}, {Lef{\`e}vre}, {Leonora},
  {Loucatos}, {Marcelin}, {Margiotta}, {Marinelli}, {Mart{\'\i}nez-Mora},
  {Mele}, {Melis}, {Michael}, {Migliozzi}, {Moussa}, {Navas}, {Nezri},
  {Organokov}, {P{\v{a}}v{\v{a}}la{\c{s}}}, {Pellegrino}, {Perrina},
  {Piattelli}, {Popa}, {Pradier}, {Quinn}, {Racca}, {Riccobene},
  {S{\'a}nchez-Losa}, {Salda{\~n}a}, {Salvadori}, {Samtleben}, {Sanguineti},
  {Sapienza}, {Sch{\"u}ssler}, {Sieger}, {Spurio}, {Stolarczyk}, {Taiuti},
  {Tayalati}, {Trovato}, {Turpin}, {T{\"o}nnis}, {Vallage}, {Van Elewyck},
  {Versari}, {Vivolo}, {Vizzocca}, {Wilms}, {Zornoza}, and
  {Z{\'u}{\~n}iga}]{Bhandari2018TheSF}
{Bhandari}, S.; {Keane}, E.F.; {Barr}, E.D.; {Jameson}, A.; {Petroff}, E.;
  {Johnston}, S.; {Bailes}, M.; {Bhat}, N.D.R.; {Burgay}, M.; {Burke-Spolaor},
  S.;  et~al.
\newblock {The SUrvey for Pulsars and Extragalactic Radio Bursts - II. New FRB
  discoveries and their follow-up}.
\newblock {\em Mon. Not. R. Astron. Soc.} {\bf 2018},
  {\em 475},~1427--1446.
  \href{http://xxx.lanl.gov/abs/1711.08110}{{
 }}
\newblock {{https://doi.org/10.1093/mnras/stx3074}}.

\bibitem[{Connor} and {Petroff}(2018)]{connor2018detecting}
{Connor}, L.; {Petroff}, E.
\newblock {On Detecting Repetition from Fast Radio Bursts}.
\newblock {\em  Astrophys. J.} {\bf 2018}, {\em 861},~L1.
  \href{http://xxx.lanl.gov/abs/1804.00896}{{
 }}
\newblock {{https://doi.org/10.3847/2041-8213/aacd02}}.

\bibitem[{Patel} {et~al.}(2018){Patel}, {Agarwal}, {Bhardwaj}, {Boyce},
  {Brazier}, {Chatterjee}, {Chawla}, {Kaspi}, {Lorimer}, {McLaughlin},
  {Parent}, {Pleunis}, {Ransom}, {Scholz}, {Wharton}, {Zhu}, {Alam}, {Caballero
  Valdez}, {Camilo}, {Cordes}, {Crawford}, {Deneva}, {Ferdman}, {Freire},
  {Hessels}, {Nguyen}, {Stairs}, {Stovall}, and {van Leeuwen}]{patel2018palfa}
{Patel}, C.; {Agarwal}, D.; {Bhardwaj}, M.; {Boyce}, M.M.; {Brazier}, A.;
  {Chatterjee}, S.; {Chawla}, P.; {Kaspi}, V.M.; {Lorimer}, D.R.; {McLaughlin},
  M.A.;  et~al.
\newblock {PALFA Single-pulse Pipeline: New Pulsars, Rotating Radio Transients,
  and a Candidate Fast Radio Burst}.
\newblock {\em  Astrophys. J.} {\bf 2018}, {\em 869},~181.
  \href{http://xxx.lanl.gov/abs/1808.03710}{{
 }}
\newblock {{https://doi.org/10.3847/1538-4357/aaee65}}.

\bibitem[{Shannon} {et~al.}(2018){Shannon}, {Macquart}, {Bannister},
  {Ekers}, {James}, {Os{\l}owski}, {Qiu}, {Sammons}, {Hotan}, {Voronkov},
  {Beresford}, {Brothers}, {Brown}, {Bunton}, {Chippendale}, {Haskins},
  {Leach}, {Marquarding}, {McConnell}, {Pilawa}, {Sadler}, {Troup}, {Tuthill},
  {Whiting}, {Allison}, {Anderson}, {Bell}, {Collier}, {G{\"u}rkan}, {Heald},
  and {Riseley}]{Shannon2018TheDR}
{Shannon}, R.M.; {Macquart}, J.P.; {Bannister}, K.W.; {Ekers}, R.D.; {James},
  C.W.; {Os{\l}owski}, S.; {Qiu}, H.; {Sammons}, M.; {Hotan}, A.W.; {Voronkov},
  M.A.;  et~al.
\newblock {The dispersion-brightness relation for fast radio bursts from a
  wide-field survey}.
\newblock {\em Nature} {\bf 2018}, {\em 562},~386--390.
\newblock {{https://doi.org/10.1038/s41586-018-0588-y}}.

\bibitem[{Farah} {et~al.}(2019){Farah}, {Flynn}, {Bailes}, {Jameson},
  {Bateman}, {Campbell-Wilson}, {Day}, {Deller}, {Green}, {Gupta}, {Hunstead},
  {Lower}, {Os{\l}owski}, {Parthasarathy}, {Price}, {Ravi}, {Shannon},
  {Sutherland}, {Temby}, {Krishnan}, {Caleb}, {Chang}, {Cruces}, {Roy},
  {Morello}, {Onken}, {Stappers}, {Webb}, and {Wolf}]{Farah2019FiveNR}
{Farah}, W.; {Flynn}, C.; {Bailes}, M.; {Jameson}, A.; {Bateman}, T.;
  {Campbell-Wilson}, D.; {Day}, C.K.; {Deller}, A.T.; {Green}, A.J.; {Gupta},
  V.;  et~al.
\newblock {Five new real-time detections of fast radio bursts with UTMOST}.
\newblock {\em Mon. Not. R. Astron. Soc.} {\bf 2019},
  {\em 488},~2989--3002.
  \href{http://xxx.lanl.gov/abs/1905.02293}{{
 }}
\newblock {{https://doi.org/10.1093/mnras/stz1748}}.

\bibitem[{Parent} {et~al.}(2020){Parent}, {Chawla}, {Kaspi}, {Agazie},
  {Blumer}, {DeCesar}, {Fiore}, {Fonseca}, {Hessels}, {Kaplan}, {Kondratiev},
  {LaRose}, {Levin}, {Lewis}, {Lynch}, {McEwen}, {McLaughlin}, {Mingyar}, {Al
  Noori}, {Ransom}, {Roberts}, {Schmiedekamp}, {Schmiedekamp}, {Siemens},
  {Spiewak}, {Stairs}, {Surnis}, {Swiggum}, and {van
  Leeuwen}]{Parent2020FirstDO}
{Parent}, E.; {Chawla}, P.; {Kaspi}, V.M.; {Agazie}, G.Y.; {Blumer}, H.;
  {DeCesar}, M.; {Fiore}, W.; {Fonseca}, E.; {Hessels}, J.W.T.; {Kaplan}, D.L.;
   et~al.
\newblock {First Discovery of a Fast Radio Burst at 350 MHz by the GBNCC
  Survey}.
\newblock {\em  Astrophys. J.} {\bf 2020}, {\em 904},~92.
  \href{http://xxx.lanl.gov/abs/2008.04217}{{
 }}
\newblock {{https://doi.org/10.3847/1538-4357/abbdf6}}.

\bibitem[{CHIME/FRB Collaboration} {et~al.}(2021){CHIME/FRB Collaboration},
  {Amiri}, {Andersen}, {Bandura}, {Berger}, {Bhardwaj}, {Boyce}, {Boyle},
  {Brar}, {Breitman}, {Cassanelli}, {Chawla}, {Chen}, {Cliche}, {Cook},
  {Cubranic}, {Curtin}, {Deng}, {Dobbs}, {Dong}, {Eadie}, {Fandino}, {Fonseca},
  {Gaensler}, {Giri}, {Good}, {Halpern}, {Hill}, {Hinshaw}, {Josephy},
  {Kaczmarek}, {Kader}, {Kania}, {Kaspi}, {Landecker}, {Lang}, {Leung}, {Li},
  {Lin}, {Masui}, {McKinven}, {Mena-Parra}, {Merryfield}, {Meyers}, {Michilli},
  {Milutinovic}, {Mirhosseini}, {M{\"u}nchmeyer}, {Naidu}, {Newburgh}, {Ng},
  {Patel}, {Pen}, {Petroff}, {Pinsonneault-Marotte}, {Pleunis},
  {Rafiei-Ravandi}, {Rahman}, {Ransom}, {Renard}, {Sanghavi}, {Scholz}, {Shaw},
  {Shin}, {Siegel}, {Sikora}, {Singh}, {Smith}, {Stairs}, {Tan}, {Tendulkar},
  {Vanderlinde}, {Wang}, {Wulf}, and {Zwaniga}]{2021arXiv210604352T}
{CHIME/FRB Collaboration}; {Amiri}, M.; {Andersen}, B.C.; {Bandura}, K.;
  {Berger}, S.; {Bhardwaj}, M.; {Boyce}, M.M.; {Boyle}, P.J.; {Brar}, C.;
  {Breitman}, D.;  et~al.
\newblock {The First CHIME/FRB Fast Radio Burst Catalog}.
\newblock {\em  Astrophys. J.} {\bf 2021}, {\em 257},~59.
  \href{http://xxx.lanl.gov/abs/2106.04352}{{
 }}
\newblock {{https://doi.org/10.3847/1538-4365/ac33ab}}.

\bibitem[{Li} {et~al.}(2021){Li}, {Wang}, {Zhu}, {Zhang}, {Zhang}, {Duan},
  {Zhang}, {Feng}, {Tang}, {Chatterjee}, {Cordes}, {Cruces}, {Dai}, {Gajjar},
  {Hobbs}, {Jin}, {Kramer}, {Lorimer}, {Miao}, {Niu}, {Niu}, {Pan}, {Qian},
  {Spitler}, {Werthimer}, {Zhang}, {Wang}, {Xie}, {Yue}, {Zhang}, {Zhi}, and
  {Zhu}]{Lidi2021}
{Li}, D.; {Wang}, P.; {Zhu}, W.W.; {Zhang}, B.; {Zhang}, X.X.; {Duan}, R.;
  {Zhang}, Y.K.; {Feng}, Y.; {Tang}, N.Y.; {Chatterjee}, S.;  et~al.
\newblock {A bimodal burst energy distribution of a repeating fast radio burst
  source}.
\newblock {\em Nature} {\bf 2021}, {\em 598},~267--271.
  \href{http://xxx.lanl.gov/abs/2107.08205}{{
 }}
\newblock {{https://doi.org/10.1038/s41586-021-03878-5}}.

\bibitem[{Xu} {et~al.}(2021){Xu}, {Niu}, {Chen}, {Lee}, {Zhu}, {Dong},
  {Zhang}, {Jiang}, {Wang}, {Xu}, {Zhang}, {Fu}, {Filippenko}, {Peng}, {Zhou},
  {Zhang}, {Wang}, {Feng}, {Li}, {Brink}, {Li}, {Lu}, {Yang}, {Caballero},
  {Cai}, {Chen}, {Dai}, {Djorgovski}, {Esamdin}, {Gan}, {Guhathakurta}, {Han},
  {Hao}, {Huang}, {Jiang}, {Li}, {Li}, {Li}, {Li}, {Li}, {Liu}, {Luo}, {Men},
  {Niu}, {Peng}, {Qian}, {Song}, {Stern}, {Stockton}, {Sun}, {Wang}, {Wang},
  {Wang}, {Wang}, {Wu}, {Xiao}, {Xiong}, {Xu}, {Xu}, {Yang}, {Yang}, {Yao},
  {Yi}, {Yue}, {Yu}, {Yu}, {Yuan}, {Zhang}, {Zhang}, {Zhang}, {Zhao}, {Zheng},
  {Zhu}, and {Zou}]{xu2021fast}
{Xu}, H.; {Niu}, J.R.; {Chen}, P.; {Lee}, K.J.; {Zhu}, W.W.; {Dong}, S.;
  {Zhang}, B.; {Jiang}, J.C.; {Wang}, B.J.; {Xu}, J.W.;  et~al.
\newblock {A fast radio burst source at a complex magnetised site in a barred
  galaxy}.
\newblock {\em arXiv} {\bf 2021}, arXiv:2111.11764.
  \href{http://xxx.lanl.gov/abs/2111.11764}{{
 }}

\bibitem[{Li} {et~al.}(2018){Li}, {Wang}, {Qian}, {Krco}, {Jiang}, {Yue},
  {Jin}, {Zhu}, {Pan}, {Nan}, and {Dunning}]{Lidi2018}
{Li}, D.; {Wang}, P.; {Qian}, L.; {Krco}, M.; {Jiang}, P.; {Yue}, Y.; {Jin},
  C.; {Zhu}, Y.; {Pan}, Z.; {Nan}, R.;  et~al.
\newblock {FAST in Space: Considerations for a Multibeam, Multipurpose Survey
  Using China's 500-m Aperture Spherical Radio Telescope (FAST)}.
\newblock {\em IEEE Microw. Mag.} {\bf 2018}, {\em 19},~112--119.
  \href{http://xxx.lanl.gov/abs/1802.03709}{{
 }}
\newblock {{https://doi.org/10.1109/MMM.2018.2802178}}.

\bibitem[{CHIME/FRB Collaboration} {et~al.}(2020){CHIME/FRB Collaboration},
  {Andersen}, {Bandura}, {Bhardwaj}, {Bij}, {Boyce}, {Boyle}, {Brar},
  {Cassanelli}, {Chawla}, {Chen}, {Cliche}, {Cook}, {Cubranic}, {Curtin},
  {Denman}, {Dobbs}, {Dong}, {Fandino}, {Fonseca}, {Gaensler}, {Giri}, {Good},
  {Halpern}, {Hill}, {Hinshaw}, {H{\"o}fer}, {Josephy}, {Kania}, {Kaspi},
  {Landecker}, {Leung}, {Li}, {Lin}, {Masui}, {McKinven}, {Mena-Parra},
  {Merryfield}, {Meyers}, {Michilli}, {Milutinovic}, {Mirhosseini},
  {M{\"u}nchmeyer}, {Naidu}, {Newburgh}, {Ng}, {Patel}, {Pen},
  {Pinsonneault-Marotte}, {Pleunis}, {Quine}, {Rafiei-Ravandi}, {Rahman},
  {Ransom}, {Renard}, {Sanghavi}, {Scholz}, {Shaw}, {Shin}, {Siegel}, {Singh},
  {Smegal}, {Smith}, {Stairs}, {Tan}, {Tendulkar}, {Tretyakov}, {Vanderlinde},
  {Wang}, {Wulf}, and {Zwaniga}]{chime2020aAndersen}
{CHIME/FRB Collaboration}; {Andersen}, B.C.; {Bandura}, K.M.; {Bhardwaj}, M.;
  {Bij}, A.; {Boyce}, M.M.; {Boyle}, P.J.; {Brar}, C.; {Cassanelli}, T.;
  {Chawla}, P.;  et~al.
\newblock {A bright millisecond-duration radio burst from a Galactic magnetar}.
\newblock {\em Nature} {\bf 2020}, {\em 587},~54--58.
  \href{http://xxx.lanl.gov/abs/2005.10324}{{
 }}
\newblock {{https://doi.org/10.1038/s41586-020-2863-y}}.

\bibitem[{Bochenek}(2021)]{bochenek2020fast}
{Bochenek}, C.
\newblock {A Fast Radio Burst Associated with a Galactic Magnetar}.
\newblock In {Proceedings of the American Astronomical Society Meeting
  Abstracts, New York,  NY, USA, 11--15 January} 2021; Volume~53, p. 236.05D. 

\bibitem[{Lin} {et~al.}(2020){Lin}, {Zhang}, {Wang}, {Gao}, {Guan}, {Han},
  {Jiang}, {Jiang}, {Lee}, {Li}, {Men}, {Miao}, {Niu}, {Niu}, {Sun}, {Wang},
  {Wang}, {Xu}, {Xu}, {Xu}, {Yang}, {Yang}, {Yu}, {Zhang}, {Zhang}, {Zhou},
  {Zhu}, {Castro-Tirado}, {Dai}, {Ge}, {Hu}, {Li}, {Li}, {Li}, {Liang}, {Jia},
  {Querel}, {Shao}, {Wang}, {Wang}, {Wu}, {Xiong}, {Xu}, {Yang}, {Zhang},
  {Zhang}, {Zheng}, and {Zou}]{lin2020no}
{Lin}, L.; {Zhang}, C.F.; {Wang}, P.; {Gao}, H.; {Guan}, X.; {Han}, J.L.;
  {Jiang}, J.C.; {Jiang}, P.; {Lee}, K.J.; {Li}, D.;  et~al.
\newblock {No pulsed radio emission during a bursting phase of a Galactic
  magnetar}.
\newblock {\em Nature} {\bf 2020}, {\em 587},~63--65.
  \href{http://xxx.lanl.gov/abs/2005.11479}{{
 }}
\newblock {{https://doi.org/10.1038/s41586-020-2839-y}}.

\bibitem[{Mereghetti} {et~al.}(2020){Mereghetti}, {Savchenko}, {Ferrigno},
  {G{\"o}tz}, {Rigoselli}, {Tiengo}, {Bazzano}, {Bozzo}, {Coleiro},
  {Courvoisier}, {Doyle}, {Goldwurm}, {Hanlon}, {Jourdain}, {von Kienlin},
  {Lutovinov}, {Martin-Carrillo}, {Molkov}, {Natalucci}, {Onori}, {Panessa},
  {Rodi}, {Rodriguez}, {S{\'a}nchez-Fern{\'a}ndez}, {Sunyaev}, and
  {Ubertini}]{mereghetti2020integral}
{Mereghetti}, S.; {Savchenko}, V.; {Ferrigno}, C.; {G{\"o}tz}, D.; {Rigoselli},
  M.; {Tiengo}, A.; {Bazzano}, A.; {Bozzo}, E.; {Coleiro}, A.; {Courvoisier},
  T.J.L.;  et~al.
\newblock {INTEGRAL Discovery of a Burst with Associated Radio Emission from
  the Magnetar SGR 1935+2154}.
\newblock {\em  Astrophys. J.} {\bf 2020}, {\em 898},~L29.
  \href{http://xxx.lanl.gov/abs/2005.06335}{{
 }}
\newblock {{https://doi.org/10.3847/2041-8213/aba2cf}}.

\bibitem[{Li} {et~al.}(2021){Li}, {Lin}, {Xiong}, {Ge}, {Li}, {Li}, {Lu},
  {Zhang}, {Tuo}, {Nang}, {Zhang}, {Xiao}, {Chen}, {Song}, {Xu}, {Liu}, {Jia},
  {Cao}, {Qu}, {Zhang}, {Gu}, {Liao}, {Zhao}, {Tan}, {Nie}, {Zhao}, {Zheng},
  {Zheng}, {Luo}, {Cai}, {Li}, {Xue}, {Bu}, {Chang}, {Chen}, {Chen}, {Chen},
  {Chen}, {Chen}, {Cui}, {Cui}, {Deng}, {Dong}, {Du}, {Fu}, {Gao}, {Gao},
  {Gao}, {Gu}, {Guan}, {Guo}, {Han}, {Huang}, {Huo}, {Jiang}, {Jiang}, {Jin},
  {Jin}, {Kong}, {Li}, {Li}, {Li}, {Li}, {Li}, {Li}, {Li}, {Liang}, {Liu},
  {Liu}, {Liu}, {Liu}, {Liu}, {Lu}, {Lu}, {Luo}, {Ma}, {Meng}, {Ou}, {Sai},
  {Shang}, {Song}, {Sun}, {Tao}, {Wang}, {Wang}, {Wang}, {Wang}, {Wang}, {Wen},
  {Wu}, {Wu}, {Wu}, {Xiao}, {Xu}, {Yang}, {Yang}, {Yang}, {Yang}, {Yi}, {Yin},
  {You}, {Zhang}, {Zhang}, {Zhang}, {Zhang}, {Zhang}, {Zhang}, {Zhang},
  {Zhang}, {Zhang}, {Zhang}, {Zhang}, {Zhang}, {Zhang}, {Zhang}, {Zhang},
  {Zhang}, {Zhou}, {Zhou}, {Zhu}, {Zhu}, and {Zhuang}]{li2021hxmt}
{Li}, C.K.; {Lin}, L.; {Xiong}, S.L.; {Ge}, M.Y.; {Li}, X.B.; {Li}, T.P.; {Lu},
  F.J.; {Zhang}, S.N.; {Tuo}, Y.L.; {Nang}, Y.;  et~al.
\newblock {HXMT identification of a non-thermal X-ray burst from SGR J1935+2154
  and with FRB 200428}.
\newblock {\em Nat. Astron.} {\bf 2021}, {\em 5},~378--384.
  \href{http://xxx.lanl.gov/abs/2005.11071}{{
 }}
\newblock {{https://doi.org/10.1038/s41550-021-01302-6}}.

\bibitem[{Ridnaia} {et~al.}(2021){Ridnaia}, {Svinkin}, {Frederiks}, {Bykov},
  {Popov}, {Aptekar}, {Golenetskii}, {Lysenko}, {Tsvetkova}, {Ulanov}, and
  {Cline}]{ridnaia2021peculiar}
{Ridnaia}, A.; {Svinkin}, D.; {Frederiks}, D.; {Bykov}, A.; {Popov}, S.;
  {Aptekar}, R.; {Golenetskii}, S.; {Lysenko}, A.; {Tsvetkova}, A.; {Ulanov},
  M.;  et~al.
\newblock {A peculiar hard X-ray counterpart of a Galactic fast radio burst}.
\newblock {\em Nat. Astron.} {\bf 2021}, {\em 5},~372--377.
  \href{http://xxx.lanl.gov/abs/2005.11178}{{
 }}
\newblock {{https://doi.org/10.1038/s41550-020-01265-0}}.

\bibitem[{Tavani} {et~al.}(2021){Tavani}, {Casentini}, {Ursi}, {Verrecchia},
  {Addis}, {Antonelli}, {Argan}, {Barbiellini}, {Baroncelli}, {Bernardi},
  {Bianchi}, {Bulgarelli}, {Caraveo}, {Cardillo}, {Cattaneo}, {Chen}, {Costa},
  {Del Monte}, {Di Cocco}, {Di Persio}, {Donnarumma}, {Evangelista}, {Feroci},
  {Ferrari}, {Fioretti}, {Fuschino}, {Galli}, {Gianotti}, {Giuliani},
  {Labanti}, {Lazzarotto}, {Lipari}, {Longo}, {Lucarelli}, {Magro},
  {Marisaldi}, {Mereghetti}, {Morelli}, {Morselli}, {Naldi}, {Pacciani},
  {Parmiggiani}, {Paoletti}, {Pellizzoni}, {Perri}, {Perotti}, {Piano},
  {Picozza}, {Pilia}, {Pittori}, {Puccetti}, {Pupillo}, {Rapisarda},
  {Rappoldi}, {Rubini}, {Setti}, {Soffitta}, {Trifoglio}, {Trois},
  {Vercellone}, {Vittorini}, {Giommi}, and {D'Amico}]{tavani2021x}
{Tavani}, M.; {Casentini}, C.; {Ursi}, A.; {Verrecchia}, F.; {Addis}, A.;
  {Antonelli}, L.A.; {Argan}, A.; {Barbiellini}, G.; {Baroncelli}, L.;
  {Bernardi}, G.;  et~al.
\newblock {An X-ray burst from a magnetar enlightening the mechanism of fast
  radio bursts}.
\newblock {\em Nat. Astron.} {\bf 2021}, {\em 5},~401--407.
  \href{http://xxx.lanl.gov/abs/2005.12164}{{
 }}
\newblock {{https://doi.org/10.1038/s41550-020-01276-x}}.

\bibitem[{Petroff} {et~al.}(2019){Petroff}, {Hessels}, and
  {Lorimer}]{petroff2019fast}
{Petroff}, E.; {Hessels}, J.W.T.; {Lorimer}, D.R.
\newblock {Fast radio bursts}.
\newblock {\em Astron. Astrophys.} {\bf 2019}, {\em 27},~4.
  \href{http://xxx.lanl.gov/abs/1904.07947}{{
 }}
\newblock {{https://doi.org/10.1007/s00159-019-0116-6}}.

\bibitem[{Zhang}(2018)]{zhang2018fast}
{Zhang}, B.
\newblock {Fast Radio Burst Energetics and Detectability from High Redshifts}.
\newblock {\em  Astrophys. J.} {\bf 2018}, {\em 867},~L21.


\bibitem[{Ravi} {et~al.}(2019){Ravi}, {Catha}, {D'Addario}, {Djorgovski},
  {Hallinan}, {Hobbs}, {Kocz}, {Kulkarni}, {Shi}, {Vedantham}, {Weinreb}, and
  {Woody}]{ravi2019fast}
{Ravi}, V.; {Catha}, M.; {D'Addario}, L.; {Djorgovski}, S.G.; {Hallinan}, G.;
  {Hobbs}, R.; {Kocz}, J.; {Kulkarni}, S.R.; {Shi}, J.; {Vedantham}, H.K.;
  et~al.
\newblock {A fast radio burst localized to a massive galaxy}.
\newblock {\em Nature} {\bf 2019}, {\em 572},~352--354.
  \href{http://xxx.lanl.gov/abs/1907.01542}{{
 }}
\newblock {{https://doi.org/10.1038/s41586-019-1389-7}}.

\bibitem[{Platts} {et~al.}(2019){Platts}, {Weltman}, {Walters}, {Tendulkar},
  {Gordin}, and {Kandhai}]{platts2019living}
{Platts}, E.; {Weltman}, A.; {Walters}, A.; {Tendulkar}, S.P.; {Gordin},
  J.E.B.; {Kandhai}, S.
\newblock {A living theory catalogue for fast radio bursts}.
\newblock {\em Phys. Rep.} {\bf 2019}, {\em 821},~1--27.
  \href{http://xxx.lanl.gov/abs/1810.05836}{{
 }}
\newblock {{https://doi.org/10.1016/j.physrep.2019.06.003}}.

\bibitem[{Yamasaki} {et~al.}(2018){Yamasaki}, {Totani}, and
  {Kiuchi}]{yamasaki2018repeatingTotani}
{Yamasaki}, S.; {Totani}, T.; {Kiuchi}, K.
\newblock {Repeating and non-repeating fast radio bursts from binary neutron
  star mergers}.
\newblock {\em Publ. Astron. Soc. Jpn.} {\bf 2018},
  {\em 70},~39.  \href{http://xxx.lanl.gov/abs/1710.02302}{{
 }}
\newblock {{https://doi.org/10.1093/pasj/psy029}}.

\bibitem[{Popov} and {Postnov}(2013)]{popov2013millisecond}
{Popov}, S.B.; {Postnov}, K.A.
\newblock {Millisecond extragalactic radio bursts as magnetar flares}.
\newblock {\em arXiv} {\bf 2013},  arXiv:1307.4924.
  \href{http://xxx.lanl.gov/abs/1307.4924}{{
  }}

\bibitem[{Metzger} {et~al.}(2019){Metzger}, {Margalit}, and
  {Sironi}]{metzger2019fastMargalit}
{Metzger}, B.D.; {Margalit}, B.; {Sironi}, L.
\newblock {Fast radio bursts as synchrotron maser emission from decelerating
  relativistic blast waves}.
\newblock {\em Mon. Not. R. Astron. Soc.} {\bf 2019},
  {\em 485},~4091--4106.
  \href{http://xxx.lanl.gov/abs/1902.01866}{{
 }}
\newblock {{https://doi.org/10.1093/mnras/stz700}}.

\bibitem[{Geng} and {Huang}(2015)]{geng2015fast}
{Geng}, J.J.; {Huang}, Y.F.
\newblock {Fast Radio Bursts: Collisions between Neutron Stars and
  Asteroids/Comets}.
\newblock {\em  Astrophys. J.} {\bf 2015}, {\em 809},~24.
  \href{http://xxx.lanl.gov/abs/1502.05171}{{
 }}
\newblock {{https://doi.org/10.1088/0004-637X/809/1/24}}.

\bibitem[{Dai} {et~al.}(2016){Dai}, {Wang}, {Wu}, and
  {Huang}]{dai2016repeating}
{Dai}, Z.G.; {Wang}, J.S.; {Wu}, X.F.; {Huang}, Y.F.
\newblock {Repeating Fast Radio Bursts from Highly Magnetized Pulsars Traveling
  through Asteroid Belts}.
\newblock {\em  Astrophys. J.} {\bf 2016}, {\em 829},~27.
  \href{http://xxx.lanl.gov/abs/1603.08207}{{
 }}
\newblock {{https://doi.org/10.3847/0004-637X/829/1/27}}.

\bibitem[{Xiao} and {Dai}(2020)]{xiao2020double}
{Xiao}, D.; {Dai}, Z.G.
\newblock {Double-peaked Pulse Profile of FRB 200428: Synchrotron Maser
  Emission from Magnetized Shocks Encountering a Density Jump}.
\newblock {\em  Astrophys. J.} {\bf 2020}, {\em 904},~L5.
  \href{http://xxx.lanl.gov/abs/2010.14787}{{
 }}
\newblock {{https://doi.org/10.3847/2041-8213/abc551}}.

\bibitem[{Geng} {et~al.}(2020){Geng}, {Li}, {Li}, {Xiong}, {Kuiper}, and
  {Huang}]{geng2020frb}
{Geng}, J.J.; {Li}, B.; {Li}, L.B.; {Xiong}, S.L.; {Kuiper}, R.; {Huang}, Y.F.
\newblock {FRB 200428: An Impact between an Asteroid and a Magnetar}.
\newblock {\em  Astrophys. J.} {\bf 2020}, {\em 898},~L55.
  \href{http://xxx.lanl.gov/abs/2006.04601}{{
 }}
\newblock {{https://doi.org/10.3847/2041-8213/aba83c}}.

\bibitem[{Geng} {et~al.}(2021){Geng}, {Li}, and {Huang}]{geng2021repeating}
{Geng}, J.; {Li}, B.; {Huang}, Y.
\newblock {Repeating fast radio bursts from collapses of the crust of a strange
  star}.
\newblock {\em  Innovation} {\bf 2021}, {\em 2},~100152,
  \href{http://xxx.lanl.gov/abs/2103.04165}{{
 }}
\newblock {{https://doi.org/10.1016/j.xinn.2021.100152}}.

\bibitem[{Li} {et~al.}(2018){Li}, {Huang}, {Geng}, and {Li}]{li2018model}
{Li}, L.B.; {Huang}, Y.F.; {Geng}, J.J.; {Li}, B.
\newblock {A model of fast radio bursts: Collisions between episodic magnetic
  blobs}.
\newblock {\em Res. Astron. Astrophys.} {\bf 2018}, {\em
  18},~061.  \href{http://xxx.lanl.gov/abs/1803.09945}{{
 }}
\newblock {{https://doi.org/10.1088/1674-4527/18/6/61}}.

\bibitem[{Zhang}(2018)]{zhang2018frb}
{Zhang}, B.
\newblock {FRB 121102: A Repeatedly Combed Neutron Star by a Nearby
  Low-luminosity Accreting Supermassive Black Hole}.
\newblock {\em  Astrophys. J.} {\bf 2018}, {\em 854},~L21.
  \href{http://xxx.lanl.gov/abs/1801.05436}{{
 }}
\newblock {{https://doi.org/10.3847/2041-8213/aaadba}}.

\bibitem[{Li} {et~al.}(2021){Li}, {Yang}, {Wang}, {Xu}, {Shao}, {Liu}, and
  {Dai}]{liyang2021}
{Li}, Q.C.; {Yang}, Y.P.; {Wang}, F.Y.; {Xu}, K.; {Shao}, Y.; {Liu}, Z.N.;
  {Dai}, Z.G.
\newblock {Periodic Activities of Repeating Fast Radio Bursts from Be/X-Ray
  Binary Systems}.
\newblock {\em  Astrophys. J.} {\bf 2021}, {\em 918},~L5,
  \href{http://xxx.lanl.gov/abs/2108.00350}{{
  }}
\newblock {{https://doi.org/10.3847/2041-8213/ac1922}}.

\bibitem[{Yang} and {Zou}(2020)]{yang2020}
{Yang}, H.; {Zou}, Y.C.
\newblock {Orbit-induced Spin Precession as a Possible Origin for Periodicity
  in Periodically Repeating Fast Radio Bursts}. 
\newblock {\em  Astrophys. J.} {\bf 2020}, {\em 893},~L31,
  \href{http://xxx.lanl.gov/abs/2002.02553}{{
 }}
\newblock {{https://doi.org/10.3847/2041-8213/ab800f}}.

\bibitem[{Levin} {et~al.}(2020){Levin}, {Beloborodov}, and
  {Bransgrove}]{levin2020precessing}
{Levin}, Y.; {Beloborodov}, A.M.; {Bransgrove}, A.
\newblock {Precessing Flaring Magnetar as a Source of Repeating FRB
  180916.J0158+65}.
\newblock {\em  Astrophys. J. Lett.} {\bf 2020}, {\em 895},~L30.
  \href{http://xxx.lanl.gov/abs/2002.04595}{{
 }}
\newblock {{https://doi.org/10.3847/2041-8213/ab8c4c}}.

\bibitem[{Tong} {et~al.}(2020){Tong}, {Wang}, and
  {Wang}]{tong2020periodicity}
{Tong}, H.; {Wang}, W.; {Wang}, H.G.
\newblock {Periodicity in fast radio bursts due to forced precession by a
  fallback disk}.
\newblock {\em Res. Astron. Astrophys.} {\bf 2020}, {\em
  20},~142.  \href{http://xxx.lanl.gov/abs/2002.10265}{{
 }}
\newblock {{https://doi.org/10.1088/1674-4527/20/9/142}}.


\bibitem[{Li} and {Zanazzi}(2021)]{li2021emission}
{Li}, D.; {Zanazzi}, J.J.
\newblock {Emission Properties of Periodic Fast Radio Bursts from the Motion of
  Magnetars: Testing Dynamical Models}.
\newblock {\em  Astrophys. J. Lett.} {\bf 2021}, {\em 909},~L25.
  \href{http://xxx.lanl.gov/abs/2101.05836}{{
 }}
\newblock {{https://doi.org/10.3847/2041-8213/abeaa4}}.

\bibitem[{Sridhar} {et~al.}(2021){Sridhar}, {Metzger}, {Beniamini},
  {Margalit}, {Renzo}, {Sironi}, and {Kovlakas}]{sridhar2021periodic}
{Sridhar}, N.; {Metzger}, B.D.; {Beniamini}, P.; {Margalit}, B.; {Renzo}, M.;
  {Sironi}, L.; {Kovlakas}, K.
\newblock {Periodic Fast Radio Bursts from Luminous X-ray Binaries}.
\newblock {\em  Astrophys. J.} {\bf 2021}, {\em 917},~13.
  \href{http://xxx.lanl.gov/abs/2102.06138}{{
  }}
\newblock {{https://doi.org/10.3847/1538-4357/ac0140}}.

\bibitem[{Zhang}(2020)]{zhang2020unexpected}
{Zhang}, B.
\newblock {Unexpected emission pattern adds to the enigma of fast radio
  bursts}.
\newblock {\em Nature} {\bf 2020}, {\em 582},~344--346.

\bibitem[{Xu} {et~al.}(2021){Xu}, {Li}, {Yang}, {Li}, {Dai}, and
  {Liu}]{xu2021periodic}
{Xu}, K.; {Li}, Q.C.; {Yang}, Y.P.; {Li}, X.D.; {Dai}, Z.G.; {Liu}, J.
\newblock {Do the Periodic Activities of Repeating Fast Radio Bursts Represent
  the Spins of Neutron Stars?}
\newblock {\em  Astrophys. J.} {\bf 2021}, {\em 917},~2.
  \href{http://xxx.lanl.gov/abs/2105.13122}{{
 }}
\newblock {{https://doi.org/10.3847/1538-4357/ac05ba}}.

\bibitem[{Thompson} {et~al.}(1980){Thompson}, {Clark}, {Wade}, and
  {Napier}]{Thompson19080vla}
{Thompson}, A.R.; {Clark}, B.G.; {Wade}, C.M.; {Napier}, P.J.
\newblock {The Very Large Array.}
\newblock {\em  Astrophys. J.} {\bf 1980}, {\em 44},~151--167.
\newblock {{https://doi.org/10.1086/190688}}.

\bibitem[{Bannister} {et~al.}(2017){Bannister}, {Shannon}, {Macquart},
  {Flynn}, {Edwards}, {O'Neill}, {Os{\l}owski}, {Bailes}, {Zackay}, {Clarke},
  {D'Addario}, {Dodson}, {Hall}, {Jameson}, {Jones}, {Navarro}, {Trinh},
  {Allison}, {Anderson}, {Bell}, {Chippendale}, {Collier}, {Heald}, {Heywood},
  {Hotan}, {Lee-Waddell}, {Madrid}, {Marvil}, {McConnell}, {Popping},
  {Voronkov}, {Whiting}, {Allen}, {Bock}, {Brodrick}, {Cooray}, {DeBoer},
  {Diamond}, {Ekers}, {Gough}, {Hampson}, {Harvey-Smith}, {Hay}, {Hayman},
  {Jackson}, {Johnston}, {Koribalski}, {McClure-Griffiths}, {Mirtschin}, {Ng},
  {Norris}, {Pearce}, {Phillips}, {Roxby}, {Troup}, and
  {Westmeier}]{bannister2017detection}
{Bannister}, K.W.; {Shannon}, R.M.; {Macquart}, J.P.; {Flynn}, C.; {Edwards},
  P.G.; {O'Neill}, M.; {Os{\l}owski}, S.; {Bailes}, M.; {Zackay}, B.; {Clarke},
  N.;  et~al.
\newblock {The Detection of an Extremely Bright Fast Radio Burst in a Phased
  Array Feed Survey}.
\newblock {\em  Astrophys. J. Lett.} {\bf 2017}, {\em 841},~L12.
  \href{http://xxx.lanl.gov/abs/1705.07581}{{
 }}
\newblock {{https://doi.org/10.3847/2041-8213/aa71ff}}.

\bibitem[{CHIME/FRB Collaboration} {et~al.}(2018){CHIME/FRB Collaboration},
  {Amiri}, {Bandura}, {Berger}, {Bhardwaj}, {Boyce}, {Boyle}, {Brar},
  {Burhanpurkar}, {Chawla}, {Chowdhury}, {Cliche}, {Cranmer}, {Cubranic},
  {Deng}, {Denman}, {Dobbs}, {Fandino}, {Fonseca}, {Gaensler}, {Giri},
  {Gilbert}, {Good}, {Guliani}, {Halpern}, {Hinshaw}, {H{\"o}fer}, {Josephy},
  {Kaspi}, {Landecker}, {Lang}, {Liao}, {Masui}, {Mena-Parra}, {Naidu},
  {Newburgh}, {Ng}, {Patel}, {Pen}, {Pinsonneault-Marotte}, {Pleunis}, {Rafiei
  Ravandi}, {Ransom}, {Renard}, {Scholz}, {Sigurdson}, {Siegel}, {Smith},
  {Stairs}, {Tendulkar}, {Vanderlinde}, and {Wiebe}]{chime2018}
{CHIME/FRB Collaboration}; {Amiri}, M.; {Bandura}, K.; {Berger}, P.;
  {Bhardwaj}, M.; {Boyce}, M.M.; {Boyle}, P.J.; {Brar}, C.; {Burhanpurkar}, M.;
  {Chawla}, P.;  et~al.
\newblock {The CHIME Fast Radio Burst Project: System Overview}.
\newblock {\em  Astrophys. J.} {\bf 2018}, {\em 863},~48.
  \href{http://xxx.lanl.gov/abs/1803.11235}{{
  }}
\newblock {{https://doi.org/10.3847/1538-4357/aad188}}.

\bibitem[{Spitler} {et~al.}(2016){Spitler}, {Scholz}, {Hessels}, {Bogdanov},
  {Brazier}, {Camilo}, {Chatterjee}, {Cordes}, {Crawford}, {Deneva}, {Ferdman},
  {Freire}, {Kaspi}, {Lazarus}, {Lynch}, {Madsen}, {McLaughlin}, {Patel},
  {Ransom}, {Seymour}, {Stairs}, {Stappers}, {van Leeuwen}, and
  {Zhu}]{spitler2016aScholz}
{Spitler}, L.G.; {Scholz}, P.; {Hessels}, J.W.T.; {Bogdanov}, S.; {Brazier},
  A.; {Camilo}, F.; {Chatterjee}, S.; {Cordes}, J.M.; {Crawford}, F.; {Deneva},
  J.;  et~al.
\newblock {A repeating fast radio burst}.
\newblock {\em Nature} {\bf 2016}, {\em 531},~202--205.
  \href{http://xxx.lanl.gov/abs/1603.00581}{{
  }}
\newblock {{https://doi.org/10.1038/nature17168}}.

\bibitem[{CHIME/FRB Collaboration} {et~al.}(2019{\natexlab{a}}){CHIME/FRB
  Collaboration}, {Amiri}, {Bandura}, {Bhardwaj}, {Boubel}, {Boyce}, {Boyle},
  {. Brar}, {Burhanpurkar}, {Cassanelli}, {Chawla}, {Cliche}, {Cubranic},
  {Deng}, {Denman}, {Dobbs}, {Fandino}, {Fonseca}, {Gaensler}, {Gilbert},
  {Gill}, {Giri}, {Good}, {Halpern}, {Hanna}, {Hill}, {Hinshaw}, {H{\"o}fer},
  {Josephy}, {Kaspi}, {Landecker}, {Lang}, {Lin}, {Masui}, {Mckinven},
  {Mena-Parra}, {Merryfield}, {Michilli}, {Milutinovic}, {Moatti}, {Naidu},
  {Newburgh}, {Ng}, {Patel}, {Pen}, {Pinsonneault-Marotte}, {Pleunis},
  {Rafiei-Ravandi}, {Rahman}, {Ransom}, {Renard}, {Scholz}, {Shaw}, {Siegel},
  {Smith}, {Stairs}, {Tendulkar}, {Tretyakov}, {Vanderlinde}, and
  {Yadav}]{chime2019aAmiriBandura}
{CHIME/FRB Collaboration}; {Amiri}, M.; {Bandura}, K.; {Bhardwaj}, M.;
  {Boubel}, P.; {Boyce}, M.M.; {Boyle}, P.J.; {. Brar}, C.; {Burhanpurkar}, M.;
  {Cassanelli}, T.;  et~al.
\newblock {A second source of repeating fast radio bursts}.
\newblock {\em Nature} {\bf 2019}, {\em 566},~235--238.
  \href{http://xxx.lanl.gov/abs/1901.04525}{{
  }}
\newblock {{https://doi.org/10.1038/s41586-018-0864-x}}.

\bibitem[{CHIME/FRB Collaboration} {et~al.}(2019{\natexlab{b}}){CHIME/FRB
  Collaboration}, {Andersen}, {Bandura}, {Bhardwaj}, {Boubel}, {Boyce},
  {Boyle}, {Brar}, {Cassanelli}, {Chawla}, {Cubranic}, {Deng}, {Dobbs},
  {Fandino}, {Fonseca}, {Gaensler}, {Gilbert}, {Giri}, {Good}, {Halpern},
  {Hill}, {Hinshaw}, {H{\"o}fer}, {Josephy}, {Kaspi}, {Kothes}, {Landecker},
  {Lang}, {Li}, {Lin}, {Masui}, {Mena-Parra}, {Merryfield}, {Mckinven},
  {Michilli}, {Milutinovic}, {Naidu}, {Newburgh}, {Ng}, {Patel}, {Pen},
  {Pinsonneault-Marotte}, {Pleunis}, {Rafiei-Ravandi}, {Rahman}, {Ransom},
  {Renard}, {Scholz}, {Siegel}, {Singh}, {Smith}, {Stairs}, {Tendulkar},
  {Tretyakov}, {Vanderlinde}, {Yadav}, and
  {Zwaniga}]{chime2019chimeAndersenBandur}
{CHIME/FRB Collaboration}.; {Andersen}, B.C.; {Bandura}, K.; {Bhardwaj}, M.;
  {Boubel}, P.; {Boyce}, M.M.; {Boyle}, P.J.; {Brar}, C.; {Cassanelli}, T.;
  {Chawla}, P.;  et~al.
\newblock {CHIME/FRB Discovery of Eight New Repeating Fast Radio Burst
  Sources}.
\newblock {\em  Astrophys. J.} {\bf 2019}, {\em 885},~L24.
  \href{http://xxx.lanl.gov/abs/1908.03507}{{
  }}
\newblock {{https://doi.org/10.3847/2041-8213/ab4a80}}.

\bibitem[{Kumar} {et~al.}(2019){Kumar}, {Shannon}, {Os{\l}owski}, {Qiu},
  {Bhandari}, {Farah}, {Flynn}, {Kerr}, {Lorimer}, {Macquart}, {Ng},
  {Phillips}, {Price}, and {Spiewak}]{kumar2019faintShannon}
{Kumar}, P.; {Shannon}, R.M.; {Os{\l}owski}, S.; {Qiu}, H.; {Bhandari}, S.;
  {Farah}, W.; {Flynn}, C.; {Kerr}, M.; {Lorimer}, D.R.; {Macquart}, J.P.;
  et~al.
\newblock {Faint Repetitions from a Bright Fast Radio Burst Source}.
\newblock {\em  Astrophys. J.} {\bf 2019}, {\em 887},~L30.
  \href{http://xxx.lanl.gov/abs/1908.10026}{{
  }}
\newblock {{https://doi.org/10.3847/2041-8213/ab5b08}}.

\bibitem[{Luo} {et~al.}(2020){Luo}, {Wang}, {Men}, {Zhang}, {Jiang}, {Xu},
  {Wang}, {Lee}, {Han}, {Zhang}, {Caballero}, {Chen}, {Chen}, {Gan}, {Guo},
  {Hao}, {Huang}, {Jiang}, {Li}, {Li}, {Li}, {Luo}, {Pan}, {Pei}, {Qian},
  {Sun}, {Wang}, {Wang}, {Wen}, {Xu}, {Xu}, {Yan}, {Yan}, {Yu}, {Yuan},
  {Zhang}, and {Zhu}]{luo2020diverse}
{Luo}, R.; {Wang}, B.J.; {Men}, Y.P.; {Zhang}, C.F.; {Jiang}, J.C.; {Xu}, H.;
  {Wang}, W.Y.; {Lee}, K.J.; {Han}, J.L.; {Zhang}, B.;  et~al.
\newblock {Diverse polarization angle swings from a repeating fast radio burst
  source}.
\newblock {\em Nature} {\bf 2020}, {\em 586},~693--696.
  \href{http://xxx.lanl.gov/abs/2011.00171}{{
  }}
\newblock {{https://doi.org/10.1038/s41586-020-2827-2}}.

\bibitem[{Zhang} and {Choi}(2008)]{zhang2008analysis}
{Zhang}, Z.B.; {Choi}, C.S.
\newblock {An analysis of the durations of Swift gamma-ray bursts}.
\newblock {\em Astron. Astrophys.} {\bf 2008}, {\em 484},~293--297.
  \href{http://xxx.lanl.gov/abs/0708.4049}{{
  }}
\newblock {{https://doi.org/10.1051/0004-6361:20079210}}.

\bibitem[{Zhang} {et~al.}(2018){Zhang}, {Zhang}, {Zhao}, {Luo}, {Jiang},
  {Wang}, {Han}, and {Terheide}]{zhang2018spectrum}
{Zhang}, Z.B.; {Zhang}, C.T.; {Zhao}, Y.X.; {Luo}, J.J.; {Jiang}, L.Y.; {Wang},
  X.L.; {Han}, X.L.; {Terheide}, R.K.
\newblock {Spectrum-energy Correlations in GRBs: Update, Reliability, and the
  Long/Short Dichotomy}.
\newblock {\em Publ. Astron. Soc. Pac.} {\bf
  2018}, {\em 130},~054202.
\newblock {{https://doi.org/10.1088/1538-3873/aaa6af}}.

\bibitem[{Zhang} {et~al.}(2020){Zhang}, {Jiang}, {Zhang}, {Zhang}, {Li}, and
  {Zhang}]{zhang2020spectral}
{Zhang}, Z.B.; {Jiang}, M.; {Zhang}, Y.; {Zhang}, K.; {Li}, X.J.; {Zhang}, Q.
\newblock {On the Spectral Peak Energy of Swift Gamma-Ray Bursts}.
\newblock {\em  Astrophys. J.} {\bf 2020}, {\em 902},~40.
  \href{http://xxx.lanl.gov/abs/2009.10258}{{
  }}
\newblock {{https://doi.org/10.3847/1538-4357/abb400}}.

\bibitem[{Luo} {et~al.}(2018){Luo}, {Lee}, {Lorimer}, and
  {Zhang}]{luo2018onLee}
{Luo}, R.; {Lee}, K.; {Lorimer}, D.R.; {Zhang}, B.
\newblock {On the normalized FRB luminosity function}.
\newblock {\em Mon. Not. R. Astron. Soc.} {\bf 2018},
  {\em 481},~2320--2337.
  \href{http://xxx.lanl.gov/abs/1808.09929}{{
  }}
\newblock {{https://doi.org/10.1093/mnras/sty2364}}.

\bibitem[{Caleb} {et~al.}(2019){Caleb}, {Stappers}, {Rajwade}, and
  {Flynn}]{caleb2019areStappers}
{Caleb}, M.; {Stappers}, B.W.; {Rajwade}, K.; {Flynn}, C.
\newblock {Are all fast radio bursts repeating sources?}
\newblock {\em Mon. Not. R. Astron. Soc.} {\bf 2019},
  {\em 484},~5500--5508.
  \href{http://xxx.lanl.gov/abs/1902.00272}{{
  }}
\newblock {{https://doi.org/10.1093/mnras/stz386}}.

\bibitem[{Li} {et~al.}(2019){Li}, {Zhang}, {Nagamine}, and {Shi}]{li2019frb}
{Li}, Y.; {Zhang}, B.; {Nagamine}, K.; {Shi}, J.
\newblock {The FRB 121102 Host Is Atypical among Nearby Fast Radio Bursts}.
\newblock {\em  Astrophys. J.} {\bf 2019}, {\em 884},~L26.
  \href{http://xxx.lanl.gov/abs/1906.08749}{{
  }}
\newblock {{https://doi.org/10.3847/2041-8213/ab3e41}}.

\bibitem[{Zhang}(2020)]{zhang2020the}
{Zhang}, B.
\newblock {The physical mechanisms of fast radio bursts}.
\newblock {\em Nature} {\bf 2020}, {\em 587},~45--53.
  \href{http://xxx.lanl.gov/abs/2011.03500}{{
  }}
\newblock {{https://doi.org/10.1038/s41586-020-2828-1}}.

\bibitem[{Ravi}(2019)]{ravi2019aCatha}
{Ravi}, V.
\newblock {The prevalence of repeating fast radio bursts}.
\newblock {\em Nat. Astron.} {\bf 2019}, {\em 3},~928--931.
  \href{http://xxx.lanl.gov/abs/1907.06619}{{
  }}
\newblock {{https://doi.org/10.1038/s41550-019-0831-y}}.

\bibitem[{Lu} {et~al.}(2020){Lu}, {Piro}, and {Waxman}]{lu2020implications}
{Lu}, W.; {Piro}, A.L.; {Waxman}, E.
\newblock {Implications of Canadian Hydrogen Intensity Mapping Experiment
  repeating fast radio bursts}.
\newblock {\em Mon. Not. R. Astron. Soc.} {\bf 2020},
  {\em 498},~1973--1982.
  \href{http://xxx.lanl.gov/abs/2003.12581}{{
  }}
\newblock {{https://doi.org/10.1093/mnras/staa2397}}.

\bibitem[{Luo} {et~al.}(2020){Luo}, {Men}, {Lee}, {Wang}, {Lorimer}, and
  {Zhang}]{luo2020frb}
{Luo}, R.; {Men}, Y.; {Lee}, K.; {Wang}, W.; {Lorimer}, D.R.; {Zhang}, B.
\newblock {On the FRB luminosity function---II. Event rate density}.
\newblock {\em Mon. Not. R. Astron. Soc.} {\bf 2020},
  {\em 494},~665--679.  \href{http://xxx.lanl.gov/abs/2003.04848}{{
  }}
\newblock {{https://doi.org/10.1093/mnras/staa704}}.

\bibitem[{Caleb} {et~al.}(2016){Caleb}, {Flynn}, {Bailes}, {Barr},
  {Hunstead}, {Keane}, {Ravi}, and {van Straten}]{caleb2016distributions}
{Caleb}, M.; {Flynn}, C.; {Bailes}, M.; {Barr}, E.D.; {Hunstead}, R.W.;
  {Keane}, E.F.; {Ravi}, V.; {van Straten}, W.
\newblock {Are the distributions of fast radio burst properties consistent with
  a cosmological population?}
\newblock {\em Mon. Not. R. Astron. Soc.} {\bf 2016},
  {\em 458},~708--717.  \href{http://xxx.lanl.gov/abs/1512.02738}{{
  }}
\newblock {{https://doi.org/10.1093/mnras/stw175}}.

\bibitem[{Niino}(2018)]{niino2018fast}
{Niino}, Y.
\newblock {Fast Radio Bursts{\textquoteright} Recipes for the Distributions of
  Dispersion Measures, Flux Densities, and Fluences}.
\newblock {\em  Astrophys. J.} {\bf 2018}, {\em 858},~4.
  \href{http://xxx.lanl.gov/abs/1801.06578}{{
  }}
\newblock {{https://doi.org/10.3847/1538-4357/aab9a9}}.

\bibitem[{Lu} and {Piro}(2019)]{lu2019implications}
{Lu}, W.; {Piro}, A.L.
\newblock {Implications from ASKAP Fast Radio Burst Statistics}.
\newblock {\em  Astrophys. J.} {\bf 2019}, {\em 883},~40.
  \href{http://xxx.lanl.gov/abs/1903.00014}{{
  }}
\newblock {{https://doi.org/10.3847/1538-4357/ab3796}}.

\bibitem[{Zhang} and {Wang}(2019)]{zhang2019energy}
{Zhang}, G.Q.; {Wang}, F.Y.
\newblock {Energy function, formation rate, and low-metallicity environment of
  fast radio bursts}.
\newblock {\em Mon. Not. R. Astron. Soc.} {\bf 2019},
  {\em 487},~3672--3678.
  \href{http://xxx.lanl.gov/abs/1906.01176}{{
  }}
\newblock {{https://doi.org/10.1093/mnras/stz1566}}.

\bibitem[{Bhattacharya} and {Kumar}(2020)]{bhattacharya2020population}
{Bhattacharya}, M.; {Kumar}, P.
\newblock {Population Modeling of Fast Radio Bursts from Source Properties}.
\newblock {\em  Astrophys. J.} {\bf 2020}, {\em 899},~124.
  \href{http://xxx.lanl.gov/abs/1902.10225}{{
  }}
\newblock {{https://doi.org/10.3847/1538-4357/aba8fb}}.

\bibitem[{Planck Collaboration} {et~al.}(2016){Planck Collaboration}, {Ade},
  {Aghanim}, {Arnaud}, {Ashdown}, {Aumont}, {Baccigalupi}, {Banday},
  {Barreiro}, {Bartlett}, {Bartolo}, {Battaner}, {Battye}, {Benabed},
  {Beno{\^\i}t}, {Benoit-L{\'e}vy}, {Bernard}, {Bersanelli}, {Bielewicz},
  {Bock}, {Bonaldi}, {Bonavera}, {Bond}, {Borrill}, {Bouchet}, {Boulanger},
  {Bucher}, {Burigana}, {Butler}, {Calabrese}, {Cardoso}, {Catalano},
  {Challinor}, {Chamballu}, {Chary}, {Chiang}, {Chluba}, {Christensen},
  {Church}, {Clements}, {Colombi}, {Colombo}, {Combet}, {Coulais}, {Crill},
  {Curto}, {Cuttaia}, {Danese}, {Davies}, {Davis}, {de Bernardis}, {de Rosa},
  {de Zotti}, {Delabrouille}, {D{\'e}sert}, {Di Valentino}, {Dickinson},
  {Diego}, {Dolag}, {Dole}, {Donzelli}, {Dor{\'e}}, {Douspis}, {Ducout},
  {Dunkley}, {Dupac}, {Efstathiou}, {Elsner}, {En{\ss}lin}, {Eriksen},
  {Farhang}, {Fergusson}, {Finelli}, {Forni}, {Frailis}, {Fraisse},
  {Franceschi}, {Frejsel}, {Galeotta}, {Galli}, {Ganga}, {Gauthier}, {Gerbino},
  {Ghosh}, {Giard}, {Giraud-H{\'e}raud}, {Giusarma}, {Gjerl{\o}w},
  {Gonz{\'a}lez-Nuevo}, {G{\'o}rski}, {Gratton}, {Gregorio}, {Gruppuso},
  {Gudmundsson}, {Hamann}, {Hansen}, {Hanson}, {Harrison}, {Helou},
  {Henrot-Versill{\'e}}, {Hern{\'a}ndez-Monteagudo}, {Herranz}, {Hildebrandt},
  {Hivon}, {Hobson}, {Holmes}, {Hornstrup}, {Hovest}, {Huang}, {Huffenberger},
  {Hurier}, {Jaffe}, {Jaffe}, {Jones}, {Juvela}, {Keih{\"a}nen}, {Keskitalo},
  {Kisner}, {Kneissl}, {Knoche}, {Knox}, {Kunz}, {Kurki-Suonio}, {Lagache},
  {L{\"a}hteenm{\"a}ki}, {Lamarre}, {Lasenby}, {Lattanzi}, {Lawrence}, {Leahy},
  {Leonardi}, {Lesgourgues}, {Levrier}, {Lewis}, {Liguori}, {Lilje},
  {Linden-V{\o}rnle}, {L{\'o}pez-Caniego}, {Lubin}, {Mac{\'\i}as-P{\'e}rez},
  {Maggio}, {Maino}, {Mandolesi}, {Mangilli}, {Marchini}, {Maris}, {Martin},
  {Martinelli}, {Mart{\'\i}nez-Gonz{\'a}lez}, {Masi}, {Matarrese}, {McGehee},
  {Meinhold}, {Melchiorri}, {Melin}, {Mendes}, {Mennella}, {Migliaccio},
  {Millea}, {Mitra}, {Miville-Desch{\^e}nes}, {Moneti}, {Montier}, {Morgante},
  {Mortlock}, {Moss}, {Munshi}, {Murphy}, {Naselsky}, {Nati}, {Natoli},
  {Netterfield}, {N{\o}rgaard-Nielsen}, {Noviello}, {Novikov}, {Novikov},
  {Oxborrow}, {Paci}, {Pagano}, {Pajot}, {Paladini}, {Paoletti}, {Partridge},
  {Pasian}, {Patanchon}, {Pearson}, {Perdereau}, {Perotto}, {Perrotta},
  {Pettorino}, {Piacentini}, {Piat}, {Pierpaoli}, {Pietrobon}, {Plaszczynski},
  {Pointecouteau}, {Polenta}, {Popa}, {Pratt}, {Pr{\'e}zeau}, {Prunet},
  {Puget}, {Rachen}, {Reach}, {Rebolo}, {Reinecke}, {Remazeilles}, {Renault},
  {Renzi}, {Ristorcelli}, {Rocha}, {Rosset}, {Rossetti}, {Roudier},
  {Rouill{\'e} d'Orfeuil}, {Rowan-Robinson}, {Rubi{\~n}o-Mart{\'\i}n},
  {Rusholme}, {Said}, {Salvatelli}, {Salvati}, {Sandri}, {Santos},
  {Savelainen}, {Savini}, {Scott}, {Seiffert}, {Serra}, {Shellard}, {Spencer},
  {Spinelli}, {Stolyarov}, {Stompor}, {Sudiwala}, {Sunyaev}, {Sutton},
  {Suur-Uski}, {Sygnet}, {Tauber}, {Terenzi}, {Toffolatti}, {Tomasi},
  {Tristram}, {Trombetti}, {Tucci}, {Tuovinen}, {T{\"u}rler}, {Umana},
  {Valenziano}, {Valiviita}, {Van Tent}, {Vielva}, {Villa}, {Wade}, {Wandelt},
  {Wehus}, {White}, {White}, {Wilkinson}, {Yvon}, {Zacchei}, and
  {Zonca}]{ade2016planck}
{Planck Collaboration}.; {Ade}, P.A.R.; {Aghanim}, N.; {Arnaud}, M.; {Ashdown},
  M.; {Aumont}, J.; {Baccigalupi}, C.; {Banday}, A.J.; {Barreiro}, R.B.;
  {Bartlett}, J.G.;  et~al.
\newblock {Planck 2015 results. XIII. Cosmological parameters}.
\newblock {\em Astron. Astrophys.} {\bf 2016}, {\em 594},~A13.
  \href{http://xxx.lanl.gov/abs/1502.01589}{{
  }}
\newblock {{https://doi.org/10.1051/0004-6361/201525830}}.

\bibitem[{Fonseca} {et~al.}(2020){Fonseca}, {Andersen}, {Bhardwaj},
  {Chawla}, {Good}, {Josephy}, {Kaspi}, {Masui}, {Mckinven}, {Michilli},
  {Pleunis}, {Shin}, {Tendulkar}, {Bandura}, {Boyle}, {Brar}, {Cassanelli},
  {Cubranic}, {Dobbs}, {Dong}, {Gaensler}, {Hinshaw}, {Landecker}, {Leung},
  {Li}, {Lin}, {Mena-Parra}, {Merryfield}, {Naidu}, {Ng}, {Patel}, {Pen},
  {Rafiei-Ravandi}, {Rahman}, {Ransom}, {Scholz}, {Smith}, {Stairs},
  {Vanderlinde}, {Yadav}, and {Zwaniga}]{fonseca2020nineAndersen}
{Fonseca}, E.; {Andersen}, B.C.; {Bhardwaj}, M.; {Chawla}, P.; {Good}, D.C.;
  {Josephy}, A.; {Kaspi}, V.M.; {Masui}, K.W.; {Mckinven}, R.; {Michilli}, D.;
  et~al.
\newblock {Nine New Repeating Fast Radio Burst Sources from CHIME/FRB}.
\newblock {\em  Astrophys. J.} {\bf 2020}, {\em 891},~L6.
  \href{http://xxx.lanl.gov/abs/2001.03595}{{
  }}
\newblock {{https://doi.org/10.3847/2041-8213/ab7208}}.

\bibitem[{Kirsten} {et~al.}(2021){Kirsten}, {Snelders}, {Jenkins}, {Nimmo},
  {van den Eijnden}, {Hessels}, {Gawro{\'n}ski}, and
  {Yang}]{kirsten2021detection}
{Kirsten}, F.; {Snelders}, M.P.; {Jenkins}, M.; {Nimmo}, K.; {van den Eijnden},
  J.; {Hessels}, J.W.T.; {Gawro{\'n}ski}, M.P.; {Yang}, J.
\newblock {Detection of two bright radio bursts from magnetar SGR 1935 + 2154}.
\newblock {\em Nat. Astron.} {\bf 2021}, {\em 5},~414--422.
  \href{http://xxx.lanl.gov/abs/2007.05101}{{
  }}
\newblock {{https://doi.org/10.1038/s41550-020-01246-3}}.

\bibitem[Bauer(1972)]{bauer1972constructing}
Bauer, D.F.
\newblock Constructing confidence sets using rank statistics.
\newblock {\em J. Am. Stat. Assoc.} {\bf 1972},
  {\em 67},~687--690.

\bibitem[Hollander {et~al.}(2013)Hollander, Wolfe, and
  Chicken]{hollander2013nonparametric}
Hollander, M.; Wolfe, D.A.; Chicken, E.
\newblock {\em Nonparametric Statistical Methods}; John Wiley \&
  Sons:  Hoboken, NJ, USA, 
 2013; \mbox{Volume 751.}

\bibitem[{Xiao} {et~al.}(2021){Xiao}, {Wang}, and {Dai}]{xiao2021physics}
{Xiao}, D.; {Wang}, F.; {Dai}, Z.
\newblock {The physics of fast radio bursts}.
\newblock {\em Sci. China Phys. Mech. Astron.} {\bf 2021},
  {\em 64},~249501.  \href{http://xxx.lanl.gov/abs/2101.04907}{{
  }}
\newblock {{https://doi.org/10.1007/s11433-020-1661-7}}.

\bibitem[Thode(2002)]{thode2002testing}
Thode, H.C.
\newblock {\em Testing for Normality}, 1st ed.; CRC Press: Boca Raton, CA, USA,  2002.

\bibitem[{Li} {et~al.}(2017){Li}, {Huang}, {Zhang}, {Li}, and
  {Li}]{li2017intensityHuang}
{Li}, L.B.; {Huang}, Y.F.; {Zhang}, Z.B.; {Li}, D.; {Li}, B.
\newblock {Intensity distribution function and statistical properties of fast
  radio bursts}.
\newblock {\em Res. Astron. Astrophys.} {\bf 2017}, {\em 17},~6.
   \href{http://xxx.lanl.gov/abs/1602.06099}{{
 }}
\newblock {{https://doi.org/10.1088/1674-4527/17/1/6}}.

\bibitem[{Nan} and {Li}(2013)]{nan2011theLi}
{Nan}, R.; {Li}, D.
\newblock {The five-hundred-meter aperture spherical radio telescope (FAST)
  project}.
\newblock \emph{IOP Conf. Ser. Mater. Sci. Eng.} 
\textbf{2013}, \emph{44},  012022.
\newblock {{https://doi.org/10.1088/1757-899X/44/1/012022}}.

\bibitem[{Li} {et~al.}(2013){Li}, {Nan}, and {Pan}]{li2013theNan}
{Li}, D.; {Nan}, R.; {Pan}, Z.
\newblock {The Five-hundred-meter Aperture Spherical radio Telescope project
  and its early science opportunities}.
\newblock In Proceedings of the Neutron Stars and Pulsars: Challenges and
  Opportunities after 80 Years, Beijing, China, 20--24 August 2012; 
 {van Leeuwen}, J., Ed.; Cambridge University Press, Cambrifge, UK, 2013;   
 \mbox{Volume 291}, pp. 325--330  \href{http://xxx.lanl.gov/abs/1210.5785}{{
  }}
\newblock {{https://doi.org/10.1017/S1743921312024015}}.

\bibitem[{Zhang} {et~al.}(2015){Zhang}, {Kong}, {Huang}, {Li}, and
  {Li}]{Zhangzb2015}
{Zhang}, Z.B.; {Kong}, S.W.; {Huang}, Y.F.; {Li}, D.; {Li}, L.B.
\newblock {Detecting radio afterglows of gamma-ray bursts with FAST}.
\newblock {\em Res. Astron. Astrophys.} {\bf 2015}, {\em
  15},~237--251.  \href{http://xxx.lanl.gov/abs/1402.6810}{{
 }}
\newblock {{https://doi.org/10.1088/1674-4527/15/2/008}}.

\bibitem[{Spitler} {et~al.}(2014){Spitler}, {Cordes}, {Hessels}, {Lorimer},
  {McLaughlin}, {Chatterjee}, {Crawford}, {Deneva}, {Kaspi}, {Wharton},
  {Allen}, {Bogdanov}, {Brazier}, {Camilo}, {Freire}, {Jenet},
  {Karako-Argaman}, {Knispel}, {Lazarus}, {Lee}, {van Leeuwen}, {Lynch},
  {Ransom}, {Scholz}, {Siemens}, {Stairs}, {Stovall}, {Swiggum},
  {Venkataraman}, {Zhu}, {Aulbert}, and {Fehrmann}]{spitler2014fastCordes}
{Spitler}, L.G.; {Cordes}, J.M.; {Hessels}, J.W.T.; {Lorimer}, D.R.;
  {McLaughlin}, M.A.; {Chatterjee}, S.; {Crawford}, F.; {Deneva}, J.S.;
  {Kaspi}, V.M.; {Wharton}, R.S.;  et~al.
\newblock {Fast Radio Burst Discovered in the Arecibo Pulsar ALFA Survey}.
\newblock {\em  Astrophys. J.} {\bf 2014}, {\em 790},~101.
  \href{http://xxx.lanl.gov/abs/1404.2934}{{
 }}
\newblock {{https://doi.org/10.1088/0004-637X/790/2/101}}.

\bibitem[{Crawford} {et~al.}(2016){Crawford}, {Rane}, {Tran}, {Rolph},
  {Lorimer}, and {Ridley}]{Crawford2016ASF}
{Crawford}, F.; {Rane}, A.; {Tran}, L.; {Rolph}, K.; {Lorimer}, D.R.; {Ridley},
  J.P.
\newblock {A search for highly dispersed fast radio bursts in three Parkes
  multibeam surveys}.
\newblock {\em Mon. Not. R. Astron. Soc.} {\bf 2016},
  {\em 460},~3370--3375.
  \href{http://xxx.lanl.gov/abs/1605.06074}{{
 }}
\newblock {{https://doi.org/10.1093/mnras/stw1233}}.

\bibitem[{Lawrence} {et~al.}(2017){Lawrence}, {Vander Wiel}, {Law}, {Burke
  Spolaor}, and {Bower}]{Lawrence2017TheNP}
{Lawrence}, E.; {Vander Wiel}, S.; {Law}, C.; {Burke Spolaor}, S.; {Bower},
  G.C.
\newblock {The Nonhomogeneous Poisson Process for Fast Radio Burst Rates}.
\newblock {\em  Astron. J.} {\bf 2017}, {\em 154},~117.
  \href{http://xxx.lanl.gov/abs/1611.00458}{{
 }}
\newblock {{https://doi.org/10.3847/1538-3881/aa844e}}.

\bibitem[{Bera} {et~al.}(2016){Bera}, {Bhattacharyya}, {Bharadwaj}, {Bhat},
  and {Chengalur}]{bera2016onBhattacharyya}
{Bera}, A.; {Bhattacharyya}, S.; {Bharadwaj}, S.; {Bhat}, N.D.R.; {Chengalur},
  J.N.
\newblock {On modelling the Fast Radio Burst population and event rate
  predictions}.
\newblock {\em Mon. Not. R. Astron. Soc.} {\bf 2016},
  {\em 457},~2530--2539.
  \href{http://xxx.lanl.gov/abs/1601.05410}{{
 }}
\newblock {{https://doi.org/10.1093/mnras/stw177}}.

\bibitem[{Aggarwal}(2021)]{Aggarwal2021}
{Aggarwal}, K.
\newblock {Observational Effects of Banded Repeating FRBs}.
\newblock {\em  Astrophys. J.} {\bf 2021}, {\em 920},~L18,
  \href{http://xxx.lanl.gov/abs/2108.04474}{{
  }}
\newblock {{https://doi.org/10.3847/2041-8213/ac2a3a}}.

\bibitem[{Zhang} {et~al.}(2018){Zhang}, {Chandra}, {Huang}, and
  {Li}]{zhang2018redshift}
{Zhang}, Z.B.; {Chandra}, P.; {Huang}, Y.F.; {Li}, D.
\newblock {The Redshift Dependence of the Radio Flux of Gamma-Ray Bursts and
  Their Host Galaxies}.
\newblock {\em  Astrophys. J.} {\bf 2018}, {\em 865},~82.
  \href{http://xxx.lanl.gov/abs/1801.00397}{{
 }}
\newblock {{https://doi.org/10.3847/1538-4357/aadc62}}.

\bibitem[{Macquart} and {Ekers}(2018)]{macquart2018frb}
{Macquart}, J.P.; {Ekers}, R.
\newblock {FRB event rate counts - II. Fluence, redshift, and dispersion
  measure distributions}.
\newblock {\em Mon. Not. R. Astron. Soc.} {\bf 2018},
  {\em 480},~4211--4230.
  \href{http://xxx.lanl.gov/abs/1808.00908}{{
 }}
\newblock {{https://doi.org/10.1093/mnras/sty2083}}.

\bibitem[{Macquart} {et~al.}(2019){Macquart}, {Shannon}, {Bannister},
  {James}, {Ekers}, and {Bunton}]{macquart2019spectral}
{Macquart}, J.P.; {Shannon}, R.M.; {Bannister}, K.W.; {James}, C.W.; {Ekers},
  R.D.; {Bunton}, J.D.
\newblock {The Spectral Properties of the Bright Fast Radio Burst Population}.
\newblock {\em  Astrophys. J.} {\bf 2019}, {\em 872},~L19.
  \href{http://xxx.lanl.gov/abs/1810.04353}{{
 }}
\newblock {{https://doi.org/10.3847/2041-8213/ab03d6}}.

\bibitem[{Salvaterra} {et~al.}(2012){Salvaterra}, {Campana}, {Vergani},
  {Covino}, {D'Avanzo}, {Fugazza}, {Ghirlanda}, {Ghisellini}, {Melandri},
  {Nava}, {Sbarufatti}, {Flores}, {Piranomonte}, and
  {Tagliaferri}]{salvaterra2012complete}
{Salvaterra}, R.; {Campana}, S.; {Vergani}, S.D.; {Covino}, S.; {D'Avanzo}, P.;
  {Fugazza}, D.; {Ghirlanda}, G.; {Ghisellini}, G.; {Melandri}, A.; {Nava}, L.;
   et~al.
\newblock {A Complete Sample of Bright Swift Long Gamma-Ray Bursts. I. Sample
  Presentation, Luminosity Function and Evolution}.
\newblock {\em  Astrophys. J.} {\bf 2012}, {\em 749},~68.
  \href{http://xxx.lanl.gov/abs/1112.1700}{{
 }}
\newblock {{https://doi.org/10.1088/0004-637X/749/1/68}}.

\bibitem[Pescalli {et~al.}(2015)Pescalli, Ghirlanda, Salvaterra, Ghisellini,
  Vergani, Nappo, Salafia, Melandri, and Götz]{pescalli2016rate}
Pescalli, A.; Ghirlanda, G.; Salvaterra, R.; Ghisellini, G.; Vergani, S.;
  Nappo, F.; Salafia, O.; Melandri, A.; Götz, D.
\newblock The rate and luminosity function of long Gamma Ray Bursts.
\newblock {\em Astron. Astrophys.} {\bf 2015}, {\em 587}.
\newblock {{https://doi.org/10.1051/0004-6361/201526760}}.

\bibitem[{Deng} {et~al.}(2016){Deng}, {Wang}, {Guo}, {Lu}, {Wang}, {Wei},
  {Wu}, and {Liang}]{deng2016cosmic}
{Deng}, C.M.; {Wang}, X.G.; {Guo}, B.B.; {Lu}, R.J.; {Wang}, Y.Z.; {Wei}, J.J.;
  {Wu}, X.F.; {Liang}, E.W.
\newblock {Cosmic Evolution of Long Gamma-Ray Burst Luminosity}.
\newblock {\em  Astrophys. J.} {\bf 2016}, {\em 820},~66.
  \href{http://xxx.lanl.gov/abs/1601.07645}{{
 }}
\newblock {{https://doi.org/10.3847/0004-637X/820/1/66}}.




\bibitem[{Gehrels}(1986)]{gehrels1986confidence}
{Gehrels}, N.
\newblock {Confidence Limits for Small Numbers of Events in Astrophysical
  Data}.
\newblock {\em  Astrophys. J.} {\bf 1986}, {\em 303},~336.
\newblock {{https://doi.org/10.1086/164079}}.



\bibitem[{James} {et~al.}(2019){James}, {Ekers}, {Macquart}, {Bannister},
  and {Shannon}]{James2019}
{James}, C.W.; {Ekers}, R.D.; {Macquart}, J.P.; {Bannister}, K.W.; {Shannon},
  R.M.
\newblock {The slope of the source-count distribution for fast radio bursts}.
\newblock {\em Mon. Not. R. Astron. Soc.} {\bf 2019}.
  {\em 483},~1342--1353.
  \href{http://xxx.lanl.gov/abs/1810.04357}{{
  }}
\newblock {{https://doi.org/10.1093/mnras/sty3031}}.

\bibitem[{Heintz} {et~al.}(2020){Heintz}, {Prochaska}, {Simha}, {Platts},
  {Fong}, {Tejos}, {Ryder}, {Aggerwal}, {Bhandari}, {Day}, {Deller},
  {Kilpatrick}, {Law}, {Macquart}, {Mannings}, {Marnoch}, {Sadler}, and
  {Shannon}]{Heintz2020}
{Heintz}, K.E.; {Prochaska}, J.X.; {Simha}, S.; {Platts}, E.; {Fong}, W.f.;
  {Tejos}, N.; {Ryder}, S.D.; {Aggerwal}, K.; {Bhandari}, S.; {Day}, C.K.;
  et~al.
\newblock {Host Galaxy Properties and Offset Distributions of Fast Radio
  Bursts: Implications for Their Progenitors}.
\newblock {\em  Astrophys. J.} {\bf 2020}, {\em 903},~152.
  \href{http://xxx.lanl.gov/abs/2009.10747}{{
 }}
\newblock {{https://doi.org/10.3847/1538-4357/abb6fb}}.

\bibitem[{Pleunis} {et~al.}(2021){Pleunis}, {Good}, {Kaspi}, {Mckinven},
  {Ransom}, {Scholz}, {Bandura}, {Bhardwaj}, {Boyle}, {Brar}, {Cassanelli},
  {Chawla}, {(Adam) Dong}, {Fonseca}, {Gaensler}, {Josephy}, {Kaczmarek},
  {Leung}, {Lin}, {Masui}, {Mena-Parra}, {Michilli}, {Ng}, {Patel},
  {Rafiei-Ravandi}, {Rahman}, {Sanghavi}, {Shin}, {Smith}, {Stairs}, and
  {Tendulkar}]{Pleunis2021}
{Pleunis}, Z.; {Good}, D.C.; {Kaspi}, V.M.; {Mckinven}, R.; {Ransom}, S.M.;
  {Scholz}, P.; {Bandura}, K.; {Bhardwaj}, M.; {Boyle}, P.J.; {Brar}, C.;
  et~al.
\newblock {Fast Radio Burst Morphology in the First CHIME/FRB Catalog}.
\newblock {\em  Astrophys. J.} {\bf 2021}, {\em 923},~1.
  \href{http://xxx.lanl.gov/abs/2106.04356}{{
 }}
\newblock {{https://doi.org/10.3847/1538-4357/ac33ac}}.

\bibitem[{Marcote} {et~al.}(2020){Marcote}, {Nimmo}, {Hessels}, {Tendulkar},
  {Bassa}, {Paragi}, {Keimpema}, {Bhardwaj}, {Karuppusamy}, {Kaspi}, {Law},
  {Michilli}, {Aggarwal}, {Andersen}, {Archibald}, {Bandura}, {Bower}, {Boyle},
  {Brar}, {Burke-Spolaor}, {Butler}, {Cassanelli}, {Chawla}, {Demorest},
  {Dobbs}, {Fonseca}, {Giri}, {Good}, {Gourdji}, {Josephy}, {Kirichenko},
  {Kirsten}, {Landecker}, {Lang}, {Lazio}, {Li}, {Lin}, {Linford}, {Masui},
  {Mena-Parra}, {Naidu}, {Ng}, {Patel}, {Pen}, {Pleunis}, {Rafiei-Ravandi},
  {Rahman}, {Renard}, {Scholz}, {Siegel}, {Smith}, {Stairs}, {Vanderlinde}, and
  {Zwaniga}]{marcote2020repeating}
{Marcote}, B.; {Nimmo}, K.; {Hessels}, J.W.T.; {Tendulkar}, S.P.; {Bassa},
  C.G.; {Paragi}, Z.; {Keimpema}, A.; {Bhardwaj}, M.; {Karuppusamy}, R.;
  {Kaspi}, V.M.;  et~al.
\newblock {A repeating fast radio burst source localized to a nearby spiral
  galaxy}.
\newblock {\em Nature} {\bf 2020}, {\em 577},~190--194.
  \href{http://xxx.lanl.gov/abs/2001.02222}{{
 }}
\newblock {{https://doi.org/10.1038/s41586-019-1866-z}}.

\bibitem[{Piro} {et~al.}(2021){Piro}, {Bruni}, {Troja}, {O'Connor},
  {Panessa}, {Ricci}, {Zhang}, {Burgay}, {Dichiara}, {Lee}, {Lotti}, {Niu},
  {Pilia}, {Possenti}, {Trudu}, {Xu}, {Zhu}, {Kutyrev}, and
  {Veilleux}]{piro2021fast}
{Piro}, L.; {Bruni}, G.; {Troja}, E.; {O'Connor}, B.; {Panessa}, F.; {Ricci},
  R.; {Zhang}, B.; {Burgay}, M.; {Dichiara}, S.; {Lee}, K.J.;  et~al.
\newblock {The fast radio burst FRB 20201124A in a star-forming region:
  Constraints to the progenitor and multiwavelength counterparts}.
\newblock {\em Astron. Astrophys.} {\bf 2021}, {\em 656},~L15.
  \href{http://xxx.lanl.gov/abs/2107.14339}{{
 }}
\newblock {{https://doi.org/10.1051/0004-6361/202141903}}.

\bibitem[{Bhardwaj} {et~al.}(2021){Bhardwaj}, {Gaensler}, {Kaspi},
  {Landecker}, {Mckinven}, {Michilli}, {Pleunis}, {Tendulkar}, {Andersen},
  {Boyle}, {Cassanelli}, {Chawla}, {Cook}, {Dobbs}, {Fonseca}, {Kaczmarek},
  {Leung}, {Masui}, {Mnchmeyer}, {Ng}, {Rafiei-Ravandi}, {Scholz}, {Shin},
  {Smith}, {Stairs}, and {Zwaniga}]{bhardwaj2021nearby}
{Bhardwaj}, M.; {Gaensler}, B.M.; {Kaspi}, V.M.; {Landecker}, T.L.; {Mckinven},
  R.; {Michilli}, D.; {Pleunis}, Z.; {Tendulkar}, S.P.; {Andersen}, B.C.;
  {Boyle}, P.J.;  et~al.
\newblock {A Nearby Repeating Fast Radio Burst in the Direction of M81}.
\newblock {\em  Astrophys. J. Lett.} {\bf 2021}, {\em 910},~L18.
  \href{http://xxx.lanl.gov/abs/2103.01295}{{
 }}
\newblock {{https://doi.org/10.3847/2041-8213/abeaa6}}.

\bibitem[{Ravi} {et~al.}(2021){Ravi}, {Law}, {Li}, {Aggarwal},
  {Burke-Spolaor}, {Connor}, {Lazio}, {Simard}, {Somalwar}, and
  {Tendulkar}]{Ravi2021}
{Ravi}, V.; {Law}, C.J.; {Li}, D.; {Aggarwal}, K.; {Burke-Spolaor}, S.;
  {Connor}, L.; {Lazio}, T.J.W.; {Simard}, D.; {Somalwar}, J.; {Tendulkar},
  S.P.
\newblock {The host galaxy and persistent radio counterpart of FRB 20201124A}.
\newblock {\em arXiv} {\bf 2021},  arXiv:2106.09710.
  \href{http://xxx.lanl.gov/abs/2106.09710}{{
 }}

\bibitem[{Bhandari} {et~al.}(2022){Bhandari}, {Heintz}, {Aggarwal},
  {Marnoch}, {Day}, {Sydnor}, {Burke-Spolaor}, {Law}, {Xavier Prochaska},
  {Tejos}, {Bannister}, {Butler}, {Deller}, {Ekers}, {Flynn}, {Fong}, {James},
  {Lazio}, {Luo}, {Mahony}, {Ryder}, {Sadler}, {Shannon}, {Han}, {Lee}, and
  {Zhang}]{Bhandari2022}
{Bhandari}, S.; {Heintz}, K.E.; {Aggarwal}, K.; {Marnoch}, L.; {Day}, C.K.;
  {Sydnor}, J.; {Burke-Spolaor}, S.; {Law}, C.J.; {Xavier Prochaska}, J.;
  {Tejos}, N.;  et~al.
\newblock {Characterizing the Fast Radio Burst Host Galaxy Population and its
  Connection to Transients in the Local and Extragalactic Universe}.
\newblock {\em  Astron. J.} {\bf 2022}, {\em 163},~69.
  \href{http://xxx.lanl.gov/abs/2108.01282}{{
 }}
\newblock {{https://doi.org/10.3847/1538-3881/ac3aec}}.

\bibitem[{Caleb} {et~al.}(2018){Caleb}, {Spitler}, and
  {Stappers}]{caleb2018one}
{Caleb}, M.; {Spitler}, L.G.; {Stappers}, B.W.
\newblock {One or several populations of fast radio burst sources?}
\newblock {\em Nat. Astron.} {\bf 2018}, {\em 2},~839--841.
  \href{http://xxx.lanl.gov/abs/1811.00360}{{
 }}
\newblock {{https://doi.org/10.1038/s41550-018-0612-z}}.

\bibitem[{Palaniswamy} {et~al.}(2018){Palaniswamy}, {Li}, and
  {Zhang}]{palaniswamy2018there}
{Palaniswamy}, D.; {Li}, Y.; {Zhang}, B.
\newblock {Are There Multiple Populations of Fast Radio Bursts?}
\newblock {\em  Astrophys. J.} {\bf 2018}, {\em 854},~L12.
  \href{http://xxx.lanl.gov/abs/1703.09232}{{
 }}
\newblock {{https://doi.org/10.3847/2041-8213/aaaa63}}.

\bibitem[{Hessels} {et~al.}(2019){Hessels}, {Spitler}, {Seymour}, {Cordes},
  {Michilli}, {Lynch}, {Gourdji}, {Archibald}, {Bassa}, {Bower}, {Chatterjee},
  {Connor}, {Crawford}, {Deneva}, {Gajjar}, {Kaspi}, {Keimpema}, {Law},
  {Marcote}, {McLaughlin}, {Paragi}, {Petroff}, {Ransom}, {Scholz}, {Stappers},
  and {Tendulkar}]{hessels2019frb}
{Hessels}, J.W.T.; {Spitler}, L.G.; {Seymour}, A.D.; {Cordes}, J.M.;
  {Michilli}, D.; {Lynch}, R.S.; {Gourdji}, K.; {Archibald}, A.M.; {Bassa},
  C.G.; {Bower}, G.C.;  et~al.
\newblock {FRB 121102 Bursts Show Complex Time-Frequency Structure}.
\newblock {\em  Astrophys. J. Lett.} {\bf 2019}, {\em 876},~L23.
  \href{http://xxx.lanl.gov/abs/1811.10748}{{
 }}
\newblock {{https://doi.org/10.3847/2041-8213/ab13ae}}.

\bibitem[{Nimmo} {et~al.}(2021){Nimmo}, {Hessels}, {Keimpema}, {Archibald},
  {Cordes}, {Karuppusamy}, {Kirsten}, {Li}, {Marcote}, and
  {Paragi}]{nimmo2021highly}
{Nimmo}, K.; {Hessels}, J.W.T.; {Keimpema}, A.; {Archibald}, A.M.; {Cordes},
  J.M.; {Karuppusamy}, R.; {Kirsten}, F.; {Li}, D.Z.; {Marcote}, B.; {Paragi},
  Z.
\newblock {Highly polarized microstructure from the repeating FRB 20180916B}.
\newblock {\em Nat. Astron.} {\bf 2021}, {\em 5},~594--603.
  \href{http://xxx.lanl.gov/abs/2010.05800}{{
 }}
\newblock {{https://doi.org/10.1038/s41550-021-01321-3}}.

\bibitem[{CHIME/FRB Collaboration} {et~al.}(2020){CHIME/FRB Collaboration},
  {Amiri}, {Andersen}, {Bandura}, {Bhardwaj}, {Boyle}, {Brar}, {Chawla},
  {Chen}, {Cliche}, {Cubranic}, {Deng}, {Denman}, {Dobbs}, {Dong}, {Fandino},
  {Fonseca}, {Gaensler}, {Giri}, {Good}, {Halpern}, {Hessels}, {Hill},
  {H{\"o}fer}, {Josephy}, {Kania}, {Karuppusamy}, {Kaspi}, {Keimpema},
  {Kirsten}, {Landecker}, {Lang}, {Leung}, {Li}, {Lin}, {Marcote}, {Masui},
  {McKinven}, {Mena-Parra}, {Merryfield}, {Michilli}, {Milutinovic},
  {Mirhosseini}, {Naidu}, {Newburgh}, {Ng}, {Nimmo}, {Paragi}, {Patel}, {Pen},
  {Pinsonneault-Marotte}, {Pleunis}, {Rafiei-Ravandi}, {Rahman}, {Ransom},
  {Renard}, {Sanghavi}, {Scholz}, {Shaw}, {Shin}, {Siegel}, {Singh}, {Smegal},
  {Smith}, {Stairs}, {Tendulkar}, {Tretyakov}, {Vanderlinde}, {Wang}, {Wang},
  {Wulf}, {Yadav}, and {Zwaniga}]{amiri2020periodic}
{CHIME/FRB Collaboration}; {Amiri}, M.; {Andersen}, B.C.; {Bandura}, K.M.;
  {Bhardwaj}, M.; {Boyle}, P.J.; {Brar}, C.; {Chawla}, P.; {Chen}, T.;
  {Cliche}, J.F.;  et~al.
\newblock {Periodic activity from a fast radio burst source}.
\newblock {\em Nature} {\bf 2020}, {\em 582},~351--355.
  \href{http://xxx.lanl.gov/abs/2001.10275}{{
 }}
\newblock {{https://doi.org/10.1038/s41586-020-2398-2}}.

\bibitem[{Rajwade} {et~al.}(2020){Rajwade}, {Mickaliger}, {Stappers},
  {Morello}, {Agarwal}, {Bassa}, {Breton}, {Caleb}, {Karastergiou}, {Keane},
  and {Lorimer}]{rajwade2020possible}
{Rajwade}, K.M.; {Mickaliger}, M.B.; {Stappers}, B.W.; {Morello}, V.;
  {Agarwal}, D.; {Bassa}, C.G.; {Breton}, R.P.; {Caleb}, M.; {Karastergiou},
  A.; {Keane}, E.F.;  et~al.
\newblock {Possible periodic activity in the repeating FRB 121102}.
\newblock {\em Mon. Not. R. Astron. Soc.} {\bf 2020},
  {\em 495},~3551--3558.
  \href{http://xxx.lanl.gov/abs/2003.03596}{{
 }}
\newblock {{https://doi.org/10.1093/mnras/staa1237}}.

\bibitem[{Cruces} {et~al.}(2021){Cruces}, {Spitler}, {Scholz}, {Lynch},
  {Seymour}, {Hessels}, {Gouiff{\'e}s}, {Hilmarsson}, {Kramer}, and
  {Munjal}]{cruces2021repeating}
{Cruces}, M.; {Spitler}, L.G.; {Scholz}, P.; {Lynch}, R.; {Seymour}, A.;
  {Hessels}, J.W.T.; {Gouiff{\'e}s}, C.; {Hilmarsson}, G.H.; {Kramer}, M.;
  {Munjal}, S.
\newblock {Repeating behaviour of FRB 121102: Periodicity, waiting times, and
  energy distribution}.
\newblock {\em Mon. Not. R. Astron. Soc.} {\bf 2021}.
  {\em 500},~448--463  \href{http://xxx.lanl.gov/abs/2008.03461}{{
 }}
\newblock {{https://doi.org/10.1093/mnras/staa3223}}.



\end{thebibliography}
\reftitle{References}

\end{adjustwidth}
\end{document}